\documentclass[12pt]{article}
\usepackage{amssymb}
\usepackage{amsmath}
\def\be{\begin{equation}}
\def\bea{\begin{eqnarray}}
\def\ee{\end{equation}}
\def\eea{\end{eqnarray}}

\def\nref#1{(\ref{#1})}
\def\half{{1\over 2}}
\def\su{{\widetilde{SU}}}

\input{tcilatex}

\begin{document}

\input epsf

\begin{titlepage}
\hfill
 \begin{flushright}
 hep-th/0509235\\
 PUPT-2172
\end{flushright}

\vspace*{20mm}
\begin{center}
{\Large {\bf  Fivebranes from gauge theory }\\ }

\vspace*{15mm}
\vspace*{1mm}
{Hai Lin$^1$ and Juan Maldacena$^2$  }

\vspace*{1cm}

${}^1$ {\it Department of Physics, Princeton University, Princeton, NJ 08544}

\vspace*{0.4cm}

${}^2$ {\it Institute for Advanced Study, Princeton, NJ 08540}

\vspace*{1cm}
\end{center}

\begin{abstract}

We study theories with sixteen supercharges and a discrete energy
spectrum. One class of theories has symmetry group $SU(2|4)$. They
arise as truncations of ${\cal N}=4$ super Yang Mills. They
include the plane wave matrix model, 2+1 super Yang Mills
on $R \times S^2$ and ${\cal N}=4$ super Yang Mills
on $R \times S^3/Z_k$. We explain how to obtain their gravity duals in a
unified way. We explore the regions of the geometry that are
relevant for the study of some  1/2 BPS and near BPS states. This
leads to a class of two dimensional  (4,4) supersymmetric sigma models with
non-zero $H$ flux, including a massive deformed WZW model. We show
how to match some features of the string spectrum with the Yang
Mills theory.

 The other class of theories are also
connected to ${\cal N}=4$ super Yang Mills and arise by making
some of the transverse scalars compact. Their vacua are
characterized by a 2d Yang Mills theory or 3d Chern Simons theory.
These theories realize  peculiar superpoincare symmetry algebras
in 2+1 or 1+1 dimensions with ``non-central" charges.
We finally discuss gravity duals of ${\cal N}=4$ super Yang Mills
on $AdS_3 \times S^1$.

 \end{abstract}

\end{titlepage}

\vskip 1cm

\section{Introduction}
\renewcommand{\theequation}{1.\arabic{equation}}
\setcounter{equation}{0}

In this paper we study an interconnected family of theories with
sixteen supercharges. All these theories share the common feature
that they have a mass gap and a discrete spectrum of excitations.
In most examples  we have a dimensionless parameter which allows
us to interpolate between weak and strong coupling. In the weakly
coupled description we have a gauge theory. These theories have
many vacua. We describe smooth gravity solutions corresponding to
all these vacua. For some particular vacua we study the 't Hooft
limit and we examine the properties of strings at large 't Hooft
coupling.

In the first part of this paper we  study theories with 16
supercharges whose symmetry algebra is an  $SU(2|4)$ supergroup.
These theories are closely related to each other. Their BPS states
can be conveniently studied by a Witten index
\cite{wittenindex,justin}. The first example is the plane-wave
matrix model \cite{bmn}. The second example is  2+1 SYM on
$R\times S^2 $ \cite{mssvr} and a third example is ${\cal N} =4$
super Yang Mills on $R \times S^3/Z_k$. We construct their gravity
duals. We give a general method for constructing the gravity
solutions, and provide a few explicit solutions.

The plane wave matrix model is a nice example of the gauge
theory/gravity correspondence because it is an ordinary quantum
mechanical system with a discrete energy spectrum. The theory has
a large number of vacua, and a correspondingly large number of
gravity solutions. Strictly speaking we can trust the gravity
approximation only for a suitable subset of solutions. The generic
solution, though formally smooth,  has curvature of the order of
the planck or string scale. In the 't Hooft large $N$ limit we can
focus on just one of these vacua at a time and ignore the
tunneling to other vacua. The properties of single trace states
(or single string states) depend on the vacuum we are expanding
around. All the theories in this family have an $SO(6)$ symmetry.
It is possible to consider half BPS states which carry $SO(6)$
angular momentum $J$. We can count these BPS states precisely in
each of these theories. In addition, we can consider near BPS
states. Their description in the weakly coupled regime is similar
to the one in four dimensional ${\cal N}=4$ super Yang-Mills and
was studied in
\cite{markvr1,kimplefka,markvr2,kimpp,plefka,fourloop}. At large
't Hooft coupling, the spectrum of large charge near BPS states is
obtained by considering pp-wave limits of the general solutions.
In the simplest case we find a IIA plane wave
\cite{iiappwave1,iiappwave2}. In general, strings in lightcone
gauge are described by a massive field theory on the worldsheet
with (4,4) supersymmetry. The details of this theory depend on the
vacuum we are expanding around. We study vacua associated to NS5
branes \cite{mssvr}. In these vacua we are led to strings
propagating in the near horizon geometry of $N_5$ fivebranes
\cite{Callan:1991at}. The field theory on the string is given by a
massive deformation of the WZW model and linear dilaton theory
that describes the near horizon region of NS5 branes. Depending on
the value of $N_5$ we get a different spectrum. We match  some
qualitative features of this spectrum with  the weakly coupled
gauge theory description. The energies of near BPS states have a
non-trivial dependence on the 't Hooft coupling. So we expect a
non-trivial interpolation between the weak and strong coupling
results. In fact, for the plane wave matrix model, at weak
coupling, this interpolating function was computed to four loops
in \cite{fourloop}. We show that this function has a physical
interpretation in the strong coupling regime as the radius of a
fivesphere in the geometry.

In the second part of our paper we consider field theories that
have 16 supercharges and $SO(4) \times SO(4)$ symmetry. These
solutions are described by droplets of an incompressible fluid as
in \cite{llm}. When this fluid lives in an infinite two
dimensional plane we find the gravity solutions corresponding to
the 1/2 BPS states of ${\cal N}=4$ super Yang Mills
\cite{Berenstein:2004kk}, \cite{Corley}, \cite{llm}. In this paper
we discuss mainly the case where this fluid lives on a two torus.
In addition we discuss the case of the cylinder. These are again
theories that have 16 supercharges and many vacua. An interesting
aspect is that these theories have Poincare supersymmetry algebras
in $2+1$ or $1+1$ dimensions which are such that the charges
appearing on the right hand side are not central, a situation that
cannot arise for Poincare superalgebras in more than three
dimensions \cite{Haag:1974qh,weinberg}.
This algebra was mentioned in the general classification in
\cite{nahm}.  This is also the symmetry algebra that is linearly
realized in the light cone gauge description of strings moving in
the maximally supersymmetric IIB plane wave \cite{Blau:2001ne}.
The theory associated to fermions on a torus gives rise to
$U(N)_K$ or $U(K)_N$ Chern Simons theory on $R \times T^2$ in the
IR. The full theory has explicit duality under $K \leftrightarrow
N$.

We also discuss another family of smooth solutions that are
obtained by doing an analytic continuation of the ansatz in
\cite{llm}. The boundary conditions are different in this case.
These solutions are associated to a certain Coulomb branch of the
${\cal N} =4$ super Yang Mills theory  on $AdS_3 \times S^1$.

This paper is organized as follows. In section two we discuss theories
with 16 supercharges and $SU(2|4)$ symmetry group. We start by discussing various
field theories and then proceed to write the gravity description for all of these
examples. We also take the large $J$ limit and analyze features of the BPS and
near BPS spectrum of single trace (or single string) states. In section three
we discuss some features of theories with sixteen supercharges that are obtained
from free fermions on a $T^2$ and also by analytic continuation of some of
the formulas in \cite{llm}.
Finally, various appendices give more details about some of the results.

\section{Theories with 16 supercharges and
${ \widetilde {SU}}(2|4)$ symmetry group.}
\renewcommand{\theequation}{2.\arabic{equation}}
\setcounter{equation}{0}

\subsection{The field theories}

In this subsection we discuss various
field theories with ${ \widetilde {SU}}(2|4)$ symmetry group.

It is convenient to start with  ${\cal N}=4$ super Yang Mills on $R \times S^3$.
This theory is dual to $AdS_5 \times S^5$ and its symmetry group is the
superconformal group $SU(2,2|4)$. The bosonic subgroup of the superconformal group is
$SO(2,4) \times SO(6)$. It is convenient to focus on an $SU(2)_L \subset SO(4) \subset
SO(2,4)$. This $SU(2)_L$ is embedded in the $SO(4)$ symmetry group that rotates the
$S^3$ on which the field theory is defined. If we take the full superconformal
algebra and we truncate it to the subset that is invariant under $SU(2)_L$ we clearly
get a new algebra. This algebra forms the supergroup ${\widetilde {SU}} (2|4)$, where the
tilde here denotes that we take its universal cover. In other words, the bosonic subgroup is
$ R \times SU(2) \times SU(4)$ \footnote{ If we replace $R$ by $U(1)$ we get the compact
from of $SU(2|4)$.}.
This is the symmetry group of the theories we are going to consider below.\footnote{ This
symmetry group also appears when we consider 1/2 BPS states in $AdS_{4,7} \times S^{7,4}$
M-theory solutions \cite{llm}. A closely related
supergroup, $SU(2,2|2)$, is the ${\cal N} =2$ superconformal group in 4 dimensions.}

We will get the theories of interest by quotienting ${\cal N} = 4$ super Yang Mills
by various subgroups of $SU(2)_L$. For example, if we quotient by the whole $SU(2)_L$
group we are left with the plane wave  matrix model \cite{plefka}.
We get a reduction to 0+1 dimensions because
all Kaluza Klein modes on $S^3$ carry $SU(2)_L$ quantum numbers except for the lowest
ones. The other theories are obtained by quotienting by $Z_k$ and $U(1)_L$ subgroups of
$SU(2)_L$.
We will discuss these theories in detail below.\footnote{Notice that
this truncation procedure is a convenient way to construct the lagrangian, but
 we cannot get the full quantum spectrum of  the plane wave matrix model by
restricting to $SU(2)_L$ invariant states of the full ${\cal N}=4$ super Yang Mills
theory.}

\begin{figure}[htb]
\begin{center}
\epsfxsize=3in\leavevmode\epsfbox{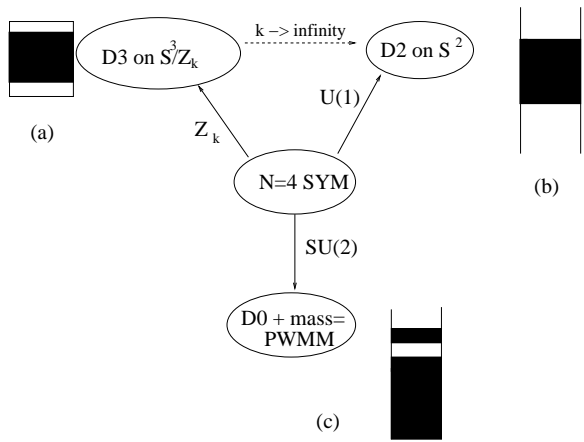}
\end{center}
\caption{ Starting from four dimensional ${\cal N}=4$ super Yang Mills
and truncating by various subgroups of $SU(2)_L$ we get various theories with
${ \widetilde{SU}}(2|4)$ symmetry. We have indicated the diagrams in the
$x_1,x_2$ space that determine their gravity solutions. The $x_1, x_2$ space
is a cylinder, with the vertical lines identified for (b) and (c)  and it is
a torus for (a).
} \label{cyltorus}
\end{figure}

Since all theories have a common symmetry group they share some properties.
One  property that we will discuss in some detail
 are  1/2 BPS states carrying
$SO(6)$ angular momenta. These are states carrying energy $E$
equal to the angular momentum under an $SO(2) \subset SO(6)$
generated by $J$. The condition $E=J$ is the BPS bound. The fact
that these 1/2 BPS states are fully protected follows from the
discussion in \cite{markvr2,kimpp}. Moreover, the arguments in
\cite{markvr2,kimpp} allow us to count precisely these BPS states.
Actually, to study BPS states it is convenient to define the index
\cite{justin} \be \label{defindfull} I(\beta_i) =Tr\left[ (-1)^F
e^{ - \mu (E - 2 S -J_1 - J_2 - J_3 )} e^{- \beta_1(E-J_1)} e^{-
\beta_2(E-J_2)} e^{- \beta_3(E-J_3)} \right] \ee where $S=S_3$ is
one of the generators of $SU(2)$, $J_1 = M_{12}$, $J_2 = M_{34}$
and $J_3 = M_{56}$ are $SO(6)$ Cartan generators. Let us explain
why \nref{defindfull} is an index. Let us consider the supercharge
$Q^\dagger = Q^\dagger_{-+++}$, where the indices indicate the
charges under $(S,J_1,J_2,J_3)$. This supercharge has $E=1/2$.
 This supercharge and its adjoint
obey the anticommutation relation \be \label{anticommf} \{
Q,Q^\dagger \} = {\cal U}  \equiv E - 2 S -(J_{1} + J_{2} + J_{3}
) \ee In addition the combinations $E-J_i$  commute with the
supercharges in \nref{anticommf}. Using the standard arguments
(see \cite{wittenindex}) any state with nonzero values of ${\cal
U}$ does not contribute to \nref{defind}. By evaluating
\nref{defindfull} we will be able to find which BPS
representations should remain as we change the coupling. The index
\nref{defindfull} contains the same information as the indices
defined in \cite{markvr2}, see \cite{justin}.
 For further discussion see
appendix \ref{indexapp}.
In order to count $1/2$ BPS states we can use a simplified version of
\nref{defindfull} obtained by taking the limit when $\beta_3 \to \infty$.
In this limit the index depends only on $q \equiv e^{-\beta_1 - \beta_2}$
\be \label{defind}
I_N(q) \equiv \sum_{J=0}^\infty D(N,J) q^{J}  =
 \lim_{\beta_3 \to + \infty} I(\beta_1,\beta_2,\beta_3)
~,~~~~~~~~~q = e^{-\beta_1 - \beta_2} \ee where $J=J_3$.  This
partition function counts the number of 1/2 BPS states $D(N,J)$ in
the system. Below we will compute \nref{defind} for
 various theories. We will not compute \nref{defindfull} in this paper, but
it could be computed using the techniques in \cite{minwalla}.

In the limit that $ 1\ll J \ll N$ we will identify these states as massless
geodesics in the geometric description. Notice that, even though
 we use some of the techniques in \cite{llm} to describe the
{\it vacua} of these theories, we do not include backreaction when
we consider 1/2 BPS states.\footnote{ The 1/2 BPS states of the
theories considered in this paper preserve less supersymmetry than
the 1/2 BPS states that were considered in \cite{llm}. In other
words, the 1/2 BPS states of \cite{llm} preserve the same amount
of supersymmetry as the {\it vacua} (which have $J=0$) of the
theories considered in this paper. Here we start with  theories
with 16 supercharges, while \cite{llm}  started with theories with
32 supercharges. } We are also going to study the near BPS limit,
with $J $ large and ${\hat E}=E-J$ finite.   For excitations along
the $S^5$   the one loop perturbative correction is the same as in
the ${\cal N}=4$ parent Yang Mills theory. On the gravity side, we
will find that, at strong 't Hooft coupling, the  result that
differs from the naive extrapolation of the weak coupling results.
This implies that there exist some interpolating functions in the
sectrums. We could similarly study other solutions with large
quantum numbers under $SO(6)$, such as the configurations
considered in e.g. \cite{Frolov,MinahanVE} (see \cite{beisert} for
a review) which have several large quantum numbers. In this theory
we could also have BPS and near BPS  configurations with large
$SO(3)$ spin, which we will not study in this paper.

In these theories we have many vacua and, in principle, we can tunnel among the
different vacua. In most of the discussion we will assume that we are in a regime
in parameter space where we can neglect the effects of tunneling. This tunneling
is suppressed in the 't Hooft regime where strings are weakly coupled.
Note that despite tunneling the vacua remain degenerate since they all contribute
positively to the   index \nref{defind}.

We will now discuss in more detail each theory individually.

\subsubsection{${\cal N} =4$ SYM on $R\times S^3/Z_k$ }
\label{sthreezk}

Here we consider $U(N)$ ${\cal N}=4$ super Yang-Mills theory on
$R\times S^3/Z_k$, with $Z_k \subset SU(2)_L$, and $SU(2)_L$ as
defined above (see also \cite{Horowitz:2001uh} for a more general
discussion). We can also obtain this theory by starting with the
free field content of  ${\cal N}=4$, projecting out all fields
which are not invariant under $Z_k$  and then considering the same
interactions for the remaining fields as the ones we had for
${\cal N}=4$. Notice that we first project the elementary fields
and we then quantize, which is not the same as retaining the
invariant states of the original full quantum  ${\cal N}=4$
theory. This is the standard procedure. The symmetry group of this
theory is ${\widetilde{SU}}(2|4)$.

This theory is parametrized by $N$, $k$, and the original Yang
Mills coupling $g^2_{YM3}$. Whereas ${\cal N}=4 $ SYM on $S^3$ has
a unique vacuum, the theory  on $S^3/Z_k$ has many supersymmetric
vacua. Let us analyze the vacua at weak coupling. Since all
excitations are massive we
 can neglect all fields except for a Wilson line of the gauge field.
More precisely, the vacua are given by the space of flat
connections on $S^3/Z_k$. This space is parametrized by giving the
holonomy of the gauge field $U$ along the non-trivial generator of
$\pi_1(S^3/Z_k)=Z_k$, up to gauge transformations. We can
therefore diagonalize $U$, with $U^k =1$. So the diagonal elements
are $k$th roots of unity. Inequivalent elements are given by
specifying  how many roots of each kind we have. So the vacuum is
specified by giving the $k$ numbers $n_1, n_2, \cdots n_k$, with
$N = \sum_{l=1}^k n_l $. Where $n_l$ specifies how many times
$e^{i 2 \pi { l \over k}}$ appears in the diagonal of $U$. We can
also view these different vacua as arising from orbifolding the
theory of D-branes on $S^3 \times R$ and applying the rules in
\cite{gmmd} with different choices for the embedding of the $Z_k$
into the gauge group. The regular representation corresponds to
$n_l = N/k$ for all $l$, and we need to take $N$ to be a multiple
of $k$.

 The
total number of vacua is then \be \label{totnumb} D(N,k) = {
(N+k-1)! \over (k-1)! N! } \ee


It is also interesting to count the total number of 1/2 BPS states with charge $J$ under
one of the $SO(6)$ generators.   These  numbers are encoded conveniently in the
partition function
\be \label{stzth}
I_{S^3/Z_k}(p,q) = \sum_{N=0}^{\infty} p^N I_N(q) =
\sum_{N,J=0}^{\infty}D_{S^3/Z_k}(N,J)p^{N}q^{J}
=[I_{{\cal N}=4} (p,q)]^{k}=\frac{1}{{\prod}^{\infty}_{n=0}(1-pq^{n})^k}
\ee
where $ I_{{\cal N}=4} (p,q)$ is the index for    ${\cal N} = 4$ super Yang-Mills.
As an aside,
note that the degeneracy of states in ${\cal N}=4$ super Yang Mills can be written
in various equivalent forms \cite{Corley,Boulatov:1991fp}\footnote{We
have not seen the last equality \nref{last_I} in recent papers, but it must be well known.}
\bea
\sum_{ N, J=0}^{\infty} p^N q^{N^2/2} q^J D(N,J) &=& \prod_{n=1}^\infty (1 + p q^{n -\half} )
\\
\sum_{J=0}^{\infty} D(N,J) q^J & =& { 1 \over \prod_{n=1}^N(1-q^n) }
\\
I_{{\cal N}=4}(p,q) \equiv  \sum_{N,J=0}^{\infty} p^N   q^J D(N,J) &=& { 1 \over \prod_{n=0}^\infty (1 - p q^{n } ) }
\label{last_I}
\eea
In the first form we express it as a system of fermions in a harmonic oscillator
potential. In the third form it looks like a system of bosons in a harmonic oscillator
potential. In writing \nref{stzth} we used the last representation in \nref{last_I}.

We see that even though we counted the vacua \nref{totnumb} at weak coupling, the
result is still valid at strong coupling since they all contribute to the Witten
index. In fact, setting $q=0$ in \nref{stzth} we recover \nref{totnumb}.

\subsubsection{ 2+1 SYM on $R\times S^2$ }
\label{d2theory}

This field theory is constructed  as follows. We start with ${\cal
N}=4$ super Yang Mills on $R\times S^3$ and we truncate the free
field theory spectrum to states that are invariant under $U(1)_L
\subset SU(2)_L $, where $SU(2)_L$ is one of the $SU(2)$ factors
in the $SO(4)$ rotation group of the $S^3$. This results in a
theory that lives in one less dimension. It is a theory living on
$R \times S^2$. This theory was already considered in \cite{mssvr}
by considering the fuzzy sphere vacuum of the plane wave matrix
model and then taking a large $N$ limit that removed the fuzzyness
and produced the theory on the ordinary sphere. Here we reproduce
it as a $U(1)_L$  truncation from ${\cal N}=4$ super Yang
Mills\footnote{We write the metric of $R \times S^3$ as $ds^{2} =
-dt^{2}+\frac{1}{4}[ d\theta ^{2}+\sin ^{2}\theta d\phi ^{2} +(
d\psi +\cos \theta d\phi ) ^{2}]$, where $\theta \in \lbrack 0,\pi
],\phi \in \lbrack 0,2\pi ],\psi \in \lbrack 0,4\pi ]$. We neglect
the $\psi$ dependence of all fields and we write the gauge field
in {\cal N}=4 SYM as $A_{{\cal N}=4}=A+\Phi(d\psi +\cos \theta
d\phi )$, where $A$ is the $2+1$ dimensional gauge field. }
\begin{eqnarray}
S &=&{\frac{1}{g_{YM2}^{2}}}\int dt{\frac{d^{2}\Omega }{\mu ^{2}}}\ \mathrm{tr}%
\left( -{\frac{1}{4}}F^{mn}F_{mn}-{\frac{1}{2}}(D_{m}X^{a})^{2}-{\frac{1}{2}}%
(D_{m}\Phi)^{2}+{\frac{i}{2}}\bar \Psi \Gamma^{m} D_{m} \Psi \right.  \notag \\
&&
+{\frac{1}{2}} \bar \Psi
\Gamma ^{a}[X^{a},\Psi ]+{%
\frac{1}{2}}\bar \Psi \Gamma^{\Phi} [\Phi,\Psi ]+{\frac{1}{4}}[X_{a},X_{b}]^{2}+{\frac{1}{%
2}}[\Phi,X^{a}]^{2} -{\frac{\mu ^{2}}{8}}X_a^2 \notag \\
&&\left. -{\frac{\mu ^{2}}{2}}{\Phi}^{2}-{\frac{%
3i\mu }{8}} \bar \Psi \Gamma ^{012\Phi}\Psi -\mu \Phi dt \wedge F
\right)    \label{dtwotheoryjm}
\end{eqnarray}
where $m= 0,1,2$, $a = 4, \cdots, 9$ and
$(\Gamma^{m},\Gamma^\Phi,\Gamma^a)$ are ten dimensional gamma
matrices. We see that out of the seven transverse scalars of the
maximally supersymmetric Yang Mills theory we select one of them,
$\Phi$, which we treat differently than the others. This breaks
the $SO(7)$ symmetry to $SO(6)$ while still preserving sixteen
supercharges.   The radius of $S^2$ has size $\mu^{-1}$ and we
have used the two dimensional metric with this radius to raise and
lower the indices in \nref{dtwotheoryjm}. For our purposes it is
convenient to set $\mu=2$, since this is the value we obtain by
doing the $U(1)_L$ truncation of ${\cal N}=4$ super Yang-Mills on
an $S^3$ of radius one.

The vacua are obtained by considering zero energy states. We write
the field strength along the directions of the sphere as $F = f
d^2 \Omega$. We then see that $\Phi$ and $f$ combine into a
perfect square in the lagrangian \be \label{sqen} -\half ( f + \mu
\Phi )^2
 \ee
For zero energy vacua this should be set to zero.
 Since the values of $f$ are
 quantized, so are the values of the $\Phi$ field at these vacua.
We can first diagonalize $\Phi$ and then we can see that its entries are integer
valued. So a vacuum is characterized by giving the value of $N$ integers
$n_1, \cdots n_N$. The number of vacua is infinite, so we will not write an index.
Nevertheless we will see that the gravity solutions reflect the existence of
these vacua.

The dimensionless parameters characterizing this theory are $N$
and the value of the 't Hooft coupling at the scale of the two
sphere $ g^2_{eff} N \equiv { 2\pi g^2_{YM2} N \over \mu }  $,
where $\mu^{-1}$ is the size of the sphere. The size of the sphere
is a dimensionful parameter which just sets the overall energy
scale. We set $\mu = 2$, so that the energy of BPS states with
angular momentum $J$ in $SO(6)$ is equal to $E=J$.

Notice that the large $k$ limit of the theory analyzed in section
\ref{sthreezk} gives us the theory analyzed here. The values of
$N$ are the same and \be \label{effcoup}
g^2_{eff} N = {2 \pi g^2_{YM2} N \over \mu}= { g_{YM3}^2
N k }
 \ee
 where $g_{YM3}$ is the Yang Mills coupling in the original
${\cal N} =4$ theory in section \ref{sthreezk}. So we see that the
limit involves taking $k \to \infty, ~g_{YM3}^2  \to 0$ while
keeping $g^2_{YM2}$ fixed.

If one takes the strong coupling limit of this theory, by taking
$g_{YM2} \to \infty$, we expect to get the theory living on M2
branes on $R \times S^2$. This theory has 32 supersymmetries and
is the familiar theory associated with $AdS_4 \times S^7$. In this
limit we find that the theory has full $SO(8) $ symmetry. When we
perform this limit we find that the energy $E$ of the theory in
this section goes over to $\Delta - \tilde J$, where $\Delta$ is
the ordinary Hamiltonian for the M2 brane theory on $R \times S^2$
and $\tilde J$ is the $SO(2)$ generator in $SO(8)$ which commutes
with the $SO(6)$ that is explicitly preserved by
\nref{dtwotheoryjm}. For a single brane, the $N=1$ case,  this can
be seen explicitly by dualizing the gauge field strength into an
eighth scalar. Then the vacua described around \nref{sqen} are
related to the 1/2 BPS states of the M2 brane theory. These should
not be confused with the 1/2 BPS sates of the 2+1 dimensional
theory \nref{dtwotheoryjm} which would be related to 1/4 BPS
states from the M2 point of view.

\subsubsection{Plane wave matrix model}

Finally we will discuss the plane wave matrix model e.g.
\cite{bmn,markvr1,kimplefka,markvr2,kimpp,mssvr,plefka,fourloop}.
This arises by truncating the ${\cal N}=4$ theory to 0+1
dimensions by keeping all free field theory states that are
invariant under $SU(2)_L$ and keeping the same interactions for
these states that we had in ${\cal N}=4$ super Yang Mills
\cite{plefka}. We keep the zero modes for $SO(6)$ scalars and
truncate the gauge field to $A_{{\cal N}=4}=X_{1}\omega
_{1}+X_{2}\omega _{2}+X_{3}\omega _{3}$, where $\omega_{i}$ are
three left invariant one-forms on $S^3$. Thus the $X_i$ are the
scalars that transform under $SO(3)$.

This theory has many vacua. These vacua are obtained by setting
the scalars $X_i$ equal  to $SU(2)$ Lie algebra generators. In
fact the vacua are in one to one correspondence with $SU(2)$
representations of dimension $N$. Suppose that we have $N(n)$
copies of the irreducible representation of dimension $n$ such
that \be \label{bmnd0} N = \sum_n N(n) n \ee Each choice of
partition of $N$ gives us a different vacuum. So the number of
vacua is equal to the partitions of $N$, $P(N)$.

 ${\cal N}=4$ super Yang Mills has a unique vacuum. On the other
 hand,  any solution
of the plane wave matrix model can be uplifted to a zero energy
solution of ${\cal N}=4$ super Yang Mills. What do  the  various
plane wave matrix model vacua correspond to  in ${\cal N}=4$ super
Yang Mills? It turns out that these are simply large gauge
transformations of the ordinary vacuum. The solutions uplift to
$A_{{\cal N}=4}=(\omega _{1}J_{1}+\omega _{2}J_{2}+\omega
_{3}J_{3})= -i\left( dg\right)g^{-1}$, were $g$ is an $SU(2)$
group element in the same representation as the $J_i$. This
$SU(2)$ group is parameterizing the $S^3$.   So they are pure
gauge transformations from $A_{{\cal N}=4}=0$. In summary, in
${\cal N}=4$ super Yang Mills these different configurations are
related by a gauge transformation. The gauge transformation is not
$SU(2)_L$ invariant, even though the actual configurations are
$SU(2)_L$ invariant. In the plane wave matrix model they are gauge
inequivalent.


As in \cite{mssvr}, it is possible to get the 2+1 theory in
subsection \ref{d2theory} from a limit in which we take $\tilde N$
copies of the representation of dimension $n$ and we take $n\to
\infty$. For finite $n$ we get a $U(\tilde N)$ theory on a fuzzy
sphere and in the $n\to \infty $ limit the fuzziness goes away
\cite{mssvr}.

One can also count
 the total number of 1/2 BPS states with $SO(6)$ charge $J$. These
 are given by the
partition function \be \label{indppw}
I_{PWMM}(p,q)=\sum_{N,J=0}^{\infty} D_{PWMM}(N,J)p^{N}q^{J}
={\prod}^{\infty}_{m=1} I_{{\cal N}=4} (p^{m},q)=
\frac{1}{{\prod}^{\infty}_{m=1}{\prod}^{\infty}_{n=0}(1-p^{m}q^{n})}
\ee Setting $q=0$ we get that the number of vacua are given by the
partitions of $N$. It is interesting to estimate the large $J$ and
$N$ behavior of this index. We obtain \be D_{PWMM}(N,J) \sim e^{
(3.189...) (NJ)^{1/3} } \ee where we assumed $J^2/N \gg 1$, $N^2/J
\gg 1$. The fact that this is symmetric under $N\leftrightarrow J$
follows from the fact that \nref{indppw} is symmetric under
$p\leftrightarrow q$ up to the $n=0$ factor.

\subsection{Dual gravity solutions}
\label{mgravity}

All the theories that we have discussed above have the same
supersymmetry group. All gravity solutions with this symmetry were
classified in  \cite{llm}. The bosonic symmetries, $R\times SO(3)
\times SO(6)$, act geometrically. The first generator implies the
existence of a Killing vector associated to shifts of a coordinate
$t$. In addition we have an $S^2 $ and an $S^5$ where the rest of
the bosonic generators act. Thus the solution depends only on
three variables $x_1,x_2, y$. The full geometry can be obtained
from a solution of the 3 dimensional Toda equation \be
\label{todaeq} ( \partial_{x_1}^2 + \partial_{x_2}^2) D +
\partial_y^2 e^D =0 \ee It turns out that $y = R_{S^2} R_{S^5}^2
\geq 0$ where $R_{S^i}$ are the radii of the two spheres. In order
to have a non-singular solution we need special boundary
conditions for the function $D$ at $y=0$.
 In fact,  the $x_1,~x_2$ plane could be divided into regions  where the
function $D$ obeys two different boundary conditions \bea
\nonumber e^D  \sim y && ~~{\rm for~~ } y \to 0 ~,~~~~~~~S^5 \to
0~,~~~~~~~~{\rm M5~ region}
\\ \label{standbc}
 \partial_y D =0 && ~~{\rm at~~} y=0 ~,~~~~~~~~~S^2 \to 0 ~,~~~~~~~~{\rm M2~region}
\eea see \cite{llm} for further details. The labels M2 and M5
indicate that in these two regions either a two sphere or a five
sphere shrinks to zero in a smooth fashion.
 There are, however,  no explicit branes in the geometry. We have a smooth solution
with fluxes. However, we
can think of these regions as arising from a set  of M2 or M5 branes that wrap
the contractible sphere.
A bounded region of each type in the $x_1,x_2$ plane implies that
we have a cycle in the geometry with a flux related to the corresponding
type of brane (see \cite{llm} for further details).

\begin{figure}[htb]
\begin{center}
\epsfxsize=2.5in\leavevmode\epsfbox{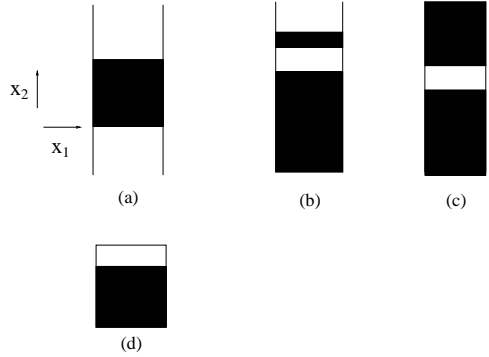}
\end{center}
\caption{
Translational invariant configurations in the $x_1,x_2$ plane which give rise
to various gravity solutions. The shaded regions indicate M2 regions and the unshaded
ones indicate M5 regions.
 The two vertical lines are identified. In (a) we
see the configuration corresponding to the vacuum of the 2+1 Yang
Mills on $R\times S^2 $ with unbroken gauge symmetry. In (b) we
consider a configuration corresponding to a vacuum of the plane
wave matrix model. In (c) we see a vacuum of the NS5 brane theory
on $R \times S^5$. Finally, in (d) we have a droplet on a two
torus in the $x_1,x_2$ plane. This corresponds to a vacuum of the
${\cal N}=4$ super Yang Mills on a $R \times S^3/Z_k$.
 }\label{transinv}
\end{figure}

The different theories discussed above are  related to different choices for
the topology of the $x_1, x_2$ plane. In addition,  for each topology the
asymptotic distribution
of M2 and M5   regions can be  different. See figure \ref{cyltorus}.
Let us consider some examples. If we choose the $x_1 , x_2$ plane to be a two torus, then
we get a solution that is dual to the vacua of the ${\cal N}=4$ super Yang Mills
on $R \times S^3/Z_k$, see figure \ref{cyltorus}(a).
If the topology is a cylinder, with $x_1$ compact and the
M2 region is localized in the $x_2$ direction, we have a solution dual to a
vacuum of the 2+1 Yang Mills theory on $R \times S^2$, see figure \ref{cyltorus}(b).
 If we
choose a cylinder and we let the M2 region extend all the way to $x_2 \to -\infty$, and
the M5 region extend to $x_2 \to + \infty $, and also there
are localized M2, M5 strips in between,
then we get a solution which is dual to a vacuum of the plane wave
 matrix model, see figure \ref{cyltorus}(c).
Finally, if we consider a cylinder and we have M5 regions that are
localized (see figure \ref{transinv}(c)) then we get a
solution that is dual to an NS5 brane theory on $R \times S^5$,
we will came back to this case later.

In principle, we could consider configurations that are not translation
invariant, as long as we consider configurations defined on a cylinder or torus as is
appropriate. In this paper we will concentrate
 on configurations that are translation invariant
along $x_1$. These will be most appropriate in the regime of parameter space
where the 11th direction is small and we can go to a IIA description. So we focus on
 the region in parameter space where  the string coupling is small and the effective
't Hooft coupling is large.
If the configuration is translation
invariant in the $x_1$ direction
we can transform the non-linear equation \nref{todaeq} to a linear equation through
the following change of variables \cite{ward}
\bea \label{vardefi}
 && y = \rho \partial_{\rho} V ~,~~~~~~~~~   x_2 = \partial_{\eta} V ~,~~~~~~~e^D = \rho^2
\\
 && { 1\over \rho} \partial_{\rho} ( \rho \partial_{\rho} V) + \partial_{\eta}^2 V =0
\label{veque} \eea
So we get the Laplace equation in three dimensions
for an axially symmetric system\footnote{The angular
direction of the three dimensional space
is not part of the 10 or 11 dimensional spacetime coordinates. }.
The fact that one can obtain solutions in this fashion was observed in \cite{llm} and
some singular solutions were explored in \cite{null}. Below we will find the
precise boundary conditions for $V$ which ensure that we have a smooth solution.

Let us now translate the  boundary conditions  \nref{standbc} at $y=0$ into certain
boundary conditions for the function $V$. In the region where
$e^D \sim y$ at $y\sim 0$,  all that we require is that $V$ is regular at $\rho=0$,
in the three dimensional sense.
On the other hand if $y=0$ but $\rho \not =0$, then we need to impose that
$\partial_y D =0 $. This is proportional to
\be
0= \frac{1}{2} \partial_{y} D= \rho { \partial \rho \over \partial y} = - \frac{
\partial_{\eta }^{2}V}{\left( \partial_{\eta }\partial_{\rho }V\right)^{2}+
\left( \partial_{\eta }^{2}V\right) ^{2}}
\ee
We conclude that $\partial_{\eta}^2 V =0$. Equation
 \nref{veque} then implies that
  $\partial_{\rho}^2 V =0$.
Therefore the curve $y =0,~ \rho \not =0$, or $\partial_{\rho} V
=0$, is at constant values of $\eta$, since the slope of the curve
defined by $\partial_{\rho} V =0$ is $\frac{\delta \eta }{\delta
\rho }=\frac{-\partial _{\rho }^{2}V}{\partial _{\eta }\partial
_{\rho }V}=0$.

\begin{figure}[htb]
\begin{center}
\epsfxsize=2in\leavevmode\epsfbox{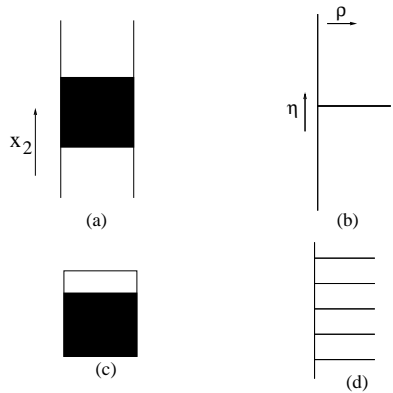}
\end{center}
\caption{
 Electrostatic problems corresponding to different droplet configurations. The shaded
 regions (M2 regions) correspond to disks and the unshaded regions
  map to $\rho =0$. Note that the $x_1$ direction in (a), (c) does not correspond
  to any variable in (b), (c).
The rest of the $\rho, \eta$ plane corresponds to $y>0$ in the
$x_2,y$ variables. In (a),(b) we see the configurations
corresponding to a vacuum of 2+1 super Yang Mills on $R\times
S^2$. In (c),(d) we see a configuration corresponding to a vacuum
of ${\cal N}=4$ super Yang Mills on $R\times S^3/Z_k$. In (d) we
have a periodic configuration of disks. The fact that it is
periodic corresponds to the fact that we have also compactified
the $x_2$ direction.
 } \label{disks}
\end{figure}

If we interpret $V$ as the potential of an electrostatics problem,
then $-\partial_{\rho} V$ is the electric field along the
$\rho$ direction. The condition that it vanishes corresponds to the
presence of a charged conducting surface.
So the problem  is reduced to an axially symmetric
electrostatic configuration in three dimensions
where we have conducting disks that are sitting at positions $\eta_i$ and have
radii $\rho_i$. See figure \ref{disks}. These disks are in an external electric field
which grows at infinity.
If we considered such conducting disks in a general configuration we would find that
the electric field would diverge as we approach the boundary of the disks.
In our case this cannot happen, otherwise the coordinate $x_2$ would be ill defined
at the rim of the disks.
So we need to impose the additional constraint that the electric field is
finite at the rim of the disks. This implies that
the charge density vanishes at the tip of the disks.
This condition relates the charge on the disks
$Q_i$ to the radii of the disks $\rho_i$. So for each disk we can only  specify two
independent parameters, its position $\eta_i$ and its total charge $Q_i$.
The precise form of the background electric  field depends on the
 theory we consider (but not on the particular vacuum) and it is
 fixed by demanding that
 the change of variable \nref{vardefi} is well defined.
The relation between the translation invariant
 droplet configurations in the $x_1,x_2$ plane and the disks can be seen
in figure \ref{disks}.

Since we are focusing on solutions which are translation invariant along
$x_1$ it is natural to compactify this direction and write the solution in
IIA variables. This procedure will make sense as long as we are in a region of
the solution where the IIA coupling is small  (see \cite{Itzhaki:1998dd} for a similar
discussion).

\begin{figure}[htb]
\begin{center}
\epsfxsize=4.5in\leavevmode\epsfbox{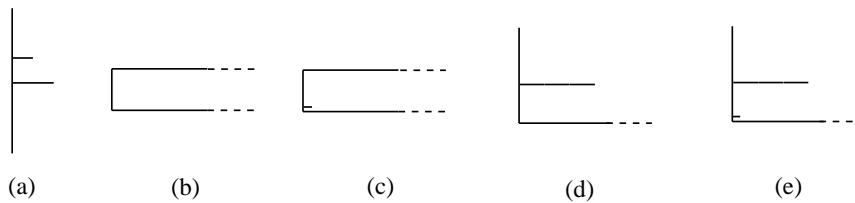}
\end{center}
\caption{  In (a) we see a configuration which corresponds to a
vacuum of 2+1 super Yang Mills on $R\times S^2$. In (b) we see the
simplest vacuum of the theory corresponding to the NS5 brane on $R
\times S^5$. In this case we have two infinite conducting disks
and only the space between them is physically meaningful. In (c)
we have another vacuum of the same theory. If the added disk is
very small and close to the the top or bottom disks the solution
looks like that of (b) with a few D0 branes added. In (d) we see a
configuration corresponding to a vacuum of the plane wave matrix
model. In this case the disk at $\eta=0$ is infinite and the
solution contains only the region with $\eta \geq 0$. In (e) we
have another vacuum of the plane wave matrix model with more
disks.} \label{bubbling}
\end{figure}

The M-theory form of the solutions can be found in \cite{llm}. We
obtain the string frame solution
 \begin{eqnarray}
ds_{10}^{2} &=& \left( \ddot V - 2 \dot V \over - V'' \right)^{1/2}
\left\{ - 4 { \ddot V \over \ddot V - 2 \dot V } dt^2 + { - 2 V'' \over \dot V} ( d\rho^2
+ d\eta^2 )  + 4 d\Omega_5^2 + 2{ V'' \dot V \over   \Delta }d\Omega_2^2
\right\} \notag \\
\label{IIA ansatz}
e^{4\Phi  } &=&{ 4 ( \ddot V - 2 \dot V)^3 \over - V'' \dot V^2
\Delta^2 }  \\
C_{1} &=&- { 2 \dot V' \, \dot V \over \ddot V - 2 \dot V } dt \\
F_{4} &=&d C_3  ,\quad \quad \quad
 \quad \quad C_{3}=- 4  { \dot V^2 V'' \over   \Delta } dt\wedge d^{2}\Omega , \\
H_{3} &=&d B_2 ~,~~~~~~~~~~~~~B_{2}   = 2 \left( { \dot V \dot V' \over   \Delta} + \eta \right) d^{2}\Omega
\\
  \Delta &\equiv& (\ddot V - 2 \dot V) V'' - ( \dot V')^2
  \label{IIA ansatz last}
\end{eqnarray}
where the dots indicate derivatives with respect to $\log \rho$
and the primes indicate derivatives with respect to $\eta$.
$V(\rho, \eta)$ is a solution of the Laplace equation
\nref{veque}. For regular solutions, we need to supplement it by
boundary condition specified by a general configuration of lines
in $(\rho, \eta)$ plane, like in figure \ref{bubbling}.

Before we get into the details of particular solutions we would like to
discuss some general properties. First note that if we take a random solution
of \nref{veque} we will get singularities.
In order to prevent them, we need to be a bit careful. As we explained above we need a
solution of an electrostatic problem involving horizontal conducting disks. In addition
we need to ensure the positive-definiteness of various
 metric components, i.e. $\Delta \leq 0$ and $ V'' \leq 0 $,
$\ddot V - 2 \dot V \geq 0$, $\dot V \geq 0$.
 This is obeyed everywhere if we choose appropriate
boundary conditions for the potential at large $\rho, \eta$. These boundary conditions
imply that there is a background
 electric field that grows as we go to large $\rho , \eta$. For example, if
 we consider a configuration such as the one in figure \ref{disks}(b), the disk is in
 the presence of a background potential of the form $V_b \sim \rho^2 - 2 \eta^2 $.
This background electric field is the same for all vacua, e.g. it is the same in
figures \ref{disks}(b) and \ref{bubbling}(a).
 For the plane wave matrix
model we have an infinite conducting surface at $\eta=0$ and only the region
$\eta\geq 0$ is physically significant. In this case the
 background potential is
 $V_b \sim \rho^2 \eta - \frac{2}{3} \eta^3$. In addition we
 have  finite size disks as seen, for example, in figure
 \ref{bubbling}(d) or \ref{bubbling}(e).
 In appendix A we show that for the configurations we talk about in
 this paper \nref{IIA ansatz}-\nref{IIA ansatz last} gives
a regular solution. We also show that the dilaton is non-singular
and that $g_{tt}$ never becomes zero for the solutions we
consider. This ensures that the solutions we have have a mass gap.
This follows from the fact that the warp factor never becomes zero
so that we cannot decrease the energy of a state by moving it into
the region where the warp factor becomes zero. In principle, this
argument does not rule out the presence of a small number of
 massless or tachyonic
modes. The latter are, of course, forbidden by supersymmetry. A
massless mode would not change the energy of the solution, so it
would preserve supersymmetry. On the other hand, once we quantize
the charges on the disks we do not have any continuous parameters
in our solutions. So we cannot have any massless modes.
Of course, this agrees with the field theory
expectations since all theories we consider have a mass gap around
any of the vacua.

Note that a rescaling of $V$ leaves the ten dimensional metric and
$B$ field invariant but rescales the dilaton and the RR fields.
This just corresponds to the usual symmetry of the IIA
supergravity theory under rescaling of the dilaton and RR fields.
There is second symmetry corresponding to rescaling $\rho, \eta$
and $V$ which corresponds to the usual scaling symmetry of gravity
which scales up the metric and the forms according to their
scaling dimensions. This allows us to put in two parameters in
\nref{IIA ansatz}-\nref{IIA ansatz last} such as an overall charge
and the value of the dilaton at its maximum.

More interestingly,
we can vary the number
of disks, their charges and the  distances between each other.
See figure \ref{bubbling}.
These parameters are related to different choices of vacua for the different
configurations.

\begin{figure}[htb]
\begin{center}
\epsfxsize=1.7in\leavevmode\epsfbox{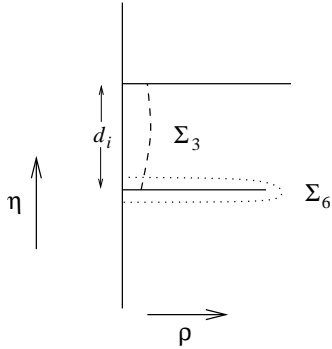}
\end{center}
\caption{ We see a configuration associated to a pair of disks.
$d_i$ indicates the distance between the two nearby disks. The
dashed line in the $\rho,\eta$ plane, together with the $S^2$ form
a three cycle $\Sigma_3$ with the topology of an $S^3$. The dotted
line, together with the $S^5$ form a six cycle $\Sigma_6$ with the
topology of an $S^6$.
 } \label{diskone}
\end{figure}

All the solutions we are discussing, contain an $S^2$ and an $S^5$
and these can shrink to zero at various locations. Using these it
is possible to construct three cycles and six cycles respectively
by tensoring the $S^2$ and $S^5$ with lines in the $\rho, \eta $
plane. These  translate into three cycles and six cycles in the
IIA geometry. See figure \ref{diskone}.
 We can then measure the flux of $H_3$ over the three cycle and call it
$N_5$ and we can measure the flux of $*{\tilde F}_4$ on the six
cycle and call it $N_2$. Using \nref{IIA ansatz}-\nref{IIA ansatz
last} or the formulas in \cite{llm}  we can write them as
\begin{eqnarray}\label{charge1}
N_{2}=\frac{1}{\pi ^{3}l_{p}^{6}}\int e^{D}dx_{2}\int
dx_{1}=\frac{2}{\pi ^{2}}  \int_{0}^{\rho _{i}}\rho ^{2}\partial
_{\rho }\left(\partial _{\eta }V|_{\eta _{i}^{+}}-\partial _{\eta
}V|_{\eta _{i}^{-}}\right) d\rho=\frac{8Q_i}{\pi ^{2} }
\end{eqnarray}
and
\begin{eqnarray}\label{charge2}
N_{5}=\frac{1}{2\pi ^{2}l_{p}^{3}}\int y^{-1}e^{D}dx_{2}\int
dx_{1}=\frac{1}{\pi}\int_{\eta _{i}+d_{i}}^{\eta _{i}}\frac{\rho
}{\partial _{\rho }V }\partial _{\eta }^{2}V|_{\rho =0}d\eta
=\frac{ 2d_{i}}{\pi }
\end{eqnarray}
In deriving \nref{charge2} we used that near $\rho \to 0$ we can
expand $V = f_0(\eta) + \rho^2 f_1(\eta) + \cdots$ and we used the
equation for $V$ \nref{veque} to relate $f_{1}(\eta)$ to $V''$. We
set $\alpha^{\prime}=1$ and $l_p =1$ for convenience.
The quantization conditions
\nref{charge1},\nref{charge2} show that $N_5$ is
proportional to the distance between neighboring disks $d_i$ and that
$N_2$ is proportional to  the total charge of each disk $Q_i$.
When we solve the electrostatic problem we need
to ensure that these parameters are quantized. Strictly speaking
the flux given by $N_2$ is quantized only after we quantize the
four form field strength.

The topology of the solutions is related to the topology of the
disk configurations.
In other words, the number of six cycles and three
cycles is related to the number of disks and the number of line
segments in between, but is independent of the size of the disks
or the distance between the disks.

As we discussed above we will be interested in   BPS
excitations with angular momentum on $S^5$. For large, but not too
large, angular momentum  these are well described by lightlike
particles moving in the background \nref{IIA ansatz}-\nref{IIA
ansatz last} with angular momentum $J$ along the $S^5$. In order
to minimize their energy, these lightlike geodesics
 want to sit at a point in the $\rho,~\eta$ space  where
\be  \label{bpsmass}
 { |g_{tt}| \over g_{55} } =  { \ddot V \over \ddot V - 2 \dot V} \geq 1
\ee
is minimized, where $\sqrt{g_{55}}$ is the radius of the five sphere.
 It turns out that this is minimized at the tip of the disks, where
 the inequality in \nref{bpsmass}
 is saturated\footnote{In the eleven dimensional description the point where \nref{bpsmass} is
 minimized lies on the $y=0$ plane at a local maximum of
 $e^D|_{y=0}$ in the $x_1,x_2$ plane.}.
This corresponds to saturating the BPS condition $E\geq |J|$. In
fact, in order to minimize \nref{bpsmass} we would like to set
$\dot V=0$. This occurs at $\rho=0$ and on the surface of the
disks. However, in these cases, also $\ddot V =0$. Expanding the
solutions near these regions we find that \nref{bpsmass} actually
diverges at $\rho =0$, this is because $S^5$ shrinks at $\rho=0$.
On the disks, \nref{bpsmass} is bigger than one, except at the tip
where it is one. See appendix A for a more detailed discussion.

\begin{figure}[htb]
\begin{center}
\epsfxsize=2.0in\leavevmode\epsfbox{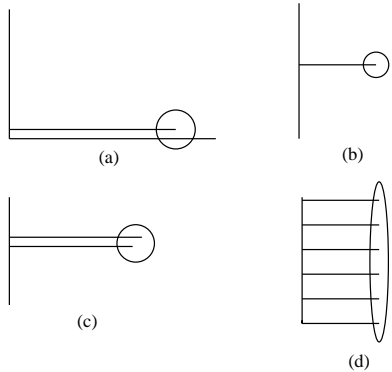}
\end{center}
\caption{ In this figure we see the expansion around the region
near the tip of the disks. In a generic situation the tip we focus
on is isolated, see (b). In other cases, there are other disks
nearby that sit close to the tip we are focusing on. In this case
we can take a limit where we include the nearby tips. We see such
situations in (a),(c) and (d). (d) corresponds to the periodic
case. We can focus on a distance that is large compared to the
period in $\eta$ but small compared to the size of the disks.
 } \label{expansion}
\end{figure}

In order to find the behavior of the solution near these geodesics  we
  expand the solution of the electrostatic problem near the tip of
the disks. Near the tip of the disks we have a simple Laplace
equation in two dimensions. Namely, we approximate the disk by an
infinite half plane.
 We can then solve the problem by doing  conformal
transformations. Actually, we can do this whenever we are
expanding around a solution at large $\rho_0$ and we are
interested in features arising at distances which are much smaller
than $\rho_0$, but could be larger than the distances between
disks, see figure \ref{expansion}. So let us first analyze this
problem in general. We can define the complex coordinate \be
\label{compl}
 z \equiv \xi + i \eta \equiv \rho - \rho_0  + i \eta
 \ee
 so that we are expanding around the point $(\rho, \eta) = (\rho_0 , 0)$.
It is actually convenient for our problem to define a complex
variable \be \label{wdefi} w(z) =  2 \partial_z V =    ({ y \over
\rho_0} - i x_2 ) \ee where we also used an approximate form of
\nref{vardefi}. Equation \nref{veque} implies that $w$ is a
holomorphic function of $z$. We see that $w$ is defined on the
right half plane: $Re(w)\geq 0$. Equation \nref{veque} is simply
the statement that the change of variables is holomorphic.
Solutions are simply given by finding a conformal transformation
that maps the $w$ half-plane into a configuration in $z$-plane
containing various cuts of lines specified by a general
configuration, like those in figure \ref{expansion}.

For example we could take $ z = w^2 $. This maps the
$w$ half-plane into the $z$ plane
with a cut running on the negative real axis. More explicitly, this leads to
$V \sim  Re(  z^{3/2} )  $. This is the solution near the tip of a disk, see figure
\ref{expansion}(b).

Once we have found this map we can go back to the general ansatz
\nref{IIA ansatz}-\nref{IIA ansatz last} and write the resulting
answer. When we do this we note that $\dot V \sim \partial_{\xi}
V/\rho_0$ and that $\ddot V \sim \partial_{\xi}^2 V/\rho_0^2$.
Since $\rho_0$ is very large in our limit we keep only the leading
order terms in $\rho_0$. After doing this we find the approximate
solution \footnote{The rest of the fields, i.e. the dilaton and
fluxes are the the same as in \nref{ppgen}-\nref{ppgen last}, with
$t=x^+$.}
\begin{eqnarray}
ds_{10}^2 &\sim & 4 \rho_0 \left\{ -   ( 1 +  { 1 \over \rho_0} f^{-1} | \partial_w z|^2
 ) dt^2 +   d\Omega_5^2  +
{ f \over \rho_0} \left[ dw d\bar w + ( { w + \bar w \over 2})^2 d\Omega_2^2 \right] \right\}
\label{ppblim}
\end{eqnarray}
where $ f = { \partial_w z + \partial_{ \bar w} {\bar z}  \over 2  ( w + \bar w) }$.

Let us first  consider the specific case where $ z = w^2 $. This
describes the configuration near the tip of the disks. In this
case we find that $f=1$ and the metric in the four dimensional
space parametrized by $w,\bar w , \Omega_2$ is flat. In addition,
we see that \nref{bpsmass} is indeed saturated at $w=0$.

Now let us go back to \nref{ppblim} and  take a general pp-wave
limit. We will take $\rho_0\to \infty$ and scale out the overall
factor $\rho_0$ away from the solution.
In other words, we parameterize $S^5$ as \be d \Omega_5^2 =
d\varphi^2 \cos^2\theta + d\theta^2 + \sin^2 \theta d\Omega_3^2
\sim d\varphi^2( 1 - {{ \vec r}^{\,\, 2} \over 4 \rho_0} ) + { 1
\over 4 \rho_0} d {\vec r}^{\, 2} \ee where we expanded around $ {
r \over \sqrt{ 4 \rho_0} } = \theta \sim 0$ and kept $\vec r$
finite in the limit. In addition, we set \bea dt &=& dx^+
~,~~~~~~~~~ d\varphi = dx^+ - { 1 \over 4 \rho_0 } dx^-
\\
- p_+ &=& E- J ~,~~~~~~~~~~~ - p_- = { J \over 4 \rho_0}  \label{rescalxm}
\\
 4 \rho_0 & =& {R^2_{S^5}  } \label{rfiverho}
\eea
where the second line tells us how the generators transform and finally the
last line is stating  that the parameter $\rho_0$ is physically the size of
the $S^5$ (we have set $\alpha'=1$).

After this pp-wave limit is taken for \nref{IIA ansatz}-\nref{IIA
ansatz last}, the solution takes the form
\begin{eqnarray}
ds_{10}^2 &=&   2 dx^+ dx^-  -   ( 4 f^{-1} | \partial_w z|^2 + \vec r^{\, 2} ) (dx^+)^2
 + d { \vec r}^{\, 2} +
  {4  f }   (  dw d\bar w + ( { w + \bar w \over 2})^2 d\Omega_2^2 ) \notag\\
e^{2 \Phi} &=& 4 f   \label{ppgen}
\\
B_2 &=& i  \left[  { (w + \bar w) \over 2 } ( \partial_w z -
\partial_{\bar w} {\bar z} )
-( { z - \bar z   }) \right] d^2 \Omega
\\
C_1 &=&  i ( w + \bar w ) { ( \partial_w z- \partial_{\bar w} { \bar z} ) \over (
\partial_w z + \partial_{\bar w} { \bar z} ) } d x^+
\\
C_3 &=& - ( w + \bar w)^3 f dx^+ \wedge d^2 \Omega
\\
f &\equiv &  { \partial_w z + \partial_{ \bar w} {\bar z}  \over 2
( w + \bar w) } \label{ppgen last}
 \end{eqnarray}
where $z$ is a holomorphic function of $w$. This is an exact solution of
IIA supergravity. When a string is
quantized in lightcone gauge on this pp wave it leads to a $(4,4)$
supersymmetric lightcone lagrangian, which will be discussed in
section \ref{sigmamodel}. One can also introduce two parameters by
rescaling $z$ and $w$. Similar classes of IIB pp-wave solutions
and their sigma models were analyzed and classified in e.g.
\cite{mm}, \cite{rt}, \cite{bs}.

For the single tip solution
\be
z = w^2 \label{singletip}
\ee
we  get
\be \label{iiapp}
 ds_{10}^2 =  - 2 dx^+ dx^-  -   ( \vec r^{\,\, 2} +4 \vec u^{\,\, 2}  ) (dx^+)^2
 + d { \vec r}^{\, \,2} + d {\vec u}^{\,\, 2}
 \ee
 where ${\vec r}$ and ${\vec u}$ each parameterize $R^4$.
This is a IIA plane wave with $SO(4) \times SO(3)$ isometry and it was
considered before in \cite{iiappwave1},\cite{iiappwave2}.

In conclusion, the expansion of the metric around the trajectories
of BPS particles locally looks like a IIA plane wave \nref{iiapp}
if the tip of the disk is far from other disks. When it is close
to other disks we need to use the more general expression
\nref{ppgen}-\nref{ppgen last}. We will analyze in detail specific
cases in section \ref{tips}. In the limit that we boost away the
$g_{++}$ component of the metric, the solution
\nref{ppgen}-\nref{ppgen last} becomes $R^{5,1}$ times a
transverse four dimensional part of the solution which is a
superposition of NS5 branes. Notice that $f$ is a solution of the
Laplace equation in the four dimensions parametrized by $w,\bar w,
\Omega_2$. This is related to the fact that we should interpret
the space between two closely spaced disks as being produced by
NS5 branes. This will become more clear after we analyze specific
solutions in e.g. section \ref{s3zk}.

The rescaling of $J$ in \nref{rescalxm}  has some physical
significance since it will appear when we express the energy of
near BPS states in terms of $J$. In other words, the light cone
hamiltonian for a string on the IIA plane wave  describes massive
particles propagating on the worldsheet. Four of the bosons have
mass 1 and the other four have mass 2.
The lightcone energy for
 each particle of momentum $n$
and mass $m$ is \be (E-J)_n = (- p_+ )_n =    \sqrt{ m^2 + {   n^2
\over  p_-^2 } } =  \sqrt{ m^2 +  { R_{S^5}^4 }
 { n^2 \over J^2} } ~,~~~~~~~\alpha'=1 \label{ppspectrum}
\ee where the masses of the worldsheet fields are $m = 1,2$
depending on the type of scalar or fermion that we consider on the
worldsheet. The subindex $n$ reminds us that this is the
contribution from a particle with a given momentum along the
string. Since the total momentum along the string should vanish,
we need to have more than one particle carrying momentum,  each
giving rise to a contribution similar to \nref{ppspectrum}.
Note
that the form of the spectrum is completely universal for all
solutions, as long as the tip is far enough from other disks.
On the the other hand the  value of  $\rho_0$ at the tip depends on
the details of the solution. It
 depends not only on the
theory we consider but also on the particular vacuum that we are
expanding around. In the following sections we will compute the
dependence of $\rho_0$ on the particular parameters of each
theory for some specific vacua.

When we can isolate a single disk we can always take pp-wave limit
of the solution to the IIA plane wave \nref{singletip},
\nref{iiapp} near the tip of this single disk. There are many
other situations when nearby disks are very close, and we need to
include also the region between disks, i.e. the region produced by
NS5 branes. In these cases, the geometry parametrized by the
second four coordinates $w,\bar w, \Omega_2$ is more complicated.
We will discuss it in following sections.

As is usual in the gravity/field theory correspondence one has to
be careful about the regime of validity of the gravity solutions,
and in our case, we should also worry about the following. In the
field theory we have many vacua. So we can have tunnelling between
the vacua. On the gravity side we have the same issue, we can
tunnel between different solutions of the system. In order to
understand this tunnelling problem it is instructive to consider
vacua whose solutions are very close to the original solution.
 Small deformations
of a given solution that still preserve all the supersymmetries
can be obtained, in the 11d language by considering small
``ripples" in the regions connecting M2 and M5 regions. In the IIA
description these become D0 branes. For very small excitations
these D0 branes sit at $\rho =0$ at the position of the disks. At
these positions it costs zero energy to add the D0 branes. In the
electrostatic description we are adding a small disk close to the
large disk, as in figure \ref{bubbling}(c). In order to estimate
the tunnelling amplitude we need to understand how we go from a
configuration with no D0 branes to a configuration with D0 branes.
In a region where we have a finite size three cycle $\Sigma_3$
(see figure \ref{diskone}) with flux $N_5$ we can create $N_5$ D0
branes via a D2 instanton that wraps the $\Sigma_3$ (see
\cite{gmjmns}). We see that such processes will be suppressed if
the string coupling in this region is small and the $\Sigma_3$ is
sufficiently large.


In the following subsections we discuss specific solutions.

\subsubsection{ Solution for NS5 brane theory on $ R \times S^5$}
\label{nsfivesec}

We start with this solution because it is the simplest from the
gravity point of view. In this case we consider two infinite disks
separated by some distance $d \sim N$, see figure
\ref{bubbling}(b). We find that the solution corresponds to $N$
IIA NS5 branes wrapping a $ R \times S^5$. The solution for $V$ is
\be \label{bessol} V  =  I_0(r ) \sin \theta  ~,~~~~~~~~~~ r = { 2
\rho \over N \alpha'} ~,~~~~\theta =
 { 2 \eta \over N \alpha' }
\ee where $I_0 (r)$ is a modified Bessel function of the first
kind. This leads to the ten dimensional
solution\footnote{$s=\sin\theta, c=\cos\theta$. We set
$\alpha^{\prime}=1$ in this paper. We used the convention in
\cite{Polchinski} that
 ${1 \over 2\pi\alpha^{\prime}} \int_{\Sigma_3} H_3 =2\pi N $, to normalize $H_3$.
 }
\begin{eqnarray}
 ds_{10}^{2} &=&N \left[ -2r \sqrt{\frac{I_{0}}{I_{2}}}dt^{2}+2r \sqrt{%
\frac{I_{2}}{I_{0}}}d\Omega _{5}^{2}+\sqrt{\frac{I_{2}}{I_{0}}}\frac{I_{0}}{%
I_{1}}(dr ^{2}+d\theta ^{2})+\sqrt{\frac{I_{2}}{I_{0}}}\frac{%
I_{0}I_{1}s^{2}}{I_{0}I_{2}s^{2}+I_{1}^{2}c^{2}}d{\Omega }_{2}^{2}\right]  \notag
\\ \label{ns5sol}
B_{2} &=& N \left[ \frac{-I_{1}^{2}cs}{I_{0}I_{2}s^{2}+I_{1}^{2}c^{2}}+\theta \right]
 d^{2}\Omega  \\
e^{\Phi } &=& g_0 N^{3/2} 2^{-1}\left( \frac{I_{2}%
}{I_{0}}\right) ^{\frac{3}{4}}\left( \frac{I_{0}}{I_{1}}\right) ^{\frac{1}{2}%
}\left( I_{0}I_{2}s^{2}+I_{1}^{2}c^{2}\right) ^{-\frac{1}{2}} \\
C_{1} &=&- g_0^{-1} { 1 \over N } 4  \frac{I_{1}^2 c}{I_{2}}dt \\
C_{3} &=&- g_0^{-1} \frac{4I_{0}I_{1}^{2}s^{3}}{I_{0}I_{2}s^{2}+I_{1}^{2}c^{2}}%
dt\wedge d^{2}\Omega  \label{ns5sol last}
\end{eqnarray}
where ${I_n} (r)$ are a series of modified Bessel functions of the
first kind.

This solution is also a limit of the a solution analyzed in
\cite{llm} using 7d gauged supergravity, except that here we
solved completely the equations. The gauged supergravity solution
in \cite{llm} describes an elliptic M5 brane droplet on the
$x_1,x_2$ plane and we can take a limit that the long axis of the
ellipse goes to infinity while keeping the short axis finite, this
becomes a single M5 strip. This then corresponds to two infinite
charged disks in the electrostatic configuration, see figures
\ref{bubbling}(b). We discuss more details of this relation in
appendix C.

The solution is dual to little string theory (see e.g.
\cite{lst1}, \cite{lst2}) on $R \times S^5$. As we go to the large
$r$ region the solution \nref{ns5sol}-\nref{ns5sol last}
asymptotes to
\begin{eqnarray}
ds_{10}^{2} &=& N \alpha ^{\prime }\left[ 2r \left(
-dt^{2}+d\Omega _{5}^{2}\right) +dr ^{2}+(d\theta^2+ \sin^2 \theta
d\Omega _{2}^{2})\right]  \notag \\
e^{\Phi } &=& g_{s} e^{-r } \label{5-brane-near}\\
H_{3}&=& 2N\alpha ^{\prime }\sin ^{2}\theta d\theta \wedge
d^{2}\Omega \notag
\end{eqnarray}
So we see that the solution asymptotes to IIA NS5 branes on $R \times S^5$.
In addition we have RR fields which are growing exponentially when we go to large $r$.
These fields break the $SO(4)$ transverse rotation symmetry of the fivebranes to $SO(3)$.
Since the coupling is also varying exponentially, it turns out that,  in the end,
the influence of the RR fields on the metric
is suppressed only by powers of $1/r$ relative to the terms that we have kept in
\nref{5-brane-near} (relative to the $H$ field terms for example).

The solution is everywhere regular. When either $S^5$ or $S^2$
shrinks, it combines with $r$ or $\theta$ to form locally $R^6$ or
$R^4$. Note that at $r=0$ the solution has a characteristic
curvature scale given by $ {\cal R} \sim {1\over \alpha' N}$ and a
string coupling of a characteristic size $g_s \sim g_0 N^{3/2}$.
The string coupling decreases as we approach the boundary. Thus,
if we take $g_s$ small and $N$ large we can trust the solution
everywhere. On the other hand if we take $g_s$ large, then we can
trust the solution for large $r$ but for small $r$ we need to go
to an eleven dimensional description, include $x_1$ dependence and
solve equation \nref{todaeq}. It is clear from the form of the
problem that for very large $g_s$ we will recover $AdS_7 \times
S^4$ in the extreme IR if we choose a suitable droplet
configuration. More precisely, as increase $g_s$ we will need to
go to the eleven dimensional description and include dependence on
$x_1$. Then we can consider a periodic array of circular droplets.
As $g_s \to \infty$ each circle becomes the isolated circle that
gives rise to $AdS_7 \times S^4$ \cite{llm}. There is also a
similar gravity picture for the relation between the 2+1 SYM on $R
\times S^2$ in section \ref{d2theory} and the 3d superconformal M2
brane theory.

In addition we could consider other solutions in the disk picture
that correspond to adding more small disks between the infinite
disks, as in figure \ref{bubbling}(c). These  correspond to
different vacua of this theory.


\subsubsection{Solution for 2+1 SYM on $ R \times S^2 $ }

This solution corresponds to a single disk, as in figure
\ref{disks}(b). This disk is in the presence of a background field
$V_b \sim \rho^2 - 2 \eta^2 $. The solution is a bit harder to
obtain. We have obtained it by combining our ansatz with the
results in \cite{Chong:2004ce}, as explained in appendix B.
The resulting 10 dimensional solution is
\begin{eqnarray}
 ds_{10}^{2} &=& \lambda^{1/3} \left[
 -8(1+r^{2})f dt^{2}+16{f}^{-1}\sin ^{2}\theta d\Omega _{5}^{2}+\frac{8r f }{%
r+(1+r^{2})\arctan r}\left( \frac{dr^{2}}{1+r^{2}}+d\theta ^{2}\right) \right.
 \notag \\
&& \left. +\frac{2r\left[r+(1+r^{2})\arctan r\right] f }{1+r\arctan r}d{\Omega }_{2}^{2}
\right]
 \label{d2sol}
\\
\quad B_{2} &=&- \lambda^{1/3} \frac{2\sqrt{2}\left[ r+(-1+r^{2})\arctan r
\right] \cos \theta }{1+r\arctan r}d^{2}\Omega  \\
e^{\Phi } &=& g_0 \lambda^{1/2}  8 r^{\frac{1}{2}}(1+r\arctan
r)^{-\frac{1}{2}} [ r+(1+r^{2})\arctan r]^{-\frac{1}{2}}f
^{-\frac{1}{2}}  \\
 C_{1} &=&- g_0^{-1} \lambda^{ - { 1 \over 3} }
\frac{\left[ r+(1+r^{2})\arctan r\right] \cos \theta }{2r}dt
\\
C_{3} &=&-g_0^{-1}
\frac{r [ r+(1+r^{2})\arctan r] ^{2}f ^{2}}{\sqrt{2}(1+r \arctan r) }dt\wedge d^{2}\Omega
\\
f &\equiv& \sqrt{\frac{2}{r}[r+ (\cos ^{2}\theta +r^{2})\arctan
r]}   \label{d2sol last}
\end{eqnarray}%
where $\lambda$ and $g_0$ are some constants. Here we have plugged
in expression (\ref{D2_toda}) in Appendix B.

This solution is dual to the vacuum of the 2+1 SYM in section
\ref{d2theory}, with $\Phi=0$ and unbroken $U(N)$ gauge symmetry.
The topology of this solution is $R \times B^3 \times
S^6$, where the boundary of $B_3$ is the $S^2$ on which the field
theory is defined. Solutions with other configurations of disks
have different topology.
  The solution is also everywhere regular. Expanding
for large $r$ we find that \nref{d2sol}-\nref{d2sol last}
approaches the D2 brane solution\footnote{Here we have D2 brane on
$R \times S^2$, where the radius of the $S^2$ is $\frac{1}{\mu}$,
and we set $\mu=2$.} \cite{Itzhaki:1998dd} on $R \times S^2$
\begin{eqnarray}
\frac {ds_{10}^{2}}{\alpha ^{\prime }} &=& ({  6 \pi g^{2}_{YM 2}N
})^{1/3} \left[
 r^{5/2}
 (-dt^{2}+\frac{1}{4}d{ {\Omega} }_{2}^{2})+  { dr^{2} \over r^{5/2} }
+ {r^{-1/2}} ( d\theta ^{2}+\sin ^{2}\theta d\Omega_{5}^{2} )  \right] \label{d2_near}
 \notag \\
e^{\Phi } &=& g_{YM2}^2 ( 6 \pi g^{2}_{YM 2}N   )^{-1/6} r^{-5/4}   \\
C_{3} &=&-g_{YM2}^{-2 } r^{5}  ( 6 \pi g^{2}_{YM 2}N   )^{-2/3} {
1 \over 4}  dt\wedge d^{2}{\Omega} \notag
\end{eqnarray}
Comparing with \nref{d2sol}-\nref{d2sol last} we can compute the
value of $\lambda$ and $g_0$ in terms of Yang Mills quantities. We
can then compute the value of the radius of $S^5$ at $r=0$,
$\theta = \pi/2$. This is the point where the BPS geodesics moving
along $S^5$ sits. We find \be \label{s5d2} { R_{S^5}^2 \over
\alpha' } =\left( \frac{6\pi^{3}g_{YM2}^{2}N}{\mu }\right) ^{1/3}
~,~~~~~~\mu=2 \ee The metric expanded around a geodesic with
momentum along $S^5$ is simply the plane wave \nref{iiapp}. We can
now insert \nref{s5d2} in the general expression \nref{ppspectrum}
to derive the spectrum of near BPS excitations with large $J$.

Note that the leading correction to ${\hat E}=E-J$ for
fluctuations in the transverse  directions in the $S^5$, which are
parametrized by $\vec r$ in \nref{iiapp},  has the form \be (E-
J)_n = 1 + {1\over 2} \left( \frac{6\pi^{3}g_{YM2}^{2}N}{\mu
}\right) ^{2/3} { n^2 \over J^2}+... \ee This is the large
coupling result from gravity approximation.

Under general principles we expect that the leading order
correction in the large $J$ limit in all regimes of the coupling
constant should go like \be (E- J)_n = 1 + f\left( { g^2_{YM2} N
\over \mu} \right) { n^2 \over J^2 } + \cdots \ee At weak coupling
we get basically the same answer we had for ${\cal N}=4$, which at
one loop order is $f\left( { g^2_{YM2} N \over \mu} \right) ={\pi
g^2_{YM2} N \over \mu} $. So we see that in this case the function
$f$ has to be non-trivial.
This is to be contrasted with the behavior in
four dimensional ${\cal N}=4$ theory where the function $f$ has the same form at weak and
strong coupling  \cite{sz}, see also \cite{Gross:2002su}.
Of course it would be very nice to compute this interpolating function from
the gauge theory side.
We will see a similar phenomenon for the plane wave matrix model
 in section \ref{bmnmm}. This phenomenon is a generic feature of the
 strong/weak coupling problem, among many
  others observed in the literature, e.g. the 3/4
 problem in the thermal Yang-Mills entropy \cite{3quaters}, and the 3-loop disagreement of the near
 plane wave string spectrum \cite{3loops}, which are results obtained in different regimes
  of couplings, and are probably explained by the presence of such
 interpolating functions.

We can have other more general solutions corresponding to multiple
disks, as in figure \ref{bubbling}(a). The different
configurations in the disk picture match the different
Higgs vacua for scalar $\Phi$ as we discussed in section
\ref{d2theory}. One can also consider strings propagating near the
tip in a multi-disk solution. In that case, the actual value of
the interpolating function $f$ in the strong coupling regime,
which is related to the position of the tip of the disk, is not
universal, in the sense that it depends on the vacuum we expand
around. What is universal, however, is the fact that the expansion
around any of the tips gives us the IIA plane wave \nref{iiapp} as
long as there are no other  nearby disks. The situation
when we consider many disks together will be discussed in the next
section.

\subsubsection{Solutions for two or  more nearby tips}
\label{tips}

If there are nearby disks, then we can expand the solution near
the tips of these disks and also include the fivebrane region
between them. Consider for example a configuration with two nearby
disks such as shown in figure \ref{expansion}(c). The holomorphic
function $z(w)$ in \nref{ppgen}-\nref{ppgen last} is given by \be
 \partial_w z = { (w- i a)(w + i b) \over w}  \label{twodisks}
 \ee
 with $a, b$ real and positive.
 We see that for $w \approx ia, -ib$ and for $w \to \infty $ we recover the results
 we expect for single disks (\ref{singletip}).
This transformation maps the $w$ right half plane (with $Re(w)\geq
0$) to the $z$ plane with two cuts. The points $w=ia, -ib$ map to
the two tips and $w=0$ maps to $Re(z) \rightarrow -\infty$ between
the two disks, which is expected to look like a fivebrane. In
fact, we can  check that the function $f$ in
\nref{ppgen}-\nref{ppgen last} is given by \be f = 1 + { a b \over
| w|^2 } \ee which means that we have a single center fivebrane
solution. The 5-branes are located at    $w=0$ as expected. One
can also check that the fivebrane charge is proportional to the
distance between two disks as in (\ref{charge2}) \be Im(
\Delta z) = Im \int_{-i b}^{ia}
\partial_w z d w  = \pi a b \ee In addition we find a contribution
to $g_{++}$ of the form \be \label{massterm} 4 f^{-1} |\partial_w
z|^2  = 4 { |w - i a |^2 | w + i b|^2 \over |w|^2 + a b } \ee

When we consider a string moving on this geometry in light cone gauge
we  find that \nref{massterm}
 appears as a potential for the worldsheet fields.
  Notice that the minima
 of the potential are  precisely at the two tips of the two disks corresponding to
$w = i a$ and $ w = - i b$ where we can take pp-wave limit.

When $a=b$ we have a symmetric situation where the two disks have
precisely the same length (same value of $\rho_i$). In this case
we see that the two minima are on the two sides of the fivebrane
at equal distance between them. Notice that the throat region of
the fivebrane corresponds to the region between the disks. This
throat region is singular in our approximation since the dilaton
blows up as $ w \to 0$. This is not physically significant since
this lies outside the range of our approximation, since $-Re(z)$
diverges. In fact, in the region between the disks we should
actually match onto the fivebrane solution
\nref{ns5sol}-\nref{ns5sol last}.

If $a \not = b$, say $a> b$ for example,
 then we have an asymmetric configuration where one disk is larger
than the other. The larger disk is the one whose tip is at $w = a$.
If $a \gg b$ then we find that the tip corresponding to the smaller disk is
in the throat region of the fivebrane while the tip corresponding to the larger
disk is in the region far from the fivebrane throat.

If we have $n$ nearby disks,
then the general solution is
\bea \label{nfivebr}
\partial_w z
&=& { ( w-ia_1) (w - i a_2) \cdots ( w- i a_n) \over
 ( w - i c_1) ( w - i c_2) \cdots ( w-i c_{n-1} ) }
\\ \notag
&&{\rm with} ~~~ a_1 < c_1 < a_2< c_2 < \cdots < c_{n-1} < a_n
\eea where $w=ia_{i}$ are the location of $n$ tips and $w=ic_{i}$
are the locations of $n-1$ sets of fivebranes. The resulting
solution \nref{ppgen}-\nref{ppgen last} describes a multi-center
configuration of  fivebranes on a plane wave. Boosting away the
$+$ components of all fields we find that we end up with a multi
centered configuration of fivebranes where the $SO(4)$ symmetry is
broken to $SO(3)$, in fact all fivebranes are sitting along a
line.


\subsubsection{Solutions for $ {\cal N} = 4 $ super Yang Mills on   $R \times S^3/Z_k $ }
\label{s3zk}

In this section we consider some aspects of the gravity solutions
describing ${\cal N}=4$ super Yang Mills on $R\times S^3/Z_k$.
This theory is particularly interesting since it is a very simple
orbifold of ${\cal N}=4$ SYM, so that one could perhaps analyze in
more detail the corresponding spin chains.

Let us start with the simplest solution, which is $AdS_5/Z_k \times S^5$.
If the orbifold is an ordinary string orbifold, then there is a $Z_k$ quantum
symmetry. On the field theory side, this orbifold corresponds to considering
a vacuum where the holonomy matrix $U$ has $n_l = N/k$ (see the notation around \nref{totnumb})
 and we need to start with
an $N$ which is a multiple of $k$. This is the configuration which corresponds
to the regular representation of the orbifold group action in the
gauge group, see \cite{Horowitz:2001uh}.
This is the simplest orbifold to consider from the string theory point of
view. Other choices for the holonomy matrix $U$, such as $U=1$, lead to an
orbifold which is not the standard string theory orbifold. Such an orbifold can
be obtained from the string theory one by turning on twisted string modes living
at the singularity.

 $AdS_5/Z_k \times S^5$ in type IIB can be dualized to   an M-theory or
IIA configuration which preserves the same supersymmetries as our
ansatz. Let us first understand the M-theory description. Let us
first single out the circle where $Z_k$ is acting. Then we lift
IIB on this circle to M-theory on $T^2$. This $T^2$ is
parametrized by the coordinates $x_1,x_2$ of the general M-theory
ansatz in \cite{llm}. The solution obtained in this fashion is
independent of $x_1,x_2$. The general solution of \nref{todaeq}
with translation symmetry along $x_1,x_2$ is\footnote{$\alpha'=1$.
This solution, if considered in the class of the analytically
continued solutions in \cite{llm}, describes $AdS_5 \times
S^5/Z_k$.} \bea
\label{unifsol} e^D &=& c_1 y + c_2 \\
c_1 &=& { g_s k \over 2} ~,~~~~~~c_2 = { {\pi g_s N\over 4} } \eea
Equivalently we can view the configuration as an electrostatic
configuration where \be \label{extp} V = - { \pi N   \over 2k}
\log \rho  + V_b ~,~~~~~~~~V_b =  { 1 \over g_s k
  }(\rho^2 - 2 \eta^2)
\ee
which means that we have a line of charge
 at the $\rho=0$ axis in the presence of the external potential $V_b$.

These solutions are singular at $y=0$ since we are not obeying
 \nref{standbc}. At $y=0$ we find that $4 \rho_0 = R^2_{S^5} = \sqrt{ 4 \pi g_s N \alpha'^2 }  $.
 In the IIB variables
this singularity is simply the $Z_k$ orbifold fixed point.
We also find that the radius of the two torus is $R_{x_1} = g_s$ and $R_{x_2} = 1/g_s$.
This is as we expect when we go from IIB to M theory.

The map between the IIB and IIA solutions is simply a T-duality
along the circle where $Z_k$ acts by a shift $\psi \sim \psi + { 4
\pi \over k}$. If $k$ is sufficiently large it is reasonable to
perform this T duality, at least for some region close to the
singularity. Once we are in the IIA variables, we can allow the
solution to depend on $\eta$. In fact, this dependence on $\eta$
allows us to resolve the singularity and get smooth solutions. The
electrostatic problem is now periodic  in the $\eta$ direction. We
have a periodic configurations of disks,  see figure
\ref{disks}(d), in the presence of an external potential of the
form $V_b$ in \nref{extp}.  Note that the external potential is
not periodic in $\eta$.
This is not a problem since the piece that determines the charge
distribution on the disks is indeed periodic in $\eta$.
Furthermore, the derivatives of $V$ that appear in \nref{IIA
ansatz}-\nref{IIA ansatz last} are all periodic in $\eta$
\footnote{ The $\eta$ dependent piece in \nref{extp} ensures that
as we go over  the period of $\eta$ we go over the period of $x_2$
which is T-dual to the circle on which the $Z_k$ acted.}. In the
IIA picture the region between the disks can be viewed as
originating from
 NS fivebranes.
  These NS fivebranes arise form the $A_{k-1}$ singularity of the IIB
solution after doing T-duality \cite{Ooguri:1995wj} (see also
\cite{Gregory:1997te}). In fact, the period of $\eta$ is
proportional to $k$, so that we have $k$ fivebranes $N_5=k$. From
this point of view the simplest situation is when all fivebranes
are coincident. This corresponds to taking the matrix $U$
proportional to the identity. On the other hand, the standard
string theory orbifold corresponds to the case that we have $k$
equally spaced disks separated by a unit distance. In other words,
the fivebranes will all be equally spaced.   In this case, since
we have single fivebranes, we do not expect the geometric
description to be accurate. Note that even though we are talking
about these fivebranes, the full solution is non-singular. These
fivebranes are a good approximation to the solution when we have
large disks that are closely spaced, as we will see in detail
below. But as we go to $\rho \to 0 $ the solution between the
disks approaches the NS5 solution \nref{ns5sol}-\nref{ns5sol
last}, which is non-singular. The different vacua \nref{totnumb}
correspond to the different ways of assigning charges $n_l$ (see
notation around \nref{totnumb}) to the disks that sit at positions
labelled by $\eta \sim l $. There are $k$ such special positions
on the circle. Only in cases where we have coincident fivebranes
can we trust the gravity description. This happens when some of
the $n_l$ are zero.

If we take the $k\to \infty $ limit, keeping $N$ finite, then the
direction $\eta $ becomes non-compact and we go back to the
configurations considered in the previous section which are
associated to the D2 brane theory (2+1 SYM) of section
\ref{d2theory}. This is also what we expected from the field
theory description.

We were not able to solve the equations explicitly in this case.
On the other hand, there are  special limits that are explicitly
solvable. These correspond to looking at the large $N$ limit so
that the disks are very large and then looking at the solution
near the tip of the disks. Let us consider the case where we have
a single disk per period of $\eta$. We can find the solution by
using \nref{nfivebr} and we get \be
\partial_w z=  i k \prod_{n= -\infty}^\infty { (w - i a n) \over   ( w - i a (n+\half) )}
= k \tanh { \pi w \over a} \label{flow} \ee where $k$ is the number of coincident
fivebranes.  When we insert this into
\nref{ppgen}-\nref{ppgen last} we find that the the solution
corresponds to a  periodic array of $k$ NS fivebranes along
spatial direction $\chi$. \bea \label{pernsf} f &=& {k \sinh r
\over 2 r ( \cosh r + \cos \chi) } = \sum_{n=-\infty}^\infty { k
\over r^2 + ( \chi +  \pi + 2 \pi n)^2 }
\\
r + i \chi &\equiv & {2 \pi \over a} w ~,~~~~~~\chi \sim \chi + 2
\pi
\\
g_{++} &=& 8 k { r \over \sinh r} ( \cosh r - \cos \chi) \eea The
rim of the disks corresponds to $w= i a $ or $r =\chi =0$ in
\nref{pernsf}. The $g_{++}$ term in the metric
\nref{ppgen}-\nref{ppgen last} implies  that the lightcone energy
is minimized by sitting at these points. These points lie  between
the fivebranes, which sit at $r=0$, $\chi = \pi$.
 In flat space the  T-dual of  an $A_{k-1}$ singularity corresponds to
 the near horizon region of a system of $k$ fivebranes on a circle.
Here we are getting a similar result in the presence of RR fields.
As $w \to \infty$ the solution \nref{ppgen}-\nref{ppgen last}
approaches the one that is the T-dual of the orbifold of a pp-wave
with $R^{4}\times R^{4}/Z_{k}$ transverse dimensions \be
\label{iibpptdual}
 ds_{10}^2 =  - 2 dx^+ dx^-  -   ( \vec r^{\,\, 2} +\vec u^{\,\, 2}  ) (dx^+)^2
 + d { \vec r}^{\, \,2} + du^{2}+\frac{u^{2}}{4}d\Omega _{2}^{2}+
\frac{k^{2}}{u^{2}}d\chi ^{2}
 \ee
At large $u$ we can T-dual this back to the $Z_k$ quotient of the
IIB plane wave, a situation studied in
\cite{pporbifold1,pporbifold2}.

Let us understand first the theory at the standard string theory
orbifold point. This corresponds to the vacuum with $n_l = N/k$,
for all $l=1,\cdots, k$. As we mentioned above, it is useful to
view the Yang Mills theory on $R \times S^3/Z_k$  as  the orbifold
of the theory on the brane according to the rules in \cite{gmmd}.
According to those rules we need to pick a representation of $Z_k$
and embed it into $U(N)$. The regular representation then gives
rise to the vacuum where all $n_l$ are equal. For this particular
choice we can use the inheritance theorem in \cite{zurab} that, to
leading order in the $1/N$ expansion,  the spectrum of $Z_k$
invariant states in the orbifold theory is exactly the same as the
spectrum of invariant states in ${\cal N}=4$. This ensures that
the matching between the string states on the orbifold and those
of the Yang Mills theory is the same as the corresponding matching
in ${\cal N}=4$. In the IIA description this regular orbifold goes
over to a picture where we have $k$ fivebranes uniformly spaced on
the circle. In this case we cannot apply our gravity solutions
near the fivebranes because we have single fivebranes.
Furthermore, we expect that the orbifold picture should be the
correct and valid description for string states even close to the
orbifold point, as long as the string coupling is small. The
spectrum of string states involving the second four dimensions
(the orbifolded ones) can be thought of as arising from $E-J=1$
excitations which get a phase of $e^{ \pm i 2 \pi/k}$ under the
generator of $Z_k$, but we choose a combination of these
excitations that is $Z_k$ invariant. This discussion is rather
similar to the one in \cite{Mukhi:2002ck}, where the $AdS_5 \times
S^5/Z_k$ orbifold (see e.g. \cite{S5orbifold}) was studied.

We can now consider other vacua. These are associated to different representations for
the Wilson line. For example, we can choose $n_k=N$ and $n_i=0$ for $i\not = k$.
In this case the IIA gravity description can be trusted when we  approach the origin
 as long as the 't Hooft coupling
is large and $k$ is large enough.
Let us describe the physics in the pp-wave limit in more detail for this case.
The pp-wave limit that we are considering consists in taking $k$
fixed and somewhat large, so that the gravity description of the
$k$ coincident fivebranes is accurate, and then taking $J$ and $ N$
to infinity  with $J^2/N$ fixed,  exactly as in ${\cal N}=4 $ super
Yang-Mills  \cite{bmn}. In fact, we find that the worldsheet theory
in the first four directions is exactly the same as for ${\cal
N}=4 $ super Yang Mills. In particular, the dispersion relation
for lightcone gauge worldsheet excitations  is precisely as in
${\cal N}=4$ super Yang Mills \cite{bmn}, with the same numerical coefficient.
 The theory in the
remaining four directions is more interesting. At large distance from the origin
 the worldsheet
field theory is just the orbifold of the standard IIB plane wave
\cite{Blau:2001ne}. This is what we had for the regular
representation vacuum that  we discussed above. A string state
whose worldsheet if far from the origin, so that its IIB
description is good, is a very excited string state. It is
reasonable to expect that the spectrum of such states is not very
sensitive to the vacuum we choose. This is what we are finding
here, since the spectrum in this region is that of the vacuum of
the regular representation we discussed above. On the other hand,
as we consider string states where the string is closer to the
minimum of its worldsheet potential  we should use the IIA
description in terms of $k$ coincident fivebranes, using the
solution in \nref{pernsf}. In this case the spectrum of
excitations on the string worldsheet is rather different than what
we had at the standard orbifold point. In this case we have
excitations of worldsheet mass $E-J = 2$ which are $Z_k$
invariant. This spectrum matches with what we naively expect from
considering impurities propagating on the string for the vacuum we
are considering. This vacuum contains only single particle gauge
theory excitations with $E - J\geq 2$ for all fields
 that could be interpreted
as excitations that are associated for the second four dimensions.
Let us be a bit more explicit.
 We can identify some of these $E-J=2$  excitations as  the Kaluza
Klein modes of $Z$ given by $\epsilon^{\dot \beta \dot \beta'}
\partial_{\alpha \dot \beta } \partial_{\alpha' \dot \beta'} Z$. This gives a singlet under
$SU(2)_L$, so that the $Z_k \subset SU(2)_L$ acts  trivially. So
this Kaluza Klein mode survives the $Z_k$ quotient.  The
$\alpha,\alpha'$ indices give rise to a spin one mode under
$SU(2)_R \subset \su (2|4)$.\footnote{The spin zero mode under
both $SU(2)_{L,R}$ vanishes due to the equation of motion. }.
There is a spin zero excitation with $E-J=2$ which comes from
the mode of the four dimensional gauge field along the $\psi$ circle,
the circle we are orbifolding.  These
elementary fields have $E-J =2$ and are associated to the $E-J =2$
excitations of the last four dimensions of the IIA plane wave. An
analysis similar to the one we will discuss for the plane wave
matrix model and 2+1 SYM in section \ref{furthersec} shows that
these excitations are exactly BPS and survive in the strong 't
Hooft coupling limit.

Other gravity solutions which are asymptotic to $AdS_5/Z_k$ were
constructed in \cite{EH}. Those solutions have a form similar to
that of the Eguchi-Hanson instanton \cite{eguchi} in the four
spatial directions.  In those solutions fermions are anti-periodic
along the $\psi$ direction. In our case, fermions are periodic in
the $\psi$ circle. So, the solutions in \cite{EH} arise when we
consider a slightly different field theory. Namely, when one
considers Yang Mills on $R\times S^3/Z_k$ but where the fermions
are antiperiodic along the circle on  which $Z_k$ acts. (One
should also restrict to $k$ even). This theory breaks
supersymmetry. The solutions in \cite{EH} describe states
(probably the lowest energy states) of these other theories. In
such cases the orbifold is  another state in the same theory,
the theory with antiperiodic fermion
boundary conditions along $\psi$. One
then expects that localized tachyon condensation, of the form
explored in \cite{aps}, makes the orbifold decay into the
solutions described in \cite{EH}.

\subsubsection{Solutions for the plane wave matrix model}
\label{bmnmm}

In this section we discuss some aspects of the gravity solutions
corresponding to the plane wave (or BMN) matrix model. In this case
we should think of the electrostatic configuration as having an
infinite disk at $\eta=0$ and the some finite number of disks of
finite size at $\eta_i >0$, see figure \ref{bubbling}(b).
 The background electric potential is
\be
 V_b = \rho^2 \eta -{ 2 \over 3} \eta^3
 \ee
The leading asymptotic form of the solution is \be \label{bmnasym}
V = V_b  + P { \eta \over ( \rho^2 + \eta^2)^{3/2} } \ee where we
have included the external potential plus the leading dipole
moment produced by the disks. The leading contribution is a dipole
moment because the conducting disk at $\eta =0$ gives an image
with the opposite charge, so that there is no monopole component
of the field at large $\rho, \eta$. The subleading terms in the
asymptotic region are higher multipoles.

We can insert this into the general ansatz \nref{IIA
ansatz}-\nref{IIA ansatz last} and we find that in the UV region
it goes over to the UV region of the solution for $N_0$ D0 branes,
where $N_0$ is proportional to the dipole moment $P$. More details
are in appendix \ref{d0match}. This dipole moment is given by \be
\label{dipolm}
 P = 2 \sum_i \eta_i Q_i \sim N_0 = \sum_i \left( \sum_{j<i} N_5^j \right) N_2^i
\ee
 where the index $i$ runs over the various disks. Notice  that
the difference between neighboring disks $d_i=\eta_{i+1} - \eta_i$
is proportional to the fivebrane charge. So the distances $d_i$
are quantized. This formula, \nref{dipolm},  should be compared to
\nref{bmnd0} by identifying $n \sim d_i$ and $N(n) = N_2^i$.

In \cite{Hai} this problem was analyzed using technique developed
by Polchinski and Strassler in \cite{P-S}, which consists in
starting with configurations of D0 branes smeared on two spheres.
In our language, this is a limit when we replace the disks by
point charges sitting at $\rho =0$. This approximation is correct
as long as the distance between the disks is much bigger than the
sizes of the disks and we look at the solution far away from the
disks\footnote{ We can make this relation more precise as follows.
Suppose that the potential in the asymptotic region behaves as $V=
\rho^2 \eta -{ 2 \over 3} \eta^3+\Delta$, where will treat
$\Delta$ as a perturbation. Then from the IIA ansatz (\ref{IIA
ansatz})-\nref{IIA ansatz last} we can write the solution as in
\cite{Hai} and find
  the warp factor $Z$ in \cite{Hai}. This gives
$Z=\frac{-1}{2\rho ^{2}\eta }
(\frac{2}{\rho}\partial_{\rho}\Delta+\partial_{\eta}^{2} \Delta)$ and
 $B_2=\frac{1}{\rho }\partial _{\rho }\left( \frac{4\eta ^{2}}{\rho }
\partial _{\rho }\Delta +2\eta \partial _{\eta }\Delta -2\Delta \right)d^2 \Omega$.
We get $\Delta= {P \eta \over ( \rho^2 + \eta^2)^{3/2} }$
 by comparing the  leading order approximation  $Z=\frac{R^7}{r^7}$ and the fluxes
$H_{3}=\alpha r^{-7}(T_{3}-\frac{7}{3}V_{3}),
~G_{6}=g_{s}^{-1}\alpha r^{-7} (\frac{1}{3}\ast
_{9}T_{3}-\frac{7}{3}\ast _{9}V_{3})$,  which are dual to the mass
terms (see \cite{Hai}), where $r=(\rho^{2}+\eta^{2})^{1/2}$, and
$\rho$ and $\eta$ are radial variables in $SO(6)$ and $SO(3)$
directions. If we replace $Z$ by the expression corresponding to
multi-center D0 branes uniformly
  smeared
on several  $S^2$s, then we get $\Delta ={\sum_i }Q_i\left[ \frac{1}{
[(\eta -\eta _{0}^{(i)})^{2}+\rho ^{2}]^{1/2}}-\frac{1}{[(\eta +\eta
_{0}^{(i)})^{2}+\rho ^{2}]^{1/2}}\right]$, this is precisely the limit when
the disks above the $\eta=0$ plane are treated as point charges.
}.

From the field theory point of view it looks like the simplest
 vacuum is the one with all $X=0$. This case
corresponds to having $N_0$ copies of the trivial (dimension one)
representation of $SU(2)$. In the gravity description this
corresponds to having a single disk at a distance of one unit from
the conducting surface at $\eta=0$, see figure \ref{expansion}(a).
Unfortunately, since this vacuum corresponds to a single
fivebrane, the gravity approximation will not be good near the
fivebrane.  We will focus on this situation in the next section.
However, we can consider vacua corresponding to many copies of
dimension $N_5$ representations of $SU(2)$. These involve $N_5$
fivebranes and we will be able to give interesting solutions, at
least in the region relevant for the description of near BPS
states. It should be possible to extrapolate these solutions to
smaller values of $N_5$ using conformal field theory. Let us now
study the case that we have only a single disk at a small distance
from the infinite disk at $\eta=0$. So we consider a situation
with $ N_5 \ll N_0 $. Based on the discussion in \cite{mssvr} we
expect that  the $N_0$ D0 branes blow up into $N_5$ NS5 branes. Of
course, our solution will be smooth, but we will see that there is
a sense in which we have $N_5$ fivebranes.
The appearance of fivebranes is probably connected with the picture
in \cite{david} for 1/2 BPS states in terms of eigenvalues that lie
on a five sphere.

Unfortunately we were not able to find the full solution of
equation \nref{veque}. Nevertheless we can expand the solution
near the tip of the disk. In fact, we can get the solution in a
simple manner by starting from the solution corresponding to the
region near the tip of two disks \nref{twodisks} and then letting
one of the disks go to infinity. After a simple rescaling this
produces \be \label{diskbmn}
\partial_w z = i {(w- i a) \over w}
\ee In this case, the function $f$ in \nref{ppgen}-\nref{ppgen
last} becomes \be f = { a \over 2|w|^2 } \ee and the contribution
to $g_{++}$ is \be \label{ggpppart} 4 f^{-1} | \partial_w z|^2 ={8
\over a} | w - i a|^2 \ee So we see that we get the near horizon
region of fivebranes. The contribution \nref{ggpppart} to the
$g_{++}$ metric component gives rise to a potential on the
lightcone gauge string worldsheet. This potential localizes
 the string at some particular position along the throat.
Writing $w=ia e^{\phi+i\theta}$, the 10 dimensional solution is
\footnote{We set $\alpha'=1$ in this paper.}
\begin{eqnarray} \label{wzw}
ds_{10}^{2} &=& -2dx^{+}dx^{-}+d{\vec r}^{\,2}-{\vec r}^{\,2}dx^{+2}-4N_5(e^{2\phi
}+1-2e^{\phi }\cos \theta )dx^{+2}  \notag \\
&&+N_5(d\phi ^{2}+d\theta ^{2}+\sin ^{2}\theta d\Omega _{2}^{2})\\
e^{\Phi } &=& g_s e^{-\phi } \\
C_{1} &=& -{ 1 \over g_s} 2 \sqrt{N_5} (e^{2\phi }-e^{\phi }\cos \theta )dx^{+} \\
C_{3} &=&{ 1\over g_s} N_5^{3/2} 2 e^{\phi }\sin ^{3}\theta dx^{+}\wedge d^{2}\Omega \\
H_{3} &=&2 N_5\sin ^{2}\theta d\theta \wedge d^{2}\Omega
\label{wzw last}
\end{eqnarray}
where $g_s$ is the value of the dilaton at the tip. By performing
a boost $x^\pm \to \lambda^{\pm 1} x^\pm$ with $\lambda \to 0$ we
set to zero all non-trivial terms involving $dx^+$ and we
  recover the usual fivebrane near horizon geometry \cite{Callan:1991at}.
  By taking a limit of small $\phi $ and $\theta$ we find
  the IIA plane-wave in (\ref{iiapp}).

An important parameter is the size of the $S^5$ in string theory units at the tip of the disks.
 This can
be approximated as \footnote{Our normalization of the action is $
S = { 1 \over g_{YM0}^2} \int {\rm Tr}\{\half ({D_0} Y_i)^2 -\half
m^2 Y_i^2 + {1 \over 4} [Y_i,Y_j]^2 + \cdots\}$ where $Y_i$ are
the $SO(6)$ scalars. The dimensionless parameter is
$g_{YM0}^2/m^3$. } (see appendix \ref{d0match}) \be \label{s5bmn}
{ R_{S^5}^2 \over \alpha'}  = 4 \pi \left( \frac{
g_{YM0}^{2}N_{2}}{ 2 m^{3}}\right) ^{1/4} ~,~~~~~~~~~N_2 = { N_0
\over N_5}, ~~~~~m=1 \ee where $m$ is the mass of the $SO(6)$
scalars and is set to 1. $N_0$ is the number of D0 branes or the
rank of the gauge group in the plane wave matrix model. Our
gravity approximation is good when we are in the regime of
interest, $ N_5 \ll N_0 $, and the size of $S^5$ in string unit is
large. From this result we can compute the spectrum of near BPS
excitations with large angular momentum $J$. For fluctuations in
the directions parametrized by $\vec r$ in \nref{wzw}  the
spectrum is \be (E- J)_n = \sqrt{ 1 + (4 \pi)^2 \left( \frac{
 g_{YM0}^{2}N_{0}}{2 m^{3} N_{5}}\right)^{1/2}
 { n^2 \over J^2} }  = 1 +  4 \pi^2
\left( \frac{2  g_{YM0}^{2}N_{0}}{m^{3} N_{5}}\right)^{1/2}
{ n^2 \over J^2}   + \cdots \label{leading}
\ee

Under general principles, in the t' Hooft limit, with $N_5$ fixed,
we expect the spectrum to be of the form \be (E- J)_n  = 1 +
f\left( { g^2_{YM0} N_{0} \over m^3 N_{5} }, N_5 \right) { n^2
\over J^2 } + \cdots \label{wholeregime} \ee in the large $J$
limit.

The $N_5=1$ case has been analyzed perturbatively up to four loops in
 \cite{fourloop}. In our conventions\footnote{
The relations between their variables and ours are $ 4\Lambda=
\left( \frac{2}{M}\right)^{3}N=\frac{g_{YM0}^{2}N_{0}}{m^{3}}$, $
8\pi ^{2}\Lambda _{r}=f$.
 } their result reads
 \bea
 f_{pert}\left({g^2_{YM0 } N_0 \over m^3},N_5=1\right) &=&
\frac {2\pi ^{2}g_{YM0}^{2}N_{0}}{m^{3}} \left[ 1-\frac{7}{8}
\frac{\ g_{YM0}^{2}N_{0}}{m^{3}}+\frac{71}{32}\left( \frac{\
g_{YM0}^{2}N_{0}}{m^{3}}\right) ^{2} \right. \notag \\
& & \left. -\frac{7767}{1024}\left( \frac{\ g_{YM0}^{2}N_{0}}{m^{3}}%
\right) ^{3}+...\right] \label{wcres} \eea
Of course we expect that the function $f$ interpolates smoothly between
the weak coupling result \nref{wcres} and the strong coupling result
\nref{leading}.

Our gravity solutions are not valid for $N_5=1$, especially in the
region relevant for this computation. On the other hand, we see
that the quantity that determines $f$ is the radius of the
fivesphere. We can think of this solution as follows. Let us first
use an approximation similar to that used by Polchinski and
Strassler \cite{P-S}, \cite{Hai}. In this case we approximate the
solution by smearing D0 branes on a fivesphere, which we interpret
as a fivebrane which carries D0 brane change. We then determine
the size of the fivebrane by coupling it to the external fields
that are responsible for inducing the mass on the D0 worldvolume.
This gives the radius of the fivebrane. In fact, this was computed
in \cite{mssvr} where the formula similar to \nref{s5bmn} was
found (the precise numerical factors were not computed in
\cite{mssvr}). So it is natural to believe that \nref{leading}
will still be the correct answer for $N_5=1$. In other words, the
coupling constant ``renormalization'' that was found in
\cite{fourloop} is interpreted here as a physical quantity giving
us the size of the fivebrane in the gravity description at strong
coupling. This is the situation for the first four coordinates.
The fact that a single fivebrane has no near horizon region also
suggests that something drastic happens to the second four
directions that are transverse to the single fivebrane. We
observed this feature for $N_5=1$ also from gauge theory side and
will explain it in the next section.

Finally, let us discuss the issue of tunneling between different
vacua. In general we can tunnel between the different vacua of the
matrix model. But the tunneling can be suppressed in some regimes.
For example, let us consider the case we discussed above where we
consider the vacuum corresponding to a single large disk at a
distance $N_5$ from the $\eta=0$ plane, see figure
\ref{expansion}(a). From the gravity point of view we can take one
unit of charge from the large disk and put some other disks.
Charge is not conserved in the process, but $N_0$ should be
conserved. Reducing the charge of the large disk by one unit we
are left with $N_5$ D0 branes to distribute in the geometry. So,
for example, we can put another disk at a distance of one unit
from the $\eta =0$ plane with $N_5$ units of charge. In the
geometry this transition is mediated by a D-brane instanton. The
geometry between the original disk and the $\eta=0$ plane can be
approximated by the solution in section \ref{nsfivesec}. That
solution contains a non-contractible $\Sigma_3$, see figure
\ref{diskone}. If we wrap a Euclidean D2 brane on this $\Sigma_3$
we find that, since there is flux $N_5$ through it, we need $N_5$
D0 branes ending on it \cite{gmjmns}. Thus, this instanton
describes the creation of $N_5$ D0 branes. Its action is
proportional to the action of the Euclidean D2 brane. This process
describes the tunneling between the vacua  in figures
\ref{bubbling}(d) and \ref{bubbling}(e). If the volume of the
$\Sigma_3$ is sufficiently large and the string coupling is
sufficiently small this process will be suppressed. In order for
this to be the case we need to arrange the field theory parameters
appropriately. Notice that there is no instanton that produces a
smaller number of D0 branes. This also agrees with the field
theory. If we start with the vacuum with many copies of the $N_5$
dimensional representation of $SU(2)$, then we can take one of
these representations and partition those $N_5$ D0 branes into
lower dimensional representations. This is basically the process
described by the above instanton. In other words, the fact that
the D-brane instanton produces $N_5$ D0 branes matches with what
we get in the field theory.

\subsection{Further analysis of near BPS states}

\label{furthersec}

In previous sections we have mainly analyzed the near BPS states
associated to string oscillations in the the first four
dimensions, which are described by free massive fields on the
worldsheet.  In this section we mainly focus on the second four
dimensions which are associated to fivebrane geometries. Since the
spectrum depends on the vacuum we expand around, we will focus on
the large $J$ near BPS excitations around some particular vacua of
the plane wave matrix model. We will consider first the $N_5=1$
vacuum and then the $N_5>1$ vacua, both from the gauge theory and
gravity points of view. We also make some remarks about the
simplest vacuum of the 2+1 super Yang Mills on $R\times S^2$.

\subsubsection{$N_5 =1$ vacua of the plane wave matrix model  }
\label{nfonesec}

\begin{figure}[htb]
\begin{center}
\epsfxsize=3.6in\leavevmode\epsfbox{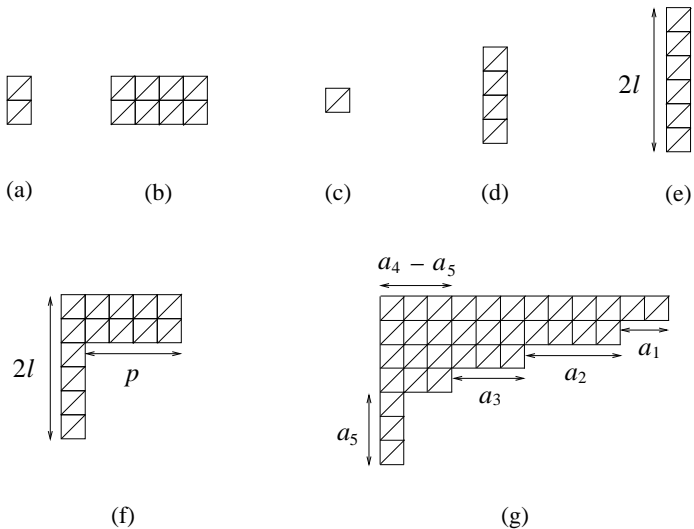}
\end{center}
\caption{Young supertableaux corresponding to various
representations of $SU(2|4)$ or $SU(2|2)$ discussed in the text.
In (e) and (f), $2(l-2)=a_5, p=a_2$ for $SU(4|2)$ Dynkin labels.
Figure (g) shows the correspondence between supertableau and
Dynkin labels for a general physically allowed representation
$(a_1,a_2,a_3|a_4|a_5)$ of $SU(4|2)$, see also
\cite{barsetal,markvr2}.} \label{young}
\end{figure}

Let us start by discussing the trivial vacuum of the matrix model,
where we expand around the classical solution where all $X=0$.
This is the vacuum we denote by $N_5=1$ and which should
correspond to a single fivebrane. When we expand around this
vacuum we have 9 bosonic and 8 fermionic excitations which form a
single representation of $\su (2|4)$, corresponding to the Young
supertableau in figure \ref{young}(a). Our notation for $\su
(2|4)$ representations follows the one in \cite{barsetal,markvr2}.
We are interested in forming single trace excitations which should
correspond to single string states in the geometry. For example,
we can consider the state created by the field $Z$ of the form
$Tr[Z^J]$ \footnote{We denote the field $Z$ and its creation
operator by the same letter.}, where $Z=Y^5+iY^6$.\footnote{In
this section  $Y^i$, $i=1, \cdots, 6$ are the scalars that
transform under $SO(6)$ and  $X^i$ are the ones transforming under
$SO(3)$ .} This state is BPS and it belongs to the doubly atypical
(or very short) representation whose Young supertableau is shown
in figure \ref{young}(b). As in \cite{bmn} we can consider near
BPS states by writing states of roughly the form $\sum_l Tr[ Y^i
Z^l Y^j Z^{J-l}] e^{ i
 2 \pi { l n \over J} }$ where $i,j =1,\cdots 4$.
  We can view each insertion of
 the field $Y^i$ as an ``impurity" that propagates along the chain formed by the
 $Z$ oscillators. These impurities are characterized by the momentum $p= n/J$ and
 a dispersion relation $\epsilon(p)$, where $\epsilon$ is the contribution of
 this impurity to $\hat E \equiv E-J$.
 Here we are thinking about a situation where we
 have an infinitely long chain where boundary effects can be neglected.
 These fields have $\epsilon(p=0) = 1$, we can think of this as the ``mass" of the
 particles. This is an exact result  and
 can be understood as a consequence of the Goldstone theorem.  Namely, when we
 pick the field $Z$ and we construct the ground state of the string
 with powers of $Z$ we are breaking $SO(6)$ to $SO(4)$.
 The excitations $Y^i$, $i=1,\cdots,4$ correspond
 to the action of the broken generators. This is a fact
 that does not even require
 supersymmetry.
 In other words, we are
 simply rotating the state $tr[Z^J]$.
It is also useful to consider the supersymmetry that is preserved
 by this chain. Out of the supergroup
 $\su (2|4)$ our choice of $Z$ leaves an $SU(2)_G \times \su (2|2)$
 subgroup\footnote{ The $_G$ subindex indicates that it is global symmetry that commutes
 with supersymmetry.}
  that acts on the excitations that propagate along the string.
The group $SU(2)_G$ together with one of the $SU(2)$ subgroups in $\su (2|2)$
forms the $SO(4)$ in $SO(6)$ that rotates the first four dimensions.  The second
$SU(2)$ subgroup of $\su (2|2)$ is the $SU(2)$ factor in $\su (2|4) $ and
rotates the three scalars $X^i$.
We can
 use $\su (2|2)$ to classify these excitations. The non-compact $U(1)$
 in $\su (2|2)$
 corresponds to the generator $\hat E = E-J$
 and gives us the mass of the particle.
 The fields $Y^i$ belong
  to the fundamental representation of $ \su (2|2)$ whose
 Young supertableau is in figure \ref{young}(c). In addition they transform in the
 spin one half representation of $SU(2)_G$.
  We can think of these excitations
 as ``quasiparticles" that propagate along the string.
 The properties of these
 quasiparticles were studied in great
 detail in \cite{fourloop} where the dispersion
 relation and particular components of the
  S-matrix were computed to four loops.
 These quasiparticles contain four bosons and four fermions.

So far we have been discussing mainly the fields $Y^i$ and the
fermions which have $E-J =1$. What about the other fields in the
theory? There are four other elementary fields which have
$\Delta-J =2$. These are the three scalars $X^i$ of $SO(3)$ and
the field $\bar Z$ plus four fermions. Naively, we might think
that these would lead to mass two impurities that propagate along
the string. This is not the case. Actually, what happens is that
they mix with the fields that we have already described and do not
lead to new quasiparticles \cite{beisertsusy}. For example an
insertion of the field $\bar Z$, such as $tr[\bar Z Z^{J+1}]$
mixes with the states $tr[Y^i Z^l Y^i Z^{J-l}]$. The result of
this mixing is such that the resulting spectrum can be fully
understood in terms of two quasiparticles of mass one that
propagate along the string. Something similar happens with the
insertion of the $SO(3)$ scalar $X^i$, which mixes with the
insertion of two fermions  of individual mass one. In fact, the
one loop Hamiltonian in this sector is a truncation of the one
loop Hamiltonian of ${\cal N}=4$ SYM in \cite{beisertnfour}. So
the results we are mentioning here follow in a direct way from the
explicit diagonalization undertaken in \cite{beisertsusy}. The
final conclusion of this discussion, is that in perturbation
theory we have a chain which contains impurities with mass one,
that transform in the fundamental of $\su (2|2)$ and fundamental
of $SU(2)_G$.
 We have four bosons and four
fermions, which can be viewed as the Goldstone modes of the
symmetries broken by the BPS operator $tr[Z^J]$. This spectrum is
compatible with the index \nref{defindfull} evaluated on single
trace states.

 Let us now discuss what happens at large 't Hooft coupling.
 The radius of the fivebrane is given by \nref{s5bmn} (with $N_5=1$).
In addition, we have seen that the near BPS states are described
by the pp-wave geometry \nref{wzw}-\nref{wzw last} which
 corresponds to the near horizon region of $N_5$ fivebranes.
 The first four
 transverse dimensions correspond to the motion of the string in the direction
 of the fivebranes and the spectrum contains particles that transform in the
 fundamental of $\su (2|2)$ and the fundamental of $SU(2)_G$
  as we had in the weak coupling analysis. The dispersion
 relation is given by the usual relativistic formula \nref{leading}.

On the other hand, when we consider the fate of the last four
transverse dimensions we run into trouble with the geometric
description. We see that the solution \nref{wzw}-\nref{wzw last}
does not make sense for $N_5=1$ since a single fivebrane is not
supposed to have a near horizon region \cite{Callan:1991at}. The
reason is that the near horizon region involves a bosonic WZW
model with level $k= N_5-2$ and this theory is unitary only if
$N_5-2\geq 0$. In our context, we also have RR fields that try to
push the string into the near horizon region. Since for $N_5=1$ we
do not have such a region, the simplest assumption
 is that the second four dimensions are somehow not present in
 our pp-wave limit.
This would agree with what we saw in the weak coupling analysis
above, where we did not have any quasiparticles propagating along
the string corresponding to the second four dimensions. Of course,
a string quantized in lightcone gauge is
  not Lorentz invariant in six dimensions. But perhaps this
is not a problem in this case, since the presence of RR fields breaks Lorentz invariance.
Nevertheless, one would like to understand the background in a more precise way in the
covariant formalism, so that one can ensure that we have a good string theory solution.

\subsubsection{$N_5>1$ vacua of the plane wave matrix model }

 In order to find a better defined string theory we
need to consider $N_5>1$. So, let us consider what happens when we
expand around the vacuum of the plane wave matrix model
corresponding to $N_5>1$. This is the vacuum where the matrices
$X_i$ are the generators of the dimension $N_5$ representation of
$SU(2)$.
 We would like to understand  the similarities and differences
between these vacua and the $N_5=1$ vacuum. When we expand around
these vacua  we find that we have $N_5$  $\su (2|4)$
supermultiplets, the ones whose Young supertableaux are given in
figure \ref{young}(e) with $l=1, \cdots, N_5$ \cite{markvr2}. We
can view them as the
 Kaluza Klein modes on a fuzzy $S^2$. The subsector of this theory where we
consider only excitations of the first Kaluza Klein mode is the same as the one
we had in the $N_5=1$ sector. In fact, the one loop Hamiltonian for these excitations
is exactly the same as the one we had for the $N_5=1$ case.  This can be seen as follows.
Since these modes are proportional to the identity matrix in the $N_5\times N_5$ space that
gives rise to the fuzzy sphere we see that their interactions are  the
same as the ones we had around the $N_5=1$ vacuum. The only difference could arise
when we consider diagrams that come from one loop propagator corrections. But the value
of these propagator corrections
 is determined by the condition that the energy of the state
$tr[Z^J]$ is not shifted, since it is a BPS state.  One
difference, relative to the expansion around the $N_5=1$ vacuum is
that the one loop Hamiltonian is proportional to $g_{YM0}^2
N_2/N_5$ as opposed to $g_{YM0}^2 N_0$ (where $N_0 = N_2 N_5$).
More precisely, we find that the function $f$ in
\nref{wholeregime} has the form \be \label{nfexpf} f({g^2_{YM0}
N_0 \over m^3 N_5 }, N_5) = 2 \pi^2 {g^2_{YM0} N_0 \over m^3 N_5^2
} + \cdots \ee for small 't Hooft coupling. We obtain this result
as follows. First we notice that $\hat E =1$ excitations are given
by diagonal matrices in the $N_5\times N_5$ blocks that produce
the fuzzy sphere. These matrices are $N_2 \times N_2$ matrices. In
other words, the relevant fields can be expressed as $Y^i =
{\mathbf 1}_{N_5 \times N_5} \otimes \tilde Y^i$ where $\tilde
Y^i$ are $N_2 \times N_2$ matrices,  with $N_2 \equiv N_0/N_5$.
Then the action truncated to the $\tilde Y^i$ fields looks like
the $N_5=1$ action except that we get an extra factor of $N_5$
from the trace over the diagonal matrix ${\mathbf 1}_{N_5\times
N_5}$. This effectively changes the coupling constant $g_{YM0}^2
\to \tilde g_{YM0}^2 = g_{YM0}^2/N_5$. Since the $\tilde Y^i$
fields are $N_2 \times N_2$ matrices we see that corrections in
this subsector will be proportional to $ \tilde g_{YM0}^2 N_2$.
Notice that \nref{nfexpf} it involves  a different combination of
$N_0$ and $N_5$ than the one that appears at strong coupling
\nref{leading}. So the interpolating function in
\nref{wholeregime} should have a non-trivial $N_5$ dependence. In
summary, at one loop, the excitations built out of impurities in
the first Kaluza Klein harmonic on the fuzzy $S^2$ give rise to
four bosonic and fermionic quasiparticles of mass $\hat E =1$  as
we had in the $N_5=1$ case.

Let us now focus on the second Kaluza Klein mode, given by the
supermultiplet of $\su (2|4)$ in figure \ref{young}(d). This
multiplet contains four bosonic and four fermionic states of mass
$\hat E = E-J = 2$. These eight states transform in the $\su
(2|2)$ representation of figure \ref{young}(a). Let us describe
how  the bosonic states arise. We expand $Z = \tilde Z + {\cal
J}^i Z_i + \cdots$ in fuzzy sphere Kaluza Klein harmonics using
the $N_5 \times N_5$ matrices ${\cal J}^i$ which give a
representation of $SU(2)$ (see \cite{fuzzysphere}). Three of the
states correspond to the impurities $Z_i$ and they are in the
$(1,0)$ representations of $SU(2)\times SU(2) \subset \su (2|2)$
and they are singlets of $SU(2)_G$.
 The fourth state, denoted by $\Phi$,
 arises when we expand $X^i = {{\cal J}^i}({\mathbf 1} + \Phi) + \cdots $.
 This has $E=2$ and spin
 zero under all $SU(2)$s.
 It gives rise to an excitation with ${\hat E}=E-J =2$ and spin zero. In addition to
 these four bosonic states we have their fermionic partners.
When we consider BPS states with $E - 2 S - \sum_i J_i =0$, the
only bosonic state that contributes is $Z_+$, which has $S=1$.
Thus the state $tr[Z_+  {\tilde Z}^J]$ is  BPS. In order to ensure
that its energy is not corrected we need to check  that it cannot
combine with other BPS states. The analysis in \cite{markvr2}
tells us which representations this could combine with. By looking
explicitly at the ones arising when we construct single trace
states we can see that these other representations are not
present. This is a result that is exact in the planar limit. In
appendix \ref{indexapp} we use the index defined in
\nref{defindfull} to prove the above statement.

What we learned is that for $N_5>1$, as opposed to the case with
$N_5=1$, we have a new quasiparticle of mass two propagating along
the string. In fact, the  same argument would go through for the
case of 2+1 SYM on $R\times S^2 $ in section \ref{d2theory},
expanded around the trivial vacuum, and the new supermultiplet
correspond to the three derivatives $D_i$, $i=0,1,2$ and the
seventh scalar $\Phi$, and their fermionic partners. They have
mass 2 and correspond to the second four coordinates of the IIA
plane wave. In all these cases we have extra quasiparticles
propagating along the string. This agrees with the fact that in
string theory we have eight transverse directions for the string.
The first four dimensions behave as we discussed above (in section
\ref{nfonesec}) and its presence is ensured by the $SO(6)$
symmetry. The details of the second four dimensions depend on the
vacuum we expand around. So let us concentrate more on these
second four dimensions.

\subsubsection{Comparison between worldsheet theory and gauge
theory} \label{sigmamodel}

We will now discuss the two dimensional field theory that
describes the second set of four transverse dimensions for a
string in light cone gauge moving in the pp-wave geometry
\nref{wzw}-\nref{wzw last}. The target space for this two
dimensional theory is
 $R \times S^3$ with an $H_3$ flux on the $S^3$ equal  to $N_5$ and a linear
 dilaton in the $R$ direction. These are the dimensions parametrized by
 $\phi, ~\theta, ~\Omega_2$ in \nref{wzw}.
 In addition we have a
  potential which localizes the string at some point along the
throat and at a point in $S^3$. This potential arises from the
$g_{++}$ component of the metric in \nref{wzw}. Ignoring the
potential for a moment we see that we have a the conformal field
theory describing the throat of $N_5$ fivebranes
\cite{Callan:1991at}. The potential breaks the $SO(4)$ rotation
symmetry of the throat region to $SO(3)$. The resulting sigma
model has (4,4) supersymmetry on the worldsheet. When the
potential is non-zero the supersymmetry in the 1+1 dimensional
worldsheet theory is of a peculiar kind \cite{nahm}.
  In ordinary global (4,4) supersymmetry the supercharges transform
under an $SU(2)\times SU(2)$ R-symmetry but those symmetries do
not appear in the right hand side of the supersymmetry
algebra\footnote{Notice that here we are talking about the global
$(4,4)$ supersymmetry. These are the modes $G^i_0$ of the
superconformal algebra generated by $G^i_{n}$. Some of the $SU(2)$
currents do appear in the anticommutators of some of the
$G^i_{n}$, $n\not =0$. }. Let us denote the supercharges by
$Q^i_{\pm}$, where  $i=1,\cdots, 4$ are $SO(4) =SU(2)\times SU(2)$
indices, and $\pm$ indicates two dimensional chirality. The
anti-commutators of these supercharges have the form \be
\label{commutsu} \{ Q^i_+,Q^j_+ \} = \delta^{ij} (E+ P)~,~~~~~ \{
Q^i_-,Q^j_- \} = \delta^{ij} (E - P) ~,~~~~~~\{Q_+^i,Q^j_-\} = m
\epsilon^{ijkl} J^{kl} \ee where $J^{kl}$ are the $SO(4)$
generators and $m$ is a dimensionful parameter which we can set to
one. This parameter is related  to the scale entering in the
potential and determines the mass of BPS particles which carry
$SO(4)$ quantum numbers. When the potential is set to zero we set
$m=0$ and
 we get the ordinary commutation relations
we expect for the usual (4,4) supersymmetry algebra. Let us denote
the algebra \nref{commutsu} by $(4,4)_m$. Notice that this is a
Poincare superalgebra which contains non-abelian charges in the
right hand side. This is possible in total spacetime dimension
$d\leq 3$ \cite{nahm} but not in $d>3$ \cite{weinberg}. This
algebra is a dimensional reduction of a Poincare superalgebra in
2+1 dimensions that we discuss in more detail in appendix
\ref{poincare}.

Note that
the potential implies that the light cone energy is minimized (and it is zero)
 when
the string sits at $\phi =\theta =0$. There is just  a finite
energy gap of the order of $N_5 |p_-|$ preventing it from going
into the region $\phi \to -\infty$ where the pp-wave
approximations leading to \nref{wzw}-\nref{wzw last} break
down\footnote{In the region $\phi \to -\infty$ we need to use the
fivebrane solution in section \ref{nsfivesec}.}. Potentials in
models preserving (4,4) supersymmetry were studied in
\cite{jourjine},\cite{freedmanag} for models based on hyperkahler
manifolds. Here we are interested in models with non-zero $H$
flux. In fact, for
 the general solution \nref{ppgen}-\nref{ppgen last} we can write down the string theory in
lightcone gauge
\begin{eqnarray}
S &=&S_{1}+S_{2}  \label{wsaction} \\
S_{1} &=&\int dt\int_{0}^{2 \pi \alpha ^{\prime }|p_{-}|}d\sigma d^{2}\theta \half%
\left[ D_{+}{R}^{i}D_{-}R^{i}+R^{i}R^{i}\right]  \\
S_{2} &=&\int dt\int_{0}^{2 \pi \alpha ^{\prime }|p_{-}|}d\sigma
d^{2}\theta
\left\{  \half f(W,\bar W)(D_{+}WD_{-}\bar{W}+D_{+}\bar{W}D_{-}W)+z(W)+\bar{z}(\bar{W}%
)+\right.   \notag \\
&&\left. +[f(W,\bar W)(W+\bar{W})^{2}g_{ij}(\Theta )+B_{ij}(\Theta ,W,\bar{W}%
)]D_{+}\Theta ^{i}D_{-}\Theta ^{j}\right\}   \label{wzwsigma}
\end{eqnarray}
where $S_1$ describes the first four coordinates and consists of
 four free massive superfields.  $S_2$  is the action describing the
second four coordinates. We have written the action in ${\cal
N}=1$ superspace, by picking one special supercharge.  Note that
this particular supercharge, say $Q^1_{\pm}$, obeys the usual
super Poincare algebra, therefore we can use the usual superspace
formalism. The $B$ field and the function $f$ are simply the ones
 in \nref{ppgen}-\nref{ppgen last}. The theory has $(4,4)_m$ supersymmetry. We have not
shown this explicitly from the lagrangian \nref{wzwsigma} but we
know this from the supergravity analysis. Compared to the usual
WZW action for
  a system of fivebranes, the only new
term is the potential term. Note that RR fields in
\nref{ppgen}-\nref{ppgen last} are such that four of the fermions
are free, which are the ones included in $S_1$, and the remaining
four are interacting and appear
 in $S_2$ in \nref{wzwsigma}.

Let us first study the theory \nref{wzwsigma} for large $N_5$. In
that case, we can expand the fields around the minimum of the
potential. If we keep only quadratic fluctuations we have four
free bosons and fermions. In order to characterize these particles
we go to their rest frame. Setting $P=0$ we find that
\nref{commutsu} reduces to the $\su (2|2)$ algebra. These
particles transform in the representation with two boxes as in
figure \ref{young}(a) (but now viewed as a representation of $\su
(2|2)$).
 In terms of $SU(2)\times SU(2)$
quantum numbers we have $(1,0)+ (1/2,1/2) + (0,0)$ where particles
with half integer spin are fermions. This is a short
representation, with energy $\hat E = 2$. In fact, if we consider
a closed string and a superposition of two such particles with
zero momentum we can form states that transform in the
representations given in figure \ref{young}(e), which are also
protected. As we make $N_5$ smaller these protected
representations have to continue having the same energy. Of
course, this argument only works perturbatively in $1/N_5$ since
$N_5$ is not a continuous parameter and we can have jumps in the
number of protected states as we change $N_5$.
 In order to figure out more precisely
which representations are protected it is convenient to introduce
an index defined by \be \label{sutwoind} {\cal I}(\gamma) = Tr[
(-1)^F 2S_3 e^{ - \hat \mu ( \hat E - S_3-\tilde S_3) } e^{-
\gamma \hat E} ] \ee where $S_3$ and $\tilde S_3$ are generators
in each of the two $SU(2)$ groups. We use the letter ${\cal I}$ to
distinguish \nref{sutwoind} from \nref{defindfull}. One can argue
that only short representations contribute and that the final
answer is independent of $\hat \mu$, see appendix \ref{indexapp}.
We can compute this for large $N_5$ using the free worldsheet
theory and we obtain \be \label{obtind} {\cal I}(\gamma)_{N_5 =
\infty} = \sum_{n=1}^\infty e^{ - 2n \gamma} \ee Since $N_5$ is
not a continuous parameter we see that as we make $N_5$ smaller
\nref{obtind} could change but only by terms that are
non-perturbative in the $1/N_5$ expansion. Thus
 for $N_5$ fixed and large we
expect that the corrections would affect only terms of the form
 $e^{- ({\rm const}) N_5  \gamma}$.

Now, let us compare this with the expectations from the gauge
theory side. In order to find protected representations on the
gauge theory side it is convenient to use the index
\nref{defindfull}. Since we are focusing on single trace states we
can compute \nref{defindfull} just for single trace states. For
the case that we expand around the vacuum corresponding to $N_2$
$SU(2)$ representations of dimension $N_5$ we get \bea I_{s.t.
~N_5} &=  & I_{s.t~N_5=1} + { e^{ - 2{N_5}(\beta_1 + \beta_2 +
\beta_3)} \over (1 - e^{ - 2 {N_5} (\beta_1 + \beta_2 + \beta_3)}
) } - { e^{ - 2(\beta_1 + \beta_2 + \beta_3)} \over (1 - e^{ -  2
(\beta_1 + \beta_2 + \beta_3)})} \label{bignf}
\\
 I_{s.t~N_5=1}&=& { e^{ - \beta_2 - \beta_1 } \over 1-e^{-\beta_2 - \beta_1} } +
{ e^{ - \beta_3 - \beta_1 } \over 1-e^{-\beta_3 - \beta_1} } +
{ e^{ - \beta_3 - \beta_2 } \over 1-e^{-\beta_3 - \beta_2} }
\eea
We describe the details of this computation in  appendix \ref{indexapp}.
Let us summarize here some of the results. In appendix \ref{indexapp} we
 show that for the
$N_5=1$ case we simply get the contributions expected from summing
over the representations in figure \ref{young}(b). These
contributions have the form expected from the BPS states on the
string theory side coming from the first four transverse
dimensions, the dimensions along the fivebrane. So we expect that
the extra contribution in \nref{bignf} should correspond to the
contribution of the second set of four dimensions. In other words,
it should be related to the BPS states in the two dimensional
field theory (\nref{wzwsigma} with \nref{diskbmn}) describing the
second four transverse dimensions. In order to extract that
contribution it is necessary to match the extra contribution we
observe in \nref{bignf} to the contributions we expect from
protected representations. In other words, we can compute the
index $I$ for various protected representations and we can then
match them \nref{bignf}. In appendix \ref{indexapp} we compute
this index for atypical (short) representations and we show that
\nref{bignf} can be reproduced by summing over
representations of the form
shown in figure \ref{young}(f). In terms of the notation
introduced in \cite{markvr2} (see figure \ref{young}(g)), which
uses the Dynkin labels,
 we expect representations with $(a_1,a_2,a_3|a_4|a_5) =
(0,p,0|a_5+1|a_5)$ with $p\geq 0$ and $a_5 = 2 (n-1)$,
$n=1,\cdots$ but $n \not =0 ~mod(N_5)$. All values of $p$ and $n$
that are allowed appear once. Representations with various values
of $p$ contribute with states that can be viewed as arising from
the product of representations of the form in figure
\ref{young}(b) and \ref{young}(e). The ones in figure
\ref{young}(b) were identified with the first four transverse
dimensions. So we interpret the sum over $p$ as producing strings
of various lengths given by the total powers of $Z$, plus the BPS
states which are associated to the first four (free) dimensions on
the string. So we conclude that the BPS states that should be
identified with the second four dimensions should be associated to
the sum over $n$. Thus we expect from gauge theory side that the
field theory on the string associated to the second four
dimensions should have an index given by \be \label{indestr} {\cal
I}_{expected} = \sum_{n=1}^\infty
 e^{ - 2 n \gamma} - \sum_{n=1}^\infty e^{ -2 n N_5 \gamma}
\ee We include the details of derivation in Appendix
\ref{indexapp}.  So we see that this differs from \nref{obtind} by
a non-perturbative terms in $1/N_5$ of the form $e^{- 2 N_5
\gamma}$.
 We view \nref{indestr} as the gauge theory prediction for BPS states
on the string theory side. Here we have checked that this matches
the string theory in a $1/N_5$ expansion, but it would be nice to
obtain the second term in \nref{indestr} (which could be viewed
as a non-perturbative correction to \nref{obtind})
from an analysis of the
two dimensional field theory based on the WZW model plus linear
dilaton theory with a potential. These theories have a large group
of symmetries and the theories with no potential are solvable. It
would be nice to see whether \nref{wzwsigma} is integrable.

\section{Theories with 16 supercharges and
$U(1) \times SO(4) \times SO(4) $ symmetry group }
\renewcommand{\theequation}{3.\arabic{equation}}
\setcounter{equation}{0}

 \label{iibtheory}

In this section we will discuss another class of theories with 16
supercharges. In this case the supersymmetry group has a $U(1)
\times SO(4) \times SO(4)$ bosonic symmetry, where the two $SO(4)$
act on the supercharges. The general form of type IIB
supergravity solutions with these symmetry was found in
\cite{llm}, and its form is rewritten in appendix
\ref{formulasllm}. Solutions depend non-trivially on three
coordinates $x_1,x_2,y$, where $y\geq 0$. The solution is
parametrized by a function $z(x_1,x_2,y)$ which obeys a linear
equation \be
 \partial_i \partial_i z + y \partial_y ( { \partial_y z \over y} ) =0
\label{zequ} \ee Regular  solutions are in one to one
correspondence with droplets of an incompressible fluid in the
$x_1 , x_2$ plane. These droplets correspond to two possible
boundary conditions $z= \pm \half$ at $y=0$ which geometrically
are associated to one of two $S^3$s   shrinking  to zero  smoothly
at $y=0$.
These solutions are much easier to obtain
than the solutions discussed in the previous sections because the
problem is precisely linear and the boundary conditions are very
simple. In the special case that the $x_1 ,x_2 $ plane is infinite
and we have finite size droplets, the solutions correspond to 1/2
BPS states in $AdS_5 \times S^5$ \cite{llm}.

\begin{figure}[htb]
\begin{center}
\epsfxsize=5in\leavevmode\epsfbox{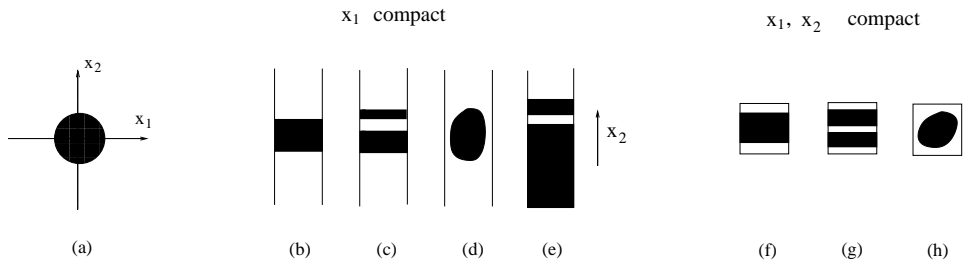}
\end{center}
\caption{ In (a) we see a circular droplet in the uncompactified
$x_1,~x_2$ plane which corresponds to the vacuum of ${\cal N}=4$
super Yang Mills. In (b,c,d) we show different vacua in the case
that we compactify the $x_1$ coordinate. This ``uplifts" ${\cal
N}=4$ super Yang Mills to a 4+1 dimensional gauge theory, or more
precisely to the $(0,2)$ six dimensional field theory  that lives
on $M5$ branes. Figure (d) shows the limit to the $M5$ brane
theory when the $x_1$ dependence recovers. If we compactify also
$x_2$, as in (f,g,h) we get a little string theory whose low
energy limit is a Chern Simons theory. If the sizes of $x_1$ and
$x_2$ are finite, we get the theory on $R\times T^2$ and figures
(f,g,h) show different vacua.
 As we take both sizes to zero, we obtain
the theory on $R^{1+2}$. The configuration in (e) corresponds to a vacuum of
the theory of
M2 branes with a mass deformation.}
\label{mcyltorus}
\end{figure}

We can now also consider cases where we compactify the $x_1,x_2$
plane. Since the asymptotic structure of the $x_1,x_2$ plane has changed, these
solutions are dual to other field theories.
 The case that $x_1$ is compact and $x_2$ is non-compact was
discussed in \cite{llm}. Let us summarize those results. If we
have a droplet that is bounded in the $x_2$ direction, like in
figure \ref{mcyltorus}(b,c,d), then the dual boundary theory can be
thought of as the theory of $N$ M5 branes on $R \times S^1 \times
S^1 \times S^3$. When one of the $S^1$s is very small we can think
of this as a theory of D4 branes
 on $R \times S^1 \times S^3$.
 A simple way to understand this theory is as follows. We consider one of the complex
 transverse scalars of ${\cal N}=4$ super Yang Mills. When the Yang Mills theory is
 on $ R \times S^3$ the lagrangian  contains a term of the form
 $ -\half (|D Z|^2  +   |Z|^2)$. We can now write $Z = e^{it}(Y + i X)$. Then the lagrangian
 becomes $ -\half (D X)^2 -\half (D Y)^2 - 2 Y D_0 X $.
 We now see that the problem
 is translational invariant in $X$. Actually,
 the problem looks like a particle in a magnetic field.\footnote{
 Note that the $(x_1,x_2)$ coordinates appearing in the gravity solution
  correspond to the coordinates $(X ,Y)$ in the field theory.}
  Note that the Hamiltonian associated to this
 Lagrangian is equal to $H' = H-J$ where $H$ is the original Hamiltonian which is
 conjugate to translations in the time direction and $J$ is the generator of $SO(6)$
 that rotates the field $Z$. If one compactifies the direction $X$, using the procedure
 in \cite{Taylor}, we get the five dimensional gauge theory living on D4 branes,
 see appendix \ref{4dand5dsym}.
 This description of the theory is
 appropriate at weak coupling or long distances on the D4 branes. The proper UV definition
 of this theory is in terms of the six dimensional $(0,2)$ theory that lives on M5 branes.
 So we have the theory on M5 branes on $ R^{1,1} \times S^1 \times S^3$.
 We could,
 of course, decompactify this theory and consider the theory of M5 branes on
 $R^{2,1} \times S^3$. These theories preserve 16 supercharges. The process of
 compactifying the coordinate $X$ broke the 32 supersymmetries to 16. When this
 theory is on $R^{1,1} \times S^1 \times S^3$ or $R^{1+2} \times S^3$, the size
 of $x_1$ should be taken to zero and the solutions correspond to those in
 figure \ref{mcyltorus}(b,c).

Let us consider the D4 theory  on $R\times S^1 \times S^3$ (or the
M5 theory  on $R\times T^2 \times S^3$ in the UV limit). This
theory has a large number of supersymmetric vacua. The structure
of these vacua is captured by the $1+1$ dimensional lagrangian
 \be \label{dflag} \int Tr[-{1 \over 4}
F^2 - {\half}(D Y)^2 - Y F ] \ee The space of vacua is the same as
the Hilbert space of 2d Yang Mills on a cylinder \cite{llm}. All these vacua
have zero energy. At first sight we might expect the
 theory on $R^{1,1} \times S^3$ to have a continuum family of vacua related to
 possible expectation values for $Y$. Note, however, that the electric field is given
 by $ E_1 \sim F_{01} + 2 Y$. For zero energy configurations $F_{01} =0$. So the quantization
 condition for the electric field will  quantize the values of $Y$.
 This is good, since, as we explain in the appendix \ref{poincare} the supersymmetry
 algebra does not allow massless particles.
In fact,  the spectrum of states around each of these vacua has a
mass gap. The explicit gravity solutions were derived in
\cite{llm} and are written in the appendix \ref{regdfour}. In
appendix \ref{regdfour}  we show  that the dilaton $\Phi$, as well
as the warp factor are bounded  in the IR region for any droplet
configuration of this type.   They never go to zero and the
solution is everywhere regular. This is related  to the fact that
the dual field theory has a mass gap.

Another configuration we can consider in the case that
 we have  a cylinder in the $x_1,x_2$ plane is shown in figure
 \ref{mcyltorus} (e). In this case
  we fill the lower half of the cylinder.
 In this case we get the M2 brane
theory in $2+1$ dimensions with a mass deformation. We get this
theory in $R^{1+2}$   after setting the radius of $x_1$ to zero
and taking the strong coupling limit (and doing the obvious
U-duality transformations\footnote{This theory via IIB/M duality
corresponds to the DLCQ of IIB plane-wave string theory
\cite{banks}, see also \cite{Sheikh-Jabbari:2004ik}.}). If the
size of $x_1$ and the string coupling are finite, then we get the
theory on $R\times T^2$. This theory was discussed in
\cite{Pope:2003jp,Bena:2004jw,llm}, and has an interesting vacuum
structure, corresponding to M2 branes polarized into M5 branes
wrapping two possible $S^3$s.

Let us discuss the situation when the $x_1,x_2$
plane is compactified into a two torus, as in figure
\ref{mcyltorus}(f,g,h). We have a 2 dimensional
array of periodic droplets\footnote{In this case, in the large $y$ region,
the  solution looks similar to the solution we would obtain if
we take the full $x_1,x_2$ plane and we consider a ``grey" configuration
 filled with a fractional density. It
shares some similarity but is different from the situation
considered in e.g. some of the references in \cite{ensemble} where
``grey" regions are finite.}. Let us start first with a
description of the gravity solution. An important first step is to
find the asymptotic behavior of the solution. The function $z$,
which obeys \nref{zequ},   goes to a constant at large $y$. We can
find the value of the constant by integrating $z$ over the two
torus at fixed $y$. The result of this integral is independent of
$y$, and we can compute it easily at $y=0$ where it is given by
the difference in areas between the two possible boundary
conditions, $z= \pm \half$. So we find $z=\frac {1}{2}
\frac{N-K}{N+K}$ asymptotically, where we used that the areas are
quantized due to the flux quantization condition \cite{llm}, so
that $N,K$ are the areas of the fermions and the holes
respectively. After doing T-dualities on both circles of the $T^2$
and an S duality we find that the
 solution is asymptotic to
\begin{eqnarray}
ds^{2}_{10} &=&-dt^{2}+du_{1}^{2}+du_{2}^{2}+N\alpha ^{\prime
}d\Omega _{3}^{2}+ K \alpha ^{\prime }d\widetilde{\Omega
}_{3}^{2}+\frac{N K }{N+K}\alpha^{\prime } d\rho ^{2} \label{nsasym} \\
e^{\Phi } &=&g_{s}\sqrt{NK}\sqrt{N+K}\alpha ^{\prime 3/2}e^{-\rho
}   \\
H_{3} &=& 2N \alpha^{\prime} d^{3} \Omega +2K \alpha^{\prime}
d^{3} \widetilde{\Omega } \label{nsasym last}
\end{eqnarray}
We can view this as a little string theory in $1+2$ dimensions.
These (asymptotic) solutions are not regular as $\rho \to - \infty
$ since the dilaton increases. In that region we should do an S
duality and then T-dualities back to the original type IIB
description. Then, once we choose a droplet configuration, the
solution is regular. This procedure works only if the coordinates
$u_1,u_2$ in \nref{nsasym} are compact. Of course, we could also
consider the situation when these coordinates are non-compact. In
that case we have Poincare symmetry in 2+1 dimensions. In fact,
such a solution appears as the near horizon limit of two
intersecting fivebranes\footnote{ One can make a change of
variables $e^{2\rho }=\sqrt{N+K}\alpha ^{\prime 1/2}r_{1}r_{2}, ~
u_{2} =\frac{\alpha ^{\prime 1/2}}{\sqrt{N+K}}(N\log r_{1}-K\log
r_{2})$, and then $r_1$ and $r_2$ become the transverse radial
directions of the two sets of fivebranes (intersecting on
$R^{1,1}$) respectively in the near horizon geometry, where the
number of supersymmetries is doubled.}
\cite{Khuri:1993ii},\cite{Gauntlett:1996pb} and was recently
studied in \cite{itzhakiseiberg}. Note that the asymptotic
geometry  \nref{nsasym}-\nref{nsasym last} is symmetric under \be
K \leftrightarrow N \ee which is associated with the symmetry $z
\leftrightarrow -z$. So we expect that this is a precise symmetry
of the field theory.

It is interesting to start from the D4 brane theory that we
discussed above and then compactify one of its transverse
directions, the direction $Y$ in \nref{dflag}. The lagrangian in
\nref{dflag} is not invariant under infinitesimal translations of
$Y$, but it is invariant under discrete translations if the period
of $Y$ is chosen appropriately. Following the standard procedure,
\cite{Taylor}, (see appendix \ref{4dand5dsym} for details) we
obtain a theory in six dimensions which can be viewed as the
theory arising on $N$  D5 branes that are wrapping an $R^{1,1}
\times S^1 \times S^3$ with $K$ units of RR 3 form flux on $S^3$.
This RR flux induces a level $K$ three dimensional Chern-Simons
term. In fact by compactifying $Y$ from (\ref{dflag}) we get a 2+1
action ${K \over 4\pi}\int Tr[ -{1 \over 4} F^2 + \omega_{cs} ]$
on $R \times T^2$ with Chern Simons term.
It turns out that the gauge coupling constant is also set by  $K$.
 Perhaps a simple way to
understand this is that the mass of the gauge bosons, which is due
to the Chern Simons term is related by supersymmetry to the mass
scale set by the radius of the threesphere, which we can set to
one. This implies that $g^2 K \sim 1$. This derivation makes sense
only when $K/N$ is large and we could be missing finite $K/N$
effects. Notice that in this limit the ${\tilde S}^3$ that is
interpreted as the worldvolume of $N$ D5 branes is larger than the
other $S^3$ in \nref{nsasym}-\nref{nsasym last}.
The gauge theory description is valid in the IR but the proper
UV definition of this theory is in terms of the little string theory
in \nref{nsasym}.
The theory has a
mass gap for propagating excitations but is governed by a $U(N)_K$
Chern Simons theory at low energies. The $U(1)$ factor is free and
it should be associated to a ``singleton" in the geometric
description. On the other hand, it seems necessary to find
formulas that are precisely symmetric under $K \leftrightarrow N$.
More precisely, in the limit $N/K$ large we get a $U(K)_N $
Chern-Simons theory by viewing the theory as coming from $K$ D5
branes wrapping the other $S^3$. Interestingly, these two Chern
Simons theories are dual to each other
\cite{levelrankdual}\footnote{Level rank duality, as analyzed in
\cite{levelrankdual},  holds up to pieces which comes from free
field correlators. This means that we have not checked whether the
$U(1)$ factor, as we introduced it here,
 leads to a completely
equivalent theory. }, which suggests that this is the precise low
energy theory for finite $N$ and $K$. Similar conclusions were
reached in \cite{itzhakiseiberg}. Of course, in our problem we do
not have just this low energy theory, we have a full massive
theory, with a mass scale set by the string scale. We do not have
an independent way to describe it other than giving the asymptotic
geometry \nref{nsasym}-\nref{nsasym last}, as is the case with
little string theories. On the other hand one can show that the
symmetry algebra implies that the theory has a mass gap, see
appendix \ref{poincare}.

We can compute the number of vacua from the gravity side. There we
have Landau levels on a torus where we have total flux $N+K$ and
we have $N$ fermions and $K$ holes. This gives a total number of
vacua \be \label{nmbgrav} D_{grav}(N,K) = { (N+K)! \over K!~N! }
\ee and the filling fraction $N \over N+K$. Actually, to be more
precise, we derive this Landau level picture as follows. We start
from the gravity solutions which are specified by giving the shape
of droplets on the torus. We should then quantize this family of
gravity solutions. This was done in \cite{liat} (see also \cite{mandaletal}),
who found that
the quantization is the same as the quantization for the
incompressible fluid we have in the lowest landau level for $N$
fermions in a magnetic field. We now simply compactify the plane
considered in \cite{liat}.
 This procedure is
guaranteed to give us the correct answer for large $N$ and $K$.
The number of vacua computed from $U(N)_K$ agrees with
\nref{nmbgrav} up to factors going like $N$, $K$ or $N+K$ which we
have not computed. These factors are related to the precise
contribution of the $U(1)$\footnote{The number of vacua for
$SU(N)_K$ Chern Simons is given by ${(N+K-1)! \over K! (N-1)!}$.}.
In order to compare the field theory answer to the gravity answer
one would have to understand properly the role of ``singletons",
which could give contributions of order $N$, $K$, etc. We leave a
precise comparison to the future but it should be noted that we
have a precise agreement for large $N$ and $K$ where the gravity
answers are valid.

We have non-singular gravity solutions if we choose simple
configurations for these fermions where they form well defined
droplets. The particle hole duality of the Landau problem is the
level rank duality in Chern-Simons theory, and is
$K\leftrightarrow N$ duality of the full configuration.

\subsection{Supersymmetry algebra}

An unusual property of all the theories we discussed above is that
their supersymmetry algebra in 2+1 (or 1+1) dimensions is rather
peculiar. In ordinary Poincare supersymmetry the generators
appearing in the right hand side of the supersymmetry algebra
commute with all other generators. This is actually a theorem for
$d\geq 4$ \cite{weinberg}. For this reason they are  called {\it
central} charges. In our case the superalgebra has anticommutators
of the form \be \label{salg} \{Q_{\alpha i},Q_{\beta j}\} =
2\tilde{\gamma}_{\alpha \beta }^{\mu }p_{\mu }\delta _{ij}+ 2m
\epsilon _{\alpha \beta }\epsilon _{ijkl}M_{kl} \ee where $m$ is a
constant of dimension of mass,
 $i,j$ are $SO(4)$ indices and $M_{ij}$ are $SO(4)$ generators.
This superalgebra appeared in the general classification in
\cite{nahm}. In this paper we have set $m=1$ for convenience. This
choice is related to the choice of mass scales (e.g. radius of
$S^3$) appearing in the various theories. These generators do not
commute with the supercharges. So this is a Poincare superalgebra
with {\it non-central} charges \footnote{This situation, is of
course, common in anti-de-Sitter superalgebras. It has also been
observed before in some deformations of Euclidean Poincare
superalgebras \cite{nsnb}.}. The $SO(4)$ that appears in
\nref{salg} is the product of an $SU(2)$ that acts on the first
$S^3$ times another $SU(2)$ which acts on the second $S^3$, where
the $S^3$s we mention here are the three spheres in the geometric
description. There are other supercharges which which transforms
under another $SO(4)$. In appendix \ref{poincare} we write down
this algebra more explicitly and we write down various lagrangians
with this symmetry. The truncation of this algebra to 1+1
dimensions is written in \nref{commutsu}.

All these theories have interesting BPS particles. Again for large $J$ we have
simple plane wave limits. In this case the plane wave geometry is basically the
one corresponding to the standard IIB plane wave. As before, it is interesting to
find out where the BPS geodesics lie in the geometry. Let us suppose that we
consider a particle carrying spin under a generator $J=J_{12}$ in the $SO(4)$ which
rotates the first sphere.
Using the metric in \cite{llm} we can see that
\be
{ E^2 \over J^2 } = { 1 \over \half +z }
\ee
In the solutions we consider here $ | z| \leq \half$. So we find that the energy
is minimized when $z=\half$. This corresponds to the regions in
the $y=0$ plane where the other
$S^3$ shrinks to zero size. In addition we have to sit at a point where
$V_i=0$. Where this point is depends on the distribution of the other droplets, but
one can see that within each droplet there is a point where $V_i=0$. This implies that
in a configuration with many droplets one will have as many BPS geodesics as droplets
of the type we are considering.
One could probably derive exact indices, or partition functions, that count these
BPS particles. These are BPS versions of the field theory objects considered in
\cite{strass}.

\subsection{Solutions with $SO(2,2) \times SO(4) \times U(1)$ symmetry }

By performing a simple analytic continuation
 of the type considered in \cite{llm} it is possible
to write an ansatz of the form
\begin{eqnarray}
ds_{10}^{2} &=&y\sqrt{\frac{2z+1}{2z-1}}ds_{AdS_{3}}^{2}+y\sqrt{\frac{2z-1}{2z+1}}%
d\tilde{\Omega}_{3}^{2}+\frac{2y}{\sqrt{4z^{2}-1}}(d\chi +V)^{2}+\frac{\sqrt{%
4z^{2}-1}}{2y}(dy^{2}+dx^{i}dx^{i})  \notag  \label{solmetric2} \\
F_{5} &=& -{1\over4}\{d[y^2 {2z+1 \over 2z-1}(d\chi +V)]+y^{3}\ast
_{3}d({\frac{z+{\frac{1}{2}}}{y^{2}}}) \}\wedge dVol_{AdS_{3}} - \notag \\
&& {1\over4}\{d[y^2 {2z-1 \over 2z+1}(d\chi +V)]+y^{3}\ast
_{3}d({\frac{z-{\frac{1}{2}}}{y^{2}}})\}\wedge
d^{3}{\tilde{\Omega}}
\\
dV &=& { 1 \over y} * d z
\end{eqnarray}%
where $z$ obeys
\be \label{zeqns}
\partial _{i}\partial _{i}z+y\partial
_{y}({\frac{\partial _{y}z}{y}})=0.
\ee

We can now look for solutions where the $AdS_3$ factor does not
shrink but where the $S^3$ factor could shrink. Regularity
requires $z=1/2$ at $y=0$. We look for solutions where $z\geq 1/2$
everywhere. In order to obtain non-trivial solutions we put
charged sources on the right hand side of \nref{zeqns}. Let us
consider a source that is localized at $y=y_0$, $\vec x = \vec
x_0$. We   take $\vec x_0 =0$ for the time being. It turns out
that if we introduce the right amount of charge, the circle
parametrized by $\chi$ shrinks in a smooth way, combining with
$y,\vec x$ to give a space that locally looks like the origin of
$R^4$. More precisely, this occurs if the function $z$ behaves
near $y=y_0$ as
\begin{equation}
z\approx \frac{y_{0}}{2\sqrt{(y-y_{0})^{2}+|{\vec x}|^{2}}}
\end{equation}%
so we we see the charge $Q_{0}$ at $y_{0}$ is equal to $y_{0}/2$.
To summarize, we have the following equation and boundary
condition for regular solutions
\begin{eqnarray}
&&\partial _{i}\partial _{i}z+y\partial _{y}({\frac{\partial _{y}z}{y}})=%
- \underset{l=1}{\overset{n}{\sum }}\frac{y_{l}}{2} (4 \pi) \delta (y-y_{l})\delta ^{2}(
\vec x - \vec x_l)  \label{famso} \\
&&z |_{y=0}={1 \over 2}
\end{eqnarray}%
where we can have several charges located at ($y_{l},\vec x_l$).
If we do not have coincident points in three dimensions the
solution is smooth.

Notice that for large $y$ and large $y_l$ these solutions reduce to the
usual Gibbons Hawking metrics \cite{ggsh} times $R^6$,
where the $R^6$ comes from the large
radius limit of $AdS_3$ and $S^3$ in \nref{solmetric2}.

\begin{figure}[htb]
\begin{center}
\epsfxsize=5in\leavevmode\epsfbox{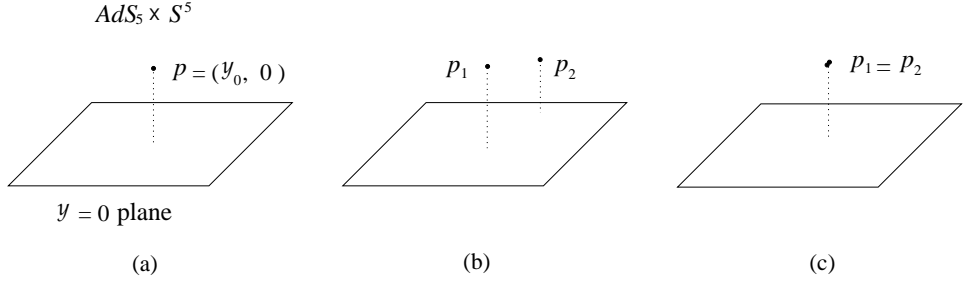}
\end{center}
\caption{ (a) In the analytically continued IIB ansatz, the $AdS_5
\times S^5$ solution corresponds to boundary conditions $z=1/2$ at
the  $y=0$ plane and a point charge $Q_0=y_0/2$ located at
$p=(y_0,0)$, away from the $y=0$ plane. In (b), there is a more
general smooth configuration where there are two point charges
located at different points $p_1=(y_1, \vec x_1)$ and $p_2 =(y_2,
\vec x_2)$. Their charges are $y_1/2,y_2/2$ respectively. In (c)
two such point charges merge at the same point, and they develop a
$R^4/Z_2$ singularity. If there were $k$ coincident such charges,
then they give rise to $R^4/Z_k$ singularity.} \label{continued}
\end{figure}

The simplest and most symmetric solution corresponds to a single
point charge  of strength $y_0/2$ at ($y_{0}, 0$), see figure
\ref{continued}(a), which corresponds to
\begin{eqnarray}
z &=&\frac{r^{2}+y_{0}^{2}+y^{2}}{2\sqrt{%
(r^{2}+y_{0}^{2}+y^{2})^{2}-4y^{2}y_{0}^{2}}}  \label{ads5} \\
V_{\phi } &=&\frac{r^{2}-y_{0}^{2}+y^{2}}{2\sqrt{%
(r^{2}+y_{0}^{2}+y^{2})^{2}-4y^{2}y_{0}^{2}}}
\end{eqnarray}
It turns out that this solution is  $AdS_5 \times S^5$. We can see
this by the coordinate change
\begin{eqnarray}
x_1+ix_2 &=& r e^{i \phi}\\
y &=&y_{0}\cosh u\cos \beta \\
r &=&y_{0}\sinh u\sin \beta \\
\psi &=&\chi -\phi /2 \\
\alpha &=&\chi +\phi /2
\end{eqnarray}
So that we get
\begin{eqnarray}
&ds_{10}^{2}=y_{0}[(\cosh ^{2}uds_{AdS_{3}}^{2}+du^{2}+\sinh ^{2}u d%
{\psi}^{2})+(\cos ^{2}\beta d\tilde{\Omega}_{3}^{2}+d\beta
^{2}+\sin ^{2}\beta d{\alpha}^{2})]  \label{m5s5} \\
&F_{5} =4y_{0}^{2}[\cosh ^{3}u\sinh udu\wedge d{\psi }\wedge
dVol_{AdS_{3}}+\cos ^{3}\beta \sin \beta d\beta \wedge d\alpha
\wedge d^{3}\tilde{\Omega}]
\end{eqnarray}%
where \be y_{0}=R_{AdS_{5}}^{2}=R_{S^{5}}^{2}=\sqrt{4\pi
g_{s}N\alpha ^{\prime 2}} \label{radiva} \ee It is $AdS_5 \times
S^5$ written with an $AdS_3 \times S^1$ slicing. This particular
solution has more symmetry and more supersymmetry than other
generic solutions in the family. It is perhaps useful to note that
the $AdS_3\times S^1$ boundary is conformally related to $R\times
S^3$
\begin{eqnarray}
ds_{AdS_{3}\times S^{1}}^{2} &=&[-\cosh ^{2}vdt^{2}+dv^{2}+\sinh
^{2}vd\varphi
^{2}+d\psi^{2}]  \label{conformal_transform} \\
&=&\frac{1}{\sin ^{2}\theta }[-dt^{2}+\cos ^{2}\theta d\varphi
^{2}+d\theta^{2}+\sin ^{2}\theta d{\psi }^{2}] \\
&=&\frac{1}{\sin ^{2}\theta }ds_{R\times S^{3}}^{2} ~;~~~~~~~~~
\sin \theta = { 1 \over \cosh v }
\end{eqnarray}%
Notice that the conformal factor blows up on a circle of $S^3$.
The blown up position corresponds to the boundary of $AdS_3$.

Now we consider situations for many point charges. Consider a
charge $Q_l$ at a position $y_l$ and integrate $F_5$ over the
$S^3$ and an $S^2$ surrounding the charge in the $(y, \vec x)$
space. From \nref{solmetric2} the result is proportional to $y_l
Q_l \sim N_l$, where we used that the flux is quantized. For
smooth solutions $Q_l = y_l/2$ and we obtain a relation
$y_{l}=\sqrt{4\pi g_{s}\alpha ^{\prime 2} N_l} $, the same as
\nref{radiva}. Notice that \nref{famso} describes a family of
solutions where we can change continuously the values of $\vec
x_l$ but we cannot change continuously the values of $y_l$ due to
the flux quantization condition.

Let us now start with a smooth solution that has two equal
charges, at $y_{1} = y_{2}$, and we take the limit when these two
charges lie on top of each other (i.e. we set $\vec x_1 = \vec
x_2$). We get a singular solution since the total charge is twice
the value that would make the solution regular. Such a solution
has a $Z_2$ singularity and near the position of the charge it
looks like $R^4/Z_2$. Similarly if we take $k$ equal charges
coincident we get a space that locally is $R^4/Z_k$ (or an
$A_{k-1}$ singularity). See figure \ref{continued}(b) and
\ref{continued}(c).

Let us now consider the corresponding field theory. These
solutions are related to ${\cal N}=4$ super Yang Mills on $AdS_3
\times S^1$. Let us focus on a complex combination, $Z$, of two of
the six scalar fields. Let us expand this field in Kaluza Klein
modes on $S^1$. The constant mode leads to a field with negative
mass on $AdS_3$. This negative mass arises from the conformal
coupling of the scalar fields\footnote{ Of course, this mass obeys
the Breitenlohner-Freedman bound \cite{brfr}.}. The lagrangian of
this theory contains the term ${1\over 2}(|D_{\psi}Z|^2-{{\cal R}
\over 6}|Z|^2)$, and the scalar curvature ${\cal R}=-6$ for
$AdS_3$ with unit radius. We can obtain a massless field on
$AdS_3$ if we take the first Kaluza Klein mode of $Z$ on the
$\psi$ circle. Namely, we can consider $Z = \hat z e^{i \psi}$. We
can take $\hat z $ to be a diagonal matrix with eigenvalues $\hat
z_l$ where the multiplicity of each eigenvalue is $N_l$, where
$\sum_l N_l = N$. It is natural to conjecture that this state is
related to the gravity configuration with $x^1_l + i x_l^2 = {\hat
z}_{l}$ and $y_{l}^2 \sim N_l $. This seems to give rise to a
picture where the symmetries match on the two sides. On the other
hand, it seems puzzling that as we take ${\hat z}_{i} \to {\hat
z}_{j}$ we do not get the same solution as the one corresponding
to the situation where we have a single eigenvalue with
multiplicity $N_i + N_j$. In fact we get a smooth configuration if
$N_i \not = N_j$ and a  singular one (with a $Z_2$ singularity) if
$N_i = N_j$. More generally we get a $Z_k$ singularity if ${\hat
z}_{1},{\hat z}_{2},...,{\hat z}_{k}$ are coincident and their
original multiplicities are the same. Since the field theory is on
$AdS_3\times S^1$ we will need boundary conditions for the fields
at the boundary of $AdS_3$, so perhaps the gauge symmetry is
broken by the boundary conditions when we have multicenter
solutions. In other words, perhaps the multicenter solutions only
exist when the gauge symmetry is already broken at the boundary.
Thus we cannot restore it by taking ${\hat z}_i \to {\hat z}_j$.
Clearly a better understanding of this point is needed. Other
gravity solutions with an $AdS_3\times S^1$ boundary were recently
considered in \cite{vazquez}. It is possible
that those solutions are related to a subset of the ones considered
in this paper. Similar, but different, solutions were analyzed in
  \cite{mannads}.

\section{Conclusions}
\renewcommand{\theequation}{4.\arabic{equation}}
\setcounter{equation}{0}

In this paper we have studied various theories with sixteen
supercharges. These theories are interesting because they have a
dimensionless parameter that allows us to interpolate continuously
between strong and weak  coupling. These theories have simple
observables, such as the spectrum of gauge invariant states. In
this respect they are rather similar to ${\cal N}=4$ Yang Mills on
$R\times S^3$. In fact, they arise as truncations of ${\cal N}=4$.
It is therefore interesting to study the similarities and
differences between these theories and ${\cal N}=4$. From some
points of view these theories are simpler. For example, the plane
wave matrix model is just an ordinary quantum mechanical theory
with a finite number of degrees of freedom. On the other hand they
are  more complicated because they have less symmetry than ${\cal
N}=4$ Yang Mills. For example, in the theories we studied here
symmetry alone does not determine the gravity solutions. These
theories have many vacua, as opposed to ${\cal N}=4$ which has
only one. In addition, the physics around different vacua can have
different qualitative features.

The first set of theories that we studied has $\su (2|4)$ symmetry
group. We discussed some features of these field theories. We gave
explicit formulas for the counting of 1/2 BPS states \nref{stzth},
\nref{indppw} and we explained how to construct an index
\nref{defindfull} carrying information about more general BPS
states.   The single trace contribution to the index was computed
in \nref{bignf}. This can, in turn, be translated into an index
for the two dimensional worldsheet theory \nref{sutwoind}
\nref{indestr} describing near BPS string states. The general form
of the gravity solutions is written in \nref{IIA ansatz}-\nref{IIA
ansatz last} (from \cite{llm}) with the boundary conditions
corresponding to an electrostatic problem  involving conducting
disks in three dimensions. For given asymptotic boundary
conditions there are many possible disk configurations. The number
of disk configurations matches with the number of expected vacua
in the field theory. Full explicit solutions were given in a
couple of cases \nref{ns5sol}-\nref{ns5sol last} and
\nref{d2sol}-\nref{d2sol last}. The solution in
\nref{d2sol}-\nref{d2sol last} is dual to
 2+1 Yang Mills theory on $R\times S^2$ for the vacuum with
 unbroken $U(N)$ gauge symmetry.
All  solutions are smooth in the IR region and they have no
horizons. We have then focused on states with large $J$, where $J$
is an $SO(6)$ generator. We treat these large $J$ states in the 't
Hooft limit, where $J$ is large but kept finite in the large $N$
limit, so that we can neglect back reaction. In this limit we can
think of the BPS states as massless geodesics moving along a
circle inside an $S^5$ and sitting at some point in the rest of
the coordinates. These geodesics sit at the points corresponding
to the tip of the disks. For a given vacuum there are as many
distinct geodesics as there are disks in the electrostatic
picture. Looking at the spacetime near these geodesics we found
the general pp-wave solution \nref{ppgen}-\nref{ppgen last}. We
then used these metrics to study the spectrum of near BPS states.
In the string theory side, at large 't Hooft coupling, we can
quantize the string in light cone gauge. Four of the transverse
dimensions are described by free massive fields. These are
associated to oscillations of the string in the $S^5$ directions.
The near BPS spectrum associated to stringy oscillations along
these directions is characterized by a single parameter which
corresponds to  the radius of the $S^5$ at the position of the BPS
geodesic. This parameter is non-universal, in the sense that it
depends on the theory we consider, the vacuum that we pick and
also the particular BPS geodesic that we are expanding around,
 e.g. see \nref{s5d2}, \nref{s5bmn}. On the other hand,  the metric
 {\it very}  close
to each massless geodesic has the form of the IIA plane wave
\nref{iiapp}. So the form of this metric is a universal feature of
the near BPS limit in these geometries. There is however an
important subtlety. Even though {\it very} near the massless
geodesic the metric behaves as in \nref{iiapp} it can happen that
the geometry has other features that are at distances comparable
to the string scale. This happens when we consider the vacua of
the plane wave matrix model that correspond to $N_5$ coincident
fivebranes (with relatively small values of $N_5$). In this case
the correct string theory description  involve a massive
deformation of the WZW model plus linear dilaton theory. This 1+1
dimensional field theory has $(4,4)_m$ supersymmetry
\nref{commutsu} which has the peculiar feature of having
non-central charges. It would be interesting to see if this theory
is integrable. We considered the weak coupling spectrum of single
trace states around various vacua of this field theory. We found
that the number of transverse oscillators depends on the vacuum.
At weak coupling, for the single NS5 vacuum (the trivial vacuum)
we have only four transverse oscillation modes while for $N_5>1$
we have  eight transverse oscillation modes. To be precise, on the
gauge theory side, we only proved that the BPS spectrum of
oscillations is consistent with eight modes, there could be more
modes that do not contribute to the index. It would be nice to
perform the complete one loop analysis of this model in order to
find out precisely how many we have. We computed the number of 1/4
BPS single trace BPS states  \nref{indestr} which lead to 1/2 BPS
states on the string worldsheet. In other words, there are 1/2 BPS
states of the field theory on the string. We did not count exactly
these BPS states independently for the 1+1 dimensional field
theory but we did show that we get the right answer for large $N_5
$.

We then considered theories that arise when we take the solutions
in \cite{llm} associated to free fermions and put them on a two
torus. If we shrink the two torus to zero we get a little string
theory \nref{nsasym}-\nref{nsasym last} with poincare invariance
in $2+1$ dimensions. If we keep the torus finite, then we get this
2+1 dimensional little string theory on a two torus. This little
string theory is characterized by two integers $N$ and $K$. In the
large $K$ limit we can argue that the low energy description is
given by a $U(N)_K$ Chern Simons theory, see \nref{5dsymcs}. We
expect that the low energy theory should be exactly that of
$U(N)_K$ Chern Simons. This low energy theory is level rank dual
to $U(K)_N$ Chern Simons\footnote{ We did not check that the
$U(1)$ factor indeed works as we are describing here.} in the
large $N,K$ limit. The $K \leftrightarrow N$ symmetry is a full
symmetry of the little string theory. The solutions that we
described, which are associated to fermion droplets on a two
torus, give a semiclassical description for the various vacua of
$U(N)_K$ Chern Simons theory on a two torus. These solutions are
relevant only when the 2+1 dimensional little string theory is on
a two torus. This theory, as well as other theories that arise in
similar ways, such as the theory living on the mass deformed M2
branes \cite{Pope:2003jp,Bena:2004jw,llm} have the supersymmetry
algebra discussed in appendix \ref{poincare} which contains
non-central terms such as \nref{noncentr}.

In addition we discussed a curious family of solutions that is obtained by doing
an analytic continuation of the ansatz in \cite{llm}. These solutions
are related to ${\cal N}=4 $ super Yang Mills on
$AdS_3 \times S^1$. They seem to correspond to a peculiar Coulomb branch
where the field $Z$ has an expectation value.
But some further study is needed to elucidate the precise
relation to the field theory.

{\bf Acknowledgements }

We would like to thank Yi Li, O. Lunin, G. Moore, M. Ro\v{c}ek and
N. Seiberg for discussions. This work was supported in part by DOE
grant \#DE-FG02-90ER40542 and NSF Grant No. PHY-0243680. J.M.
thanks the hospitality of the KITP at UCSB where part of this work
was done.

\appendix

\section{   Detailed analysis of
 the regularity of the solutions}
 \renewcommand{\theequation}{A.\arabic{equation}}
\setcounter{equation}{0}

In this appendix we prove general properties of the solutions we consider in this paper.
We can show,
 using equations for $V$,  that $ \ddot V V'' - (\dot
V')^2 = - [ \rho^2 (V'')^2 +(\dot V')^2] $ is always negative. The dot denotes
$\rho \partial_{\rho}$ and the prime denotes $\partial_{\eta}$.
We
also need that everywhere $\dot V =y \geq 0$ and $V'' \leq 0$,
$\ddot V -2\dot V \geq 0$.
 In this section it is a bit more convenient
conceptually to formulate the problem in terms of a new variable
$Z = \dot V$. We see that we can express all the functions in
\nref{IIA ansatz}-\nref{IIA ansatz last} in terms of $Z$, $Z'$ and
$\dot Z$ since we can express $V''$ in terms of $\ddot V$ using
the Laplace equation. The variable $Z$ obeys the equation \be
\label{laplequ} Z'' + \rho
\partial_{\rho} ( { \partial_{\rho} Z \over \rho }) =0
 \ee
 This equation has the same form as  the equation in the
 IIB solution \nref{zequation} when we have an additional
 isometry. Note, however, that we have different boundary conditions.
  Notice that we want to show that $Z$ is non-negative.
 The boundary conditions at $\rho=0$ and on the disks imply that there $Z$ is zero.
The positivity conditions constrain the
allowed asymptotic boundary
conditions. They   allow only $V \sim \rho^2 - 2 \eta^2$, or $Z \sim \rho^2$
we have the whole $\rho$, $\eta$ plane. If we only have $\eta>0$, we could
also allow $V = \eta \rho^2 - { 2\over 3} \eta^3 $, or $Z = \eta \rho^2$.
Note that $Z$ needs to grow at infinity in both cases, since we want to impose, in
addition,  that $\dot Z \geq 0$. Notice that the structure of the equation
\nref{laplequ} is such that if $Z$ is zero at the disks and $Z$ is positive
at infinity, then $Z$ is positive everywhere. Now we want to ensure that
$\dot Z$ is positive everywhere. For this purpose it is convenient to
define $ Y = \dot Z/\rho^2 = (\partial_{\rho} Z)/\rho = - V''$, which obeys the same
equation as the variable $V$ itself. The boundary conditions are such that $Y$ is
positive far away. In addition, $Y$ it is zero on the disks.
At the origin $Y$ is
required to be regular. So it is intuitively clear that it will be positive
everywhere, except on the disks.

In order to find a non-singular solution we need an additional condition. We need to
ensure that
\be \label{condddv}
 0 \leq  { 2 \dot V \over \ddot V} \leq 1
\ee
The first inequality is obeyed automatically and is a strict inequality away
from the disks and the origin. The second inequality can be analyzed as follows.
Choose a function $U = (\partial_{\rho} V)/\rho  $. Then we need to show that
$\partial_{\rho }U \geq 0$. The equation that $U$ obeys has the form
\be
  U'' +{ 1 \over \rho^3} \partial_{\rho} ( \rho^3 \partial_{\rho} U )=0
 \ee
It it the Laplace equation in five dimensions for a system that is
$SO(4)$ rotationally symmetric. The boundary conditions at
infinity are such that $U$ is positive. At $\rho=0$ and on the
disks we have that $U$ is zero. As long as we have some finite
disks, then we see that $\dot U$ must be strictly positive
everywhere, except on the disks and possibly at $\rho=0$. Note
that in the case that we have only an infinite disk at $\eta=0$,
so that we have the solution that corresponds to the 11
dimensional plane wave, with no excitations, then we get that
$\dot U =0$ and the solution \nref{IIA ansatz}-\nref{IIA ansatz
last} is singular. This is expected since we are doing a reduction
on a circle that is null everywhere.

Our discussion so far has ensured that the solution is
non-singular and the dilaton is finite everywhere except, possibly
at $\rho =0$ and on the disks where the various inequalities that
we have discussed are saturated. In order to show that the
solution is non-singular also in these regions, we need a more
detailed analysis. For example, near $\rho=0$, we have that $\dot
V \sim \rho^2 $ and that $\ddot V - 2 \dot V \sim a \rho^4$, with
$a>0$ and $-V'' >0$ due to our previous arguments. These
conditions ensure that the dilaton stays finite and that the
solution \nref{IIA ansatz}-\nref{IIA ansatz last} is non-singular.

Doing a similar analysis near a disk we also find that the solution is regular
at the disk positions.

\subsection{Regularity of the solutions coming from D4 on $R^{1,1} \times S^3$}
\label{regdfour}

Here we analyze the regularity property of the D4 brane solutions.
In order to characterize the solution we need to give the numbers
$a_j , b_j$ which obey $ a_j < b_j < a_{j+1} \cdots$. These
numbers are the values of $x_2$ at the boundaries of the black
strips, see figure \ref{mcyltorus}(b,c). We have a black strip
between $a_j$ and $b_j$. Then the solution is given by \cite{llm}
\bea 2z &=& -1 + \sum_j { x-a_j \over \sqrt{(x-a_j)^2 + y^2} } - {
x - b_j \over \sqrt{ (x-b_j)^2 + y^2 } }
\\
 2y V_1 &=& \sum_j { y \over \sqrt{(x-a_j)^2 + y^2} } - { y \over
\sqrt{ (x-b_j)^2 + y^2 } }
\\
2z + i 2 y V_1 &=& -1 + \sum_j  (w_j - z_j)  \label{complzv}
\\ \label{formwz}
w_j &=& {  x - a_j  + i y \over \sqrt{(x-a_j)^2 + y^2} } ~,~~~~~~~~z_j
={  x - b_j  + i y \over \sqrt{(x-b_j)^2 + y^2} }
\eea
We see that the complex numbers $w_j$ and $z_j$ lie on the unit circle in the upper half
plane.

The ten dimensional solution is
 \bea ds^2_{IIA} &=& e^{2 \Phi} ( - dt^2 + d\tilde x_1^2)
+ { \sqrt{ 1 - 4 z^2} \over 2 y} (
dy^2 + dx_2^2 ) + y\sqrt{\frac{1+2z}{1-2z}}d\Omega_{3}^{2}+y\sqrt{\frac{1-2z}{1+2z}}%
d\tilde{\Omega}_{3}^{2}  \notag
\\ \label{dilf}
e^{-2 \Phi} &=& { 1 - 4 z^2 - 4 y^2 V_1^2 \over 2 y \sqrt{ 1-4 z^2} }
\\
F_{4} &=& - { e^{- 2 \Phi}  \over 4}\left[ {(1-2z)^{3/2}\over (1+
2 z)^{3/2} } *_2 d \left( y^2 { 1+ 2z \over 1- 2 z} \right) \wedge
d \tilde \Omega_3 + {(1+2z)^{3/2}\over (1- 2 z)^{3/2} } *_2 d
\left( y^2 { 1- 2z \over 1+ 2 z} \right) \wedge d  \Omega_3\right]
\notag
\\
B_2 &=& -{ 4 y^2 V_1 \over  1 - 4 z^2 - 4 y^2 V_1^2 }dt\wedge
d{\tilde x}_1 \eea Note that $g_{00}$ is determined in terms of
the dilaton. This is related to the fact that the eleven
dimensional lift of this solution is lorentz invariant in $2+1$
dimensions. We will now show that $e^{-2 \Phi}$ remains finite and
non-zero in the IR region. Of course, in the UV region $\Phi \to
\infty$ and we need to go to the eleven dimensional description.
Note that away from $y=0$ the denominator in \nref{dilf} is
non-zero. The fact that the numerator is nonzero follows from the
representation \nref{complzv} and the fact that $w_j , z_j$ in
\nref{formwz} are ordered points on the unit circle on the upper
half plane, so the norm $|2z + i 2 y V_1|<1$. As we take the $y\to
0 $ limit we see that both the numerator and denominator in
\nref{dilf} vanish. We can then expand in powers of $y$ and check
that indeed we get a finite, non-zero result, both for $ a_j < x_2
< b_j$ and $x_2 = a_j, ~b_j$.

\section{  Derivation of the D2 solution}
\renewcommand{\theequation}{B.\arabic{equation}}
\setcounter{equation}{0}

In this appendix we explain how we  obtained  the solution
\nref{d2sol}-\nref{d2sol last}. We start with the configuration of
four dimensional gauged supergravity in \cite{Chong:2004ce}, which
has
 four commuting
angular momenta in $SO(8)$. We consider the special case when only
one of the angular momenta is nonzero. This is a half BPS state of
M-theory on $AdS_4 \times S^7$ and, as such, it can be described
in terms of the general M-theory ansatz in \cite{llm}. By
comparing the expressions in \cite{Chong:2004ce} and \cite{llm} we
find that the solution corresponds to an elliptical droplet of M2
boundary conditions \nref{standbc}. The solution can be written in
the following parametric form \footnote{ We can also write the
solution corresponding to the 1/2 BPS extremal one-charge limit of
the
 $AdS_4$ black hole,
e.g. \cite{Duff:1999gh}, in the Toda form. This solution and the
solution for M2 strip both belong to the more general solution:
$e^{D}= 4\sin^2 \theta  ( 1 + z^2 H(z))/{\tilde F}(z)^2$, $ x_2 +
i x_1=(e^{- \half \varphi }{\cos }{\phi }+ i e^{\half \varphi
}\sin \phi )
 {\tilde F}{ \cos \theta }$, $y=z \sin ^{2}\theta$, where
 $\partial _{z} {\tilde F} = \frac{z {\tilde F} }{2(1 + z^2 H(z))}\cosh\varphi$ and
 $H$ obeys \nref{eqnH}. The solution that corresponds    to the
extremal $AdS_4$ black hole is $H=1+\frac{2Q}{z}$.  We can plug this in the previous
equations  and integrate to get
$\log {\tilde F}=\int \frac{z }{2(1 + z^2 H(z))}dz $.
$z$ is related to the
radius $r$ of $AdS_4$ black hole as $z=2r$.}
\begin{eqnarray}
e^{D} &=& 4 {\sin}^{2} \theta  ( 1 + z^2 H(z)) \sinh \varphi \\
x_2 + i x_1  &=& (e^{- \half \varphi } {\cos } {\phi }+ i e^{\half
\varphi } \sin \phi )
 { \cos \theta \over \sqrt{ \sinh \varphi}} \\
y &=&z \sin ^{2}\theta \\
\partial _{z}\varphi &=& \frac{-z\sinh \varphi }{1 + z^2 H(z)}, \qquad \partial _{z}(zH(z))=\cosh \varphi
\label{lasteqn1} \eea The last two first order equations
(\ref{lasteqn1}) are equivalent to a second order equation for
$zH(z)$
\be (z^{-1}+z H(z))\partial^{2}_{z}(z H(z))=1-(\partial
_{z}(z H(z)))^2 \label{eqnH} \ee Note that we still  need to
solve this equation to find a full solution.

The elliptic droplet in the $x_{1,}x_{2}$ plane is
\begin{equation}
{ x_1^2 \over a^2 } + {x_2^2 \over b^2 }=1~,~~~~~~~~~
a={1 \over \sqrt{ \sinh \varphi}} e^{\frac{\varphi }{2}}\mid _{z=0},\qquad b={1 \over \sqrt{ \sinh \varphi}}e^{-
\frac{\varphi }{2}}\mid _{z=0}
\end{equation}

If we take a limit that $a \rightarrow \infty$, $b=1$, we can let
\begin{equation}
 \sinh \varphi \approx {1 \over 2}e^{ \varphi } \approx \cosh \varphi
\end{equation}%
and  $\varphi _{0}=\varphi (z=0)$ goes to infinity.
This limit corresponds to dropping the 1 in the right hand side of (\ref{eqnH}).

If we expand around $\cos \phi \approx 1,$ we can neglect
the dependence of $x_{1}$ and we find  a solution of the 2d
Toda equation. We can now write $\varphi = \varphi_0 + \tilde \varphi$ where $\tilde \varphi$ is
stays constant in the limit. We can remove the $\varphi_0$ dependence performing
a simple symmetry transformation of the Toda equation which does not affect the eleven
dimensional solution: $e^{D}\rightarrow e^{-2\varphi _{0}}e^{D},x_2 + i x_1\rightarrow
(x_2 + i x_1)e^{\varphi _{0}},y\rightarrow y$. Now we have
the solution corresponding to a single M2 brane strip in $x_1,x_2$ plane.
\begin{eqnarray}
e^{D} &=&4\sin ^{2}\theta (1+z^2 H(z))e^{\tilde \varphi}  ,\qquad x_{2}=
e^{-\tilde \varphi} \cos \theta ,\qquad y=z\sin ^{2}\theta
\label{singleM2solution-potential1}
\end{eqnarray}%
where $\tilde \varphi$ is defined through
$ e^{\tilde \varphi} = 2 \partial_z( z H(z) )$, with $H(z)$ obeying equation
\nref{eqnH} without the one in the right hand side.
We also have a boundary condition $e^{\tilde \varphi(0)} = 2 C$.

One can see that the single strip of M2 branes in the electrostatic problem corresponds to a single charged
conducting disk in the external potential $\frac{\rho ^{2}-2\eta ^{2}}{8\beta}$.
 The solution for the whole potential is
\begin{eqnarray}
V &=&-z+ \sin ^{2}\theta \left(  z H(z)e^{-\tilde \varphi} + { z \over 2}\right) \label{singleM2solution-potential2}\\
\rho &=&2  \sin \theta \sqrt {(1+{z^2}H(z))e^{ \tilde \varphi} },\qquad \eta =-2zH(z)\cos \theta ,
\end{eqnarray}

The $S^5$ shrinks when $\sin\theta=0$, which corresponds to
 the $\rho=0$ axis, and the $S^2 $ shrinks when $z=0$,
which is a disk in $\eta=0$ plane centered around the origin and extends to finite $\rho_0$.
The size of the disk is
\begin{equation}
\rho _{0}=2\sqrt{2C}
\end{equation}
The charge density $\sigma(\rho)$ on the disk is proportional to the
jump of the $\eta $ component of the electric field
$-\partial_{\eta } V =-x_2$, so we have
\begin{equation}
\sigma(\rho) =\frac{1}{4\pi C\rho_{0}}\sqrt{\rho _{0}^{2}-\rho ^{2}}  \label{charge_density_singleM2}
\end{equation}%
which has maximum at the center and vanishes at the edge.
The full potential can be expressed in integral form
\begin{equation}
V=\frac{\rho ^{2}-2\eta ^{2}}{8\beta}+\int_{0}^{2\pi
}\int_{0}^{\rho _{0}}\frac{\frac{1}{4\pi C\rho _{0}}\sqrt{\rho _{0}^{2}-r^{2}%
}rdrd\phi }{\sqrt{\rho ^{2}+\eta ^{2}-2\rho r\cos \phi +r^{2}}}
\label{V_int}
\end{equation}
up to a constant shift.

Now we will compare the two expressions (\ref{singleM2solution-potential2}),(\ref{V_int}) and
solve the equation (\ref{eqnH}) in the limit that we drop the 1 in the right hand
side of \nref{eqnH}.
Let us look at $V$ along the $\rho =0$ line above the $\eta =0$ plane. This
corresponds to $\cos \theta =-1$.  We now integrate \nref{V_int} at $\rho=0$
and we impose the condition that $z=0$ when $\eta=0$. Comparing with
\nref{singleM2solution-potential2}
we find a  relation between $z$ and $\eta$
\begin{equation}
z=\frac{\eta ^{2}}{4\beta}-\frac{1}{4C\rho _{0}}[(\rho _{0}^{2}+\eta
^{2})\arctan {\frac{\rho _{0}}{\eta }}-\rho _{0}\eta -\frac{\pi }{2}\rho
_{0}^{2}]  \label{x_eta}
\end{equation}%
This is for $\rho =0,\eta >0$. In addition,  we know the expression for $zH(z)$ in terms
of $\eta $ from (\ref{singleM2solution-potential2})
\begin{equation}
zH(z)=\frac{\eta }{2}  \label{F_eta}
\end{equation}%
Thus (\ref{x_eta}) and (\ref{F_eta}) give a solution of $H(z)$ in a parametric form.
We also find   the relation between $\beta$ and $C$
\begin{equation}
\beta=\frac{4\sqrt{2}}{\pi }C^{3/2}  \label{D2_kC}
\end{equation}
One may now write the solution in a simpler form by
 introducing $r=\eta/\rho _{0}$
\begin{eqnarray}
zH(z) =\sqrt{2C} r, \qquad z = \frac{1}{\sqrt{2C}}[r+(1+r^{2})\arctan r]
\end{eqnarray}
where $r$ ranges from 0 to $\infty $.
This is a  solution for equation (\ref{eqnH}) when we drop the 1 in the right hand side.

We end up with the solution corresponding to single strip of M2 branes
\begin{eqnarray}
e^{D} &=&8C(1+r^{2})\sin ^{2}\theta  \notag \\
x_{2} &=&\frac{1}{2C}[1+r\arctan r]\cos \theta, ~~~~~~
y = \frac{1}{\sqrt{2C}}[r+(1+r^{2})\arctan r]\sin ^{2}\theta
\label{D2_toda}
\end{eqnarray}
and $C$ is a simple rescaling parameter that is associated to the charge of the solution.

\section{ Solution for NS5 branes on $R \times S^5$}
\renewcommand{\theequation}{C.\arabic{equation}}
\setcounter{equation}{0}

In this appendix, we write the solution for
 for NS5 branes   on $R \times S^5$ in the Toda form.
This is not necessary for anything we did in this paper but connects it to
the gauged supergravity solution of \cite{llm}.
Let us start with
 the 7d gauged-supergravity solution which corresponds to an elliptic M5
  droplet in $x_1,x_2$ plane \cite{llm}.
This solution for the 3d Toda equation is \cite{llm} \footnote{The
extremal limit of $AdS_7$ black hole, e.g. \cite{Cvetic:1999ne},
\cite{Liu:1999ai}, can be
 written in the Toda form: We start from a more general solution $e^{D}=m^{2}r^{2}f/{\tilde
F}^2, y=m^{2}r^{2}\sin \theta, x_{2}+ix_{1}=\left( e^{-\rho }\cos
\phi +ie^{\rho }\sin \phi \right) {\tilde F}{\cos \theta}$, where
$\partial _{r}\tilde{F}(r)=\frac{2m^{2}r\tilde{F}(r) }{f}\cosh
2\rho$, see \cite{llm}. The extremal $AdS_7$ black hole
corresponds to solution $F=x+Q$, we can integrate to get $\log
{\tilde F}=\int \frac{2m^{2}r }{f} dr$, and plug in these into the
more general solution.}
\begin{eqnarray}
e^{D} &=&m^{2}r^{2}f \sinh 2\rho, \quad \quad y=m^{2}r^{2}\sin \theta \\
x_{2}+ix_{1}&=&\left( e^{-\rho }\cos \phi +ie^{\rho }\sin \phi \right)
\frac {\cos \theta} {\sqrt{\sinh 2\rho}} \\
\cosh 2\rho &=&F^{\prime }, \quad\quad f=1+\frac{F}{2\sqrt{x}}, \quad\quad x \equiv 4m^{4}r^{4} \label{xr}\\
(2\sqrt{x}+F)F^{\prime \prime }&=&1-(F^{\prime })^{2} \label{eqnF}
\end{eqnarray}
where prime is the derivative with respect to $x$.
The elliptic droplet in the $x_{1,}x_{2}$ plane has axis $a=\frac{e^{\rho }}{\sqrt{\sinh 2\rho}}\mid _{r=0},
b=\frac{e^{-\rho }}{\sqrt{\sinh 2\rho}}\mid_{r=0}$ and we take the limit similar to appendix B, that $a\rightarrow \infty, b=1$.
Then we can approximate
 $\sinh 2\rho \approx \frac{1}{2} e^{2\rho} \approx \cosh 2\rho$ and $\rho _{0}=\rho|_{r=0}$ will goes to infinity.
This is equivalent to dropping the 1 in (\ref{eqnF}).
After a simple rescaling we find the solution to the 2d Toda equation
\begin{eqnarray}
e^{D} &= &m^{2}r^{2}fe^{2\rho }\approx (\sqrt{x}+\frac{1}{2}F)F^{^{\prime }},\qquad \\
x_{2} &= &e^{-2\rho }\cos \theta \approx (2F^{^{\prime }})^{-1}\cos
\theta ,\qquad \\
y &=&m^{2}r^{2}\sin \theta =\frac{\sqrt{x}}{2}\sin \theta .
\end{eqnarray}
where $F$ obeys equation \nref{eqnF} without the 1. Comparing this
to \nref{bessol}, \nref{vardefi} we can write
\begin{equation}
x=\frac{1}{4}C^{-2}\rho ^{2}I_{1}^{2}(\rho),\qquad
F=\frac{1}{2}C^{-1}\rho ^{2}I_{2}(\rho),\qquad F^{\prime
}=CI_{0}^{-1}(\rho) \label{scale_solution}
\end{equation}%
where $C$ is a trivial overall scale. This gives a solution to
\nref{eqnF} (without the 1) in a parametric form.

\section{Charge and asymptotics of the D0 brane solutions}
\renewcommand{\theequation}{D.\arabic{equation}}
\setcounter{equation}{0}
 \label{d0match}

In this appendix, we discuss the charge $N_2$ and $N_5$ and
asymptotic matching of the solutions dual to vacua of the plane
wave matrix model. We then discuss the interpolating function $f$
in the leading gravity approximation in section \ref{bmnmm}.

We now consider the boundary conditions that correspond to the solutions
dual to the plane wave matrix model and we consider a vacuum corresponding to
a single large disk at distance $d \sim N_5$ from the $\eta =0$ plane. These
are the vacua corresponding to $N_5$ fivebranes.
 We write the leading solution of the potential in asymptotic
 region
\begin{equation} \label{ottt}
V=\alpha (\rho ^{2}\eta -{\frac{2}{3}}\eta ^{3})+\tilde{\Delta}~,~~~~~~~%
\tilde{\Delta}={\frac{P\eta }{(\eta ^{2}+\rho ^{2})^{3/2}}}
\end{equation}
Using the coordinate $r=4\sqrt{\rho ^{2}+\eta ^{2}}$ and $t=x_0$
we find that the leading order solution at large $r$  in \nref{IIA
ansatz}-\nref{IIA ansatz last} is the standard D0 brane solution
\cite{Itzhaki:1998dd} at large $r$, with warp factor
\begin{eqnarray}
Z &=&{\frac{2^{8}15P}{r^{7}\alpha }},~~~~~~~\alpha
=\frac{8}{g_{s}}
\end{eqnarray}

We now need to compute $P$. We compute the charge and the
distance. Since we have images we have $P=2dQ$. The distance is
given in terms of $N_5$ by \nref{charge2}. In order to compute the
charge we note that if we have a large disk with a size $\rho_0
\gg N_5$ then the configuration  at large distances looks like a
single conducting disk at $\eta =0$ with some extra sources
localized near $(\rho,\eta) = (\rho_0,0)$. We can thus approximate
the induced charge on the disk to be the induced charge we would
have on the conducting plane at $\eta=0$ if we had not introduced
the disk. This induced charge is given simply by  the external
potential which is the first term in \nref{ottt}. We can thus
approximate \bea Q = {\frac{1}{4\pi }}\int \partial _{\eta
}V_{ext}={\frac{\alpha \rho _{0}^{4}}{8}},~~~~~~~~~~~ d={\frac{\pi
}{2}}N_{5}\eea
 Now we can
go back to the expression for $Z$ and write it as
\begin{equation}
Z={\frac{2^{5}15\pi \rho _{0}^{4}N_{5}}{r^{7}}}
\end{equation}
where we are in the regime where the disk is very close to the
$\eta=0$ plane.

We can now compare with the result in \cite{Itzhaki:1998dd}
\begin{equation}
Z={\frac{2^{7}\pi ^{9/2}\Gamma (7/2)g_{YM0}^{2}N_{0}}{r^{7}}}={\frac{%
2^{4}15\pi ^{5}g_{YM0}^{2}N_{0}}{r^{7}}}
\end{equation}
Comparing the two we find
\begin{equation}
\rho _{0}^{4}={\frac{1}{2}}\pi ^{4}g_{YM0}^{2}N_{2}
\end{equation}
We find also that the function $f$ in section \ref{bmnmm} is
\begin{equation}
E -J\sim 1+f{\frac{n^{2}}{J^{2}}}~,~~~~~~~f={\frac{1}{2}}%
R_{S^{5}}^{4}~,~~~~~R_{S^{5}}^{2}~=4\rho _{0}
\end{equation}
Finally we obtain
\begin{eqnarray}
\frac{~R_{S^{5}}^{2}}{\alpha ^{\prime }} &=&\ 4\left( \frac{\pi
^{4}g_{YM0}^{2}N_{2}}{2m^{3}}\right) ^{1/4}\ ~ \\
f &=&4\pi ^{2} \left({2g_{YM0}^{2}N_{2} \over m^3}\right)^{1/2}
\end{eqnarray}%
in the strong coupling regime.

\section{ Poincare super algebras with \emph {non central} charges}
\renewcommand{\theequation}{E.\arabic{equation}}
\setcounter{equation}{0}
 \label{poincare}

In this appendix we discuss two Poincare superalgebras with mass
deformations \cite{nahm} which appeared in our discussion. First we present an
algebra with 8 supercharges and then an algebra with 16
supercharges.

Let us define $(\gamma ^{\mu })_{\alpha }^{\,\beta }$ as
\begin{equation}
\gamma ^{0}=i\sigma ^{2}~,~~~~\gamma ^{1}=\sigma
^{1},~~~~~~~~~~\gamma ^{2}=\sigma ^{3}
\end{equation}%
where $\sigma^i$ are Pauli matrices. We also define
\begin{equation}
\tilde{\gamma}_{\alpha \beta }^{\mu }=(\gamma ^{\mu })_{\alpha
}^{\gamma }\epsilon _{\gamma \beta
}~,~~~~~~\tilde{\gamma}^{0}=-\delta ^{\alpha \beta
},~~~~\tilde{\gamma}^{1}=-\sigma
^{3},~~~~~~~~~\tilde{\gamma}^{2}=\sigma ^{1}
\end{equation}%
and we see that $(\tilde{\gamma}^{\mu })_{\alpha \beta }$ is
symmetric in the indices $\alpha ,\beta $.

\subsection{Superalgebras with 8 supercharges}

  We define supercharges $Q_{\alpha i}$ with $i$ an $SO(4)$ index and $\alpha$ is
  the 2+1 Lorentz index (spinor of $SO(2,1)$).
We can impose the reality condition
 $Q_{\alpha i}^{\dagger }=Q_{\alpha i}$.

We start by considering a superalgebra with 8 supercharges  given
by
 \begin{eqnarray}
\{Q_{\alpha i},Q_{\beta j}\} &=&2\tilde{\gamma}_{\alpha \beta
}^{\mu }p_{\mu }\delta _{ij}+ 2 m \epsilon _{\alpha \beta
}\epsilon _{ijkl}M_{kl} \label{noncentr}
\\
\lbrack p_{\mu },Q_{\alpha i}] &=&0,\qquad \lbrack p_{\mu },p_{\nu }]=0, \\
\lbrack \Sigma _{\mu \nu },Q_{\alpha i}] &=&\frac{1}{2}\left( \tilde{\gamma}%
_{\mu \nu }\right) _{\alpha }^{\beta }Q_{\beta i} \\
\lbrack M_{jl},Q_{\alpha i}] &=&i(\delta _{ij}Q_{\alpha l}-\delta
_{il}Q_{\alpha j})   \\
\lbrack M_{ij},M_{kl}] &=&i(\delta _{ik}M_{jl}+\delta
_{jl}M_{ik}-\delta
_{jk}M_{il}-\delta _{il}M_{jk}) \\
\lbrack \Sigma _{\mu \nu },p_{\lambda }] &=&i(\eta _{\nu \lambda
}p_{\mu
}-\eta _{\mu \lambda }p_{\nu }) \\
\lbrack \Sigma _{\mu \nu },\Sigma _{\lambda \rho }] &=&i(\eta
_{\nu \lambda }\Sigma _{\mu \rho }+\eta _{\mu \rho }\Sigma _{\nu
\lambda }-\eta _{\mu
\lambda }\Sigma _{\nu \rho }-\eta _{\nu \rho }\Sigma _{\mu \lambda }) \\
\qquad \lbrack M_{jl},p_{\mu }] &=&0,\qquad \lbrack M_{jl},\Sigma
_{\mu \nu }]=0 \label{superalg_last}
\end{eqnarray}%
where $M_{ij}$ are $SO(4)$ generators. $M_{ij}$ are non-central
charges in the superpoincare algebra in 2+1 dimensions. $\Sigma
_{\mu \nu }$ is the Lorentz generator in $SO(2,1)$. Notice that
the first line is the only non-obvious commutator and is the one
stating that we have {\it non-central} charges.

In order to check the closure of the superalgebra we need to check
the Jacobi identity. The identities involving one bosonic
generator will be automatically obeyed since they are just simply
stating that objects transform covariantly under the appropriate
symmetries.

So the only non-trivial identity that we need to check is the one
involving three odd generators. The Jacobi identity is
\begin{eqnarray}
&&[Q_{\alpha i},\{Q_{\beta j},Q_{\gamma l}\}]+[Q_{\beta
j},\{Q_{\gamma
l},Q_{\alpha i}\}]+[Q_{\gamma l},\{Q_{\alpha i},Q_{\beta j}\}]  \notag \\
&=&i\epsilon _{\beta \gamma }\epsilon _{jlab}(\delta
_{bi}Q_{\alpha a}-\delta _{ai}Q_{\alpha b})+i\epsilon _{\gamma
\alpha }\epsilon _{liab}(\delta _{bj}Q_{\beta a}-\delta
_{aj}Q_{\beta b})+i\epsilon _{\alpha \beta }\epsilon
_{ijab}(\delta _{bl}Q_{\gamma a}-\delta _{al}Q_{\gamma b})
\notag \\
&=&-i\epsilon _{ijla}[\epsilon _{\beta \gamma }Q_{\alpha
a}+\epsilon _{\gamma \alpha }Q_{\beta a}+\epsilon _{\alpha \beta
}Q_{\gamma a}]-i\epsilon _{ijlb}[\epsilon _{\beta \gamma
}Q_{\alpha b}+\epsilon _{\gamma \alpha }Q_{\beta b}+\epsilon
_{\alpha \beta }Q_{\gamma b}] \equiv 0
\end{eqnarray}

It is interesting to study the particle spectrum for theories
based on this superalgebra \nref{noncentr}. This theory cannot
have massless propagating particles. This can be seen as follows.
We   assume that the massless particle has $p_-=0$, $p_+ \not =0$
and $p_2=0$. In this case the supersymmetry algebra implies that
$Q^{i}_{-}$  and  $M_{ij}$ annihilate all states in the
supermultiplet.
 On the other hand the $Q^{i}_{+}$ generators arrange themselves into creation
and annihilation operators and change the $SO(4)$ quantum numbers
in the multiplet. Thus we reached a contradiction. This argument
allows Chern Simons interactions since that is a topological
theory. So all propagating particles are massive. Let us go to the
rest frame of the massive
 particle, with $p_1=p_2=0$. Then the ``little group''
  (i.e. the truncation of \nref{noncentr} to the generators that
 leave this choice of momenta invariant) is
 the  ${\widetilde {SU}}(2|2)$
supergroup. The tilde represents the fact that we take the
corresponding $U(1)$ to be non-compact. The representation theory
of this algebra was studied in \cite{barsetal,otherrep,kaccharac}.
As usual, there are short representations when the BPS bound is
obeyed when the mass of the particle is  $M = 2 m (j_1 + j_2)$,
where $m$ is the mass parameter in \nref{noncentr}.

The superalgebra \nref{noncentr}-\nref{superalg_last} can be
reduced to 1+1 dimensions in a trivial fashion, we just set
$p_2=0$ and remove two of the Lorentz generators. This is the
symmetry algebra \nref{commutsu} of the sigma model considered in
\nref{wzwsigma}. The reason this superalgebra arises is the
following. Suppose we start with  a theory with supergroup
$\widetilde{SU}(2|4)$ and we pick a 1/2 BPS state with charge $J$
under generator $J$ in $SO(2) \subset SO(6) $. The supercharges
that annihilate this state form the supergroup
$\widetilde{SU}(2|2)$. The lightcone string lagrangian
\nref{wzwsigma} describes small fluctuations around these BPS
states so that the supergroup $\widetilde{SU}(2|2)$ should act on
them linearly. Since the worldsheet action is boost invariant
along the worldsheet, we find that this supergroup should be
extended to \nref{noncentr}.

Let us give some further examples of theories with this
superalgebra. We can construct a 1+1 dimensional SYM with this
superalgebra from the plane wave matrix model via matrix theory
compactification techniques \cite{Taylor} (also \cite{BFSS}). In
fact this 1+1 SYM was constructed in this way by e.g.
\cite{iiappwave1},\cite{iiappwave2},\cite{jeremy}. Here we will
reproduce this result and we will use $SO(9,1)$ gamma matrices and
the fermions are $SO(9,1)$ spinors\footnote{Our convention is
different from that of \cite{BFSS} or \cite{bmn}, which use
$SO(9)$ gamma matrices.}. We will then compactify a scalar of the
1+1 SYM and get a 2+1 super Yang Mills Chern Simons theory
satisfying the above superalgebra.

One starts from the plane wave matrix model whose mass terms for
the $SO(6)$ scalars takes the form $-\tfrac{1}{2} \,(X_{a})^{2}$,
where $a=1,2,...,6$. We have set the mass for the $SO(6)$ scalar
to 1. We should write the action so that it is translation
invariant in one of the transverse scalars. We can make a
field-redefinition for two $SO(6)$ scalars $X_{1}+iX_{2}=e^{i t
}(Y+i\phi)$ and for fermions $\Psi = e^{{1 \over 2} \Gamma_{12} t}
\theta $. Then the action of plane wave matrix model is
\begin{eqnarray}
S &=&\frac{1}{g_{YM0}^{2}}\int dx_{0}\mathrm{Tr} \left(
-\frac{1}{2}(D_{0 }X_{I})^{2}-\frac{1}{2}(D_{0
}Y)^{2}-\frac{1}{2}(D_{0 }{\phi})^{2}-\frac{i}{2} \bar{\theta
}\Gamma ^{0 }D_{0 }\theta -\frac{1}{2}\bar{\theta} \Gamma
_{I}\,[X_{I},\theta
] \right. \notag \\
&&-\frac{1}{2}\bar{\theta}\Gamma_{1}\,[\phi,\theta]-\frac{1}{2}\bar{\theta}\Gamma
_{2}\,[Y,\theta]+\frac{1}{2}[\phi,X_{I}]^{2}+\frac{1}{2}[\phi,Y]^{2}+\frac{1}{2}[Y,X_{I}]^{2}
+\frac{1}{4}[X_{I},X_{J}]^{2}\,-\frac{1}{2}\,(X_{a})^{2} \notag \\
&&\left. -\frac{1}{2}2^{2}\,(X_{i})^{2}
+\frac{3}{2}i\,\bar{\theta}\Gamma _{789}\theta+
2i\epsilon^{ijk}\,X_{i}\,X_{j}\,X_{k} - \half i\,\bar
{\theta}\Gamma _{0}\,\Gamma _{12}\theta - 2Y D_{0}{\phi} \right)
\label{bmnmodel}
\end{eqnarray}
We have 3+4+2 scalars, where the first seven scalars with indices
$I=3,4,...,9$ are split into $a=3,4,5,6$ and $i=7,8,9$ and the
rest two scalars are $Y$ and $\phi$.

Then the action becomes translation invariant in the  $\phi$
direction. We now compactify $\phi$ by replacing $\phi$ with gauge
covariant derivative $\phi \rightarrow i\frac{\partial }{\partial
x_{1}}+A_{1}, -i[\phi ,O]\rightarrow
\partial _{1}O-i[A_{1},O]$ \cite{Taylor} (also \cite{BFSS}).
Plugging this into the original action \nref{bmnmodel} one get the
1+1 dimensional super Yang Mills on $R^{1,1}$ with a mass
deformation
\begin{eqnarray}
S &=&\frac{1}{g_{YM1}^{2}}\int \!dx_{0}dx_{1}\mathrm{Tr}\left(
-\frac{1}{4}F_{\mu \nu }^{2}-\frac{1}{2}(D_{\mu
}X_{I})^{2}-\frac{1}{2}(D_{\mu }Y)^{2}-\frac{i}{2}\bar{\theta
}\Gamma ^{\mu }D_{\mu }\theta -\frac{1}{2}\bar{\theta}\Gamma
_{I}\,[X_{I},\theta ] \right.
 \notag \\
&& -\frac{1}{2}\bar{\theta}\Gamma _{2}\,[Y,\theta
]+\frac{1}{2}[Y,X_{I}]^{2}+\frac{1}{4}[X_{I},X_{J}]^{2}\,-\frac{1}{2}\,(X_{a})^{2}-\frac{1}{2}2^{2}\,(X_{i})^{2}
+\frac{3}{2}i\,\bar{\theta}\Gamma _{789}\theta \notag \\
&&\left. +2 i\epsilon^{ijk}\,X_{i}\,X_{j}\,X_{k} - \half
i\,\bar{\theta} \Gamma _{012}\theta - Y\epsilon ^{\mu \nu }F_{\mu
\nu } \right) \label{2dsym}
\end{eqnarray}
We have 3+4+1 scalars, with the seven scalars whose indices are
$I=3,4,...,9$, where $a=3,4,5,6$ and $i=7,8,9$ and another scalar
$Y$, and $\mu=0,1$. The theory has super-poincare algebra on
$R^{1,1}$ with $SU(2) \times SU(2) $ R symmetry. The first $SU(2)$
rotates the first three scalars $i=7,8,9$ and the second $SU(2)$
is one of the $SU(2)$ factors in the $SO(4)$ rotating the four
scalars $a=3,4,5,6$. In addition, the theory has an $SU(2)$ global
symmetry, which is the second $SU(2)$ factor in the $SO(4)$ we
have just mentioned. Compactifying along $x_{1}$ and taking the
compactification size to zero we get back to the plane wave matrix
model which has a larger symmetry group. The parameters in the two
theories are related by $ {g_{YM1}^{2}}=2\pi
R_{x_{1}}{g_{YM0}^{2}}$, where $R_{x_{1}}$ is the radius of the
$x_{1}$ circle. The 1+1 SYM constructed from the plane wave matrix
model coincides with the DLCQ of the IIA plane wave
\cite{iiappwave1},\cite{iiappwave2}, which was first obtained by
\cite{iiappwave1},\cite{iiappwave2} from reduction of the
supermembrane action under kappa-symmetry fixing condition on 11d
maximal plane wave. The action we reproduce here \nref{2dsym} is
written manifestly Lorentz invariant in 1+1 dimensions.

We pointed out that this theory can be uplifted again  making $Y$
periodic.  We make the replacement $Y\rightarrow i\frac{\partial
}{\partial x_{2}}-A_{2}, -i[Y,O]\rightarrow \partial
_{2}O+i[A_{2},O] $ \cite{Taylor}. The coupling  $Y F_{01},$
becomes a Chern-Simons term in 2+1 dimensions. The quantization of
the level of the Chern Simons action implies that the
compactification radius of $Y$ is quantized. This quantization
condition also follows from the fact that the coupling $YF_{01}$
is not invariant under arbitrary shifts of $Y$, and $e^{iS}$  is
periodic only if we shift $Y$ by the right amount. Finally we get
the 2+1 dimensional super Yang Mills Chern Simons theory
 \bea S &=& \frac{k }{4\pi
}\left\{ \int \mathrm{Tr} \{-\frac{1}{2}F\wedge \ast F+ A\wedge
dA+\frac{2}{3}A\wedge A\wedge A-{i \over 12}
\bar{\psi}\Gamma_{\mu\nu\lambda} \psi
dx^{\mu}\wedge dx^{\nu} \wedge dx^{\lambda}\} \right. \notag \\
&& +\int d^{3}x\mathrm{Tr}\{ -\frac{1}{2}(D_{\mu }X_{I})^{2}
-\frac{i}{2} \bar{\psi }\Gamma ^{\mu }D_{\mu }\psi
-\frac{1}{2}\bar{\psi} \Gamma^{I}\,[X_{I},\psi
]+\frac{1}{4}[X_{I},X_{J}]^{2}
\notag \\
&& \left. -\frac{1}{2}\,(X_{a})^{2}  -\frac{1}{2}
2^{2}\,(X_{i})^{2} +2i\epsilon^{ijk}\,X_{i}\,X_{j}\,X_{k}+
\frac{i}{4} \epsilon^{ijk}\bar{\psi}\Gamma _{ijk}\psi \} \right\}
 \label{3dYM-CS}
\eea where we have $3+4$ scalars with indices $I=3,4,...,9$ split
into $a=3,4,5,6$ and $i,j,k=7,8,9$,  and the worldvolume indices
are $\mu,\nu,\lambda=0,1,2$. The coupling constants is related to
the 1+1 SYM by ${g_{YM2}^{2}}=2\pi R_{x_{2}}{g_{YM1}^{2}} $,
$\frac{k}{4\pi }= \frac{ 1 }{ g^2_{YM 2} } $ and $k\in {\mathbb
Z}$. So we see that $k$ is the only coupling constant in the
theory. When $k$ is large the theory is weakly coupled.

\subsection{Superalgebras with 16 supercharges}

Finally, let us turn our attention to the superalgebra for
theories with 16 supercharges. Now we have two $SO(4)$ groups and
a second set of supercharges $\widetilde{Q}_{\alpha m}$. We add
the anticommutators
\begin{equation}
\{\widetilde{Q}_{\alpha m},\widetilde{Q}_{\beta
n}\}=2\tilde{\gamma}_{\alpha
\beta }^{\mu }p_{\mu }\delta _{mn}+2 m'\epsilon _{\alpha \beta }\epsilon _{mnrs}%
\widetilde{M}_{rs}
\end{equation}%
where $\widetilde{M}_{rs}$ is  generator of the second $SO(4)$.
The anticommutator of $Q_{\alpha i}$ with $\widetilde{Q}_{\alpha
m}$ is zero. The rest of the algebra is rather obvious and is just
given by the covariance properties of the indices as in
\nref{noncentr}. In principle we can have $m' \not = m$ in
\nref{noncentr}. In the theories studied here we have $m' =m$. If
we want to have BPS states under both $Q$ and $\tilde Q$ then we
need that $m/m'$ to be a rational number. Note that the little
group for a massive particle is $\su (2|2) \times \su (2|2)$.

Let us be a little more precise about these $SO(4)$ groups. The
ansatz in \cite{llm} above has two three spheres on which two
$SO(4)$ group act. Let us call them $SO(4)_{i}$ with $i=1,2$. Each
of these two groups are $SO(4)_i =SU(2)_{Li} \times SU(2)_{Ri} $.
The supercharges $Q_{\alpha i}$ in \nref{noncentr} transform under
$SU(2)_{L1} \times SU(2)_{L2}$. The supercharges $\tilde Q_{\alpha
i}$ transform under $SU(2)_{R1} \times SU(2)_{R2}$. If we quotient
any of these theories by a $Z_k$ in $SU(2)_{Ri}$, we get a theory
that only has 8 generators as in \nref{noncentr}.

This algebra with 16 generators is the one that appeared on the
worldvolume of theories related to the IIB constructions of
section \ref{iibtheory}. In the case of the M2 brane theory the
two $SO(4)$s are global R-symmetries of the theory. In the case
that  we consider an M5 on $R^{2,1} \times S^3$ one of the $SO(4)$
groups is a symmetry acting on the worldvolume. When the size of
$S^{3}$ becomes infinity, the $SO(4)$ that acts on the worldvolume
is contracted to ISO(3) and only the translation generators remain
in the right hand side of the supersymmetry algebra. Thus, we do
not get into trouble with the Haag-Lopuszanski-Sohnius theorem
\cite{Haag:1974qh} in total spacetime dimension $d \geq 4$.

The dimensional reduction of this algebra to 1+1 dimensions gives
the linearly realized symmetries on the lightcone worldsheet of a
string moving in the maximally supersymmetric IIB plane wave
\cite{Blau:2001ne}.

\section {  4+1 d SYM and 5+1 d SYM with Chern-Simons term from ${\cal N}=4$ SYM }
\renewcommand{\theequation}{F.\arabic{equation}}
\setcounter{equation}{0}

\label{4dand5dsym}

In this section we discuss in more detail the lagrangian on the D4
brane and the D5 brane that we obtained by starting from ${\cal
N}=4$ super Yang Mills and compactifying the transverse scalars.
The procedure is identical to the one used in appendix
\ref{poincare}.

We start from the ${\cal N}=4$ SYM on $R \times S^3$ with mass
terms for the 6 scalars $-\frac{1}{2}\mu ^{2}X_{a}^{2}$, where
$a=4,5,6,7,8,9$. We redefine two of the scalars $X_{4}+i
X_{5}=e^{i \mu t} (Y+\phi)$ and fermions $\Psi_{old}=e^{\half \mu
\Gamma_{45} t} \Psi$. We then make the replacement $ \phi
\rightarrow i\frac{\partial }{\partial x_{4}}+A_{4}, -i[\phi
,O]\rightarrow
\partial _{4}O-i[A_{4},O]$ \cite{Taylor}. We obtain a 4+1 super
Yang-Mills theory on $R^{1,1}\times S^{3}$ with a mass deformation
\begin{eqnarray}
S &=&\frac{2}{g_{YM4}^{2}}\int d^{2}x d^{3}\Omega \mathrm{Tr}\left(-\frac{1}{4}%
F_{MN}F^{MN}-\frac{1}{2}D_{M}X_{a}D^{M}X_{a}-\frac{i}{2}\bar{\Psi}
\Gamma ^{M}D_{M}\Psi -\frac{1}{2}\bar{\Psi}\Gamma ^{a}[X_{a},\Psi ] \right. \notag \\
&&\left. -\frac{1}{2} \bar{\Psi} \Gamma_{5}[Y,\Psi
]+\frac{1}{4}[X_{a},X_{b}]^{2}+\frac{1}{2} [Y,X_{a}]^{2}
+\tfrac{\mu }{2}i\,\bar{\Psi}\Gamma _{045}\Psi -\frac{1}{2 }\mu
^{2}X_{a}^{2}-2\mu YF_{04} \right) \label{5dsym}
\end{eqnarray}
where $M,N=0,1,...,4$; $a=6,7,8,9$; $\Gamma_{M}, \Gamma_5,
\Gamma_{a}$ are ten dimensional gamma matrices. The theory does
not have poincare invariance in 4+1 dimensions, but it has
poincare invariance in the $R^{1,1}$ subspace $x_0, x_4$. It has
$SO(4)$ R symmetry. When it is truncated by keeping states that
are invariant only under the $SU(2)_L$ which acts on $S^{3}$ it
gives the 1+1 SYM in (\ref{2dsym}). When we
 reduce it  on $S^{1}$, it gives back the ${\cal N}=4$ super Yang Mills. The
parameters are related by ${g_{YM4}^{2}}=2\pi
R_{x_{1}}{g_{YM3}^{2}}$.

This  4+1 super Yang Mills
 can be uplifted again by compactifying $Y$, and making replacement $Y\rightarrow i\frac{\partial
}{\partial x_{5}}-A_{5}, -i[Y,O]\rightarrow \partial
_{5}O+i[A_{5},O] $, similar to appendix \ref{poincare}. The
 uplifted
action is a 5+1 super Yang Mills
 on $R^{2,1}\times S^{3}$ which contains a Chern-Simons term for the 2+1 dimensional
 gauge fields
\begin{eqnarray}
S &=&\frac{2}{g_{YM5}^{2}}\int\mathrm{Tr}
[-\frac{1}{4}F_{MN}F^{MN}] +\frac{K}{ 4\pi }\int
\mathrm{Tr}[A\wedge dA+\frac{2}{3}A\wedge A\wedge A]
\wedge { d^{3}\Omega \over Vol_{S^{3}}}  \notag \\
&&+\frac{2}{g_{YM5}^{2}}\int d^{3}x d^{3}\Omega \mathrm{Tr}\left[ -%
\frac{1}{2}D_{M}X_{a}D^{M}X_{a}+\frac{1}{4}[X_{a},X_{b}]^{2}-\frac{1}{2}\mu
^{2}X_{a}^{2}\right. \notag \\
  && \left. -\frac{i}{2}\bar{\Psi}
\Gamma ^{M}D_{M}\Psi -\frac{1}{2}\bar{\Psi}\Gamma ^{a}[X_{a},\Psi
] +\tfrac{\mu }{2}i\,\bar{\Psi}\Gamma _{045}\Psi \right]
\label{5dsymcs}
\end{eqnarray}
where $M,N=0,1,...,4,5$; $a=6,7,8,9$. The coupling constant is
${g_{YM5}^{2}}={2\pi R_{x_{2}}}{g_{YM4}^{2}} $, $\frac{K}{4\pi
}=\frac{2\mu Vol_{S^{3}}}{ g_{YM5}^2}$, $K \in {\mathbb Z}$, where
$Vol_{S^{3}}$ is the volume of the $S^{3}$. This is the $S^3$
 on which the original ${\cal N}=4$ is defined. Notice that the coupling constant is
 given in terms of $K$. This implies the weak coupling limit corresponds to large $K$.
 The theory only has Poincare invariance on $R^{2,1}$ subspace $x_0,x_4,x_5$.
Truncating this theory by keeping only states invariant under the
$SU(2)_L$ that acts on
 $S^{3}$ we recover the 2+1 dimensional Yang-Mills Chern-Simons in (\ref{3dYM-CS}).
This 5+1 d theory can also be reduced to a  4+1 dimensional super
Yang Mills on $R^{2,1}\times S^{2}$ if we replace $S^3$ with
$S^3/Z_k$ and reduce on the fiber direction of the latter in the
similar way in section \ref{d2theory}.

Similarly, the 4+1 dimensional Yang Mills (\ref{5dsym}) can be
reduced to a 3+1 dimensional Yang Mills theory on $R^{1,1} \times
S^2$ by truncating by $U(1)_L \subset SU(2)_L $ which acts on
$S^3$. Alternatively, this theory can be obtained through the
uplifting procedure applied to the D2 brane theory on $R\times
S^2$ that we discussed in section \ref{d2theory}.

\section{Computation of the index counting BPS states}
\setcounter{equation}{0}
\renewcommand{\theequation}{G.\arabic{equation}}
\label{indexapp}

In this appendix we compute the index \nref{defindfull} for
various situations. We are interested in computing this index for
single trace states in the 't Hooft $N=\infty$ limit. Since the
index is basically a counting problem we can use Polya theory, as
explained in \cite{polyakov,minwalla}. What we want to do is the
following. We have a set of ``letters'' which are the various
oscillator modes. This set depends on the vacuum we expand around.
We define the single particle  partition function \be z =
\sum_{bosons} e^{- \beta_i Q_i} - \sum_{fermions} e^{-\beta_i Q_i}
\ee where $Q_i$ are various charges. The single trace states are
counted by \be   \label{singletrace}
 Z_{s.t} =- \sum_{n=1}^{\infty} {\varphi(n) \over n} \log[1-z(n\beta_i )]
\ee Where $\varphi(n)$ is the Euler Phi function which counts
number of integers less than $n$ that are relatively prime with
$n$. $\varphi(1)\equiv 1$, $\varphi(2)=1$, $\varphi(3)=2$, etc.

When we compute \nref{defindfull} only states with ${\cal U}
\equiv H_4 \equiv E - 2 S - \sum_i J_i =0$ contribute. Let us first
consider the states in the first Kaluza Klein mode, which is in
the representation in figure \ref{young}(a). For convenience we
will use $Y_j$ to denote $SO(6)$ scalars and $X_i$ for $SO(3)$
scalars in this appendix. The bosons that contribute are given by
$Y^j+ i Y^{j+1}$, $j=1,3,5$ and $X^+ = X^1 + i X^2$. The fermions
that contribute have the indices $\psi_{+,++-}$, $\psi_{+,+-+}$,
$\psi_{+,-++}$ where the indices indicate the charges under
$(S,J_1,J_2,J_3)$ and $S$ is one of the generators of $SU(2)
\subset SU(2|4)$. Then we find \bea z_1 &=& e^{- \beta_2 -
\beta_3} +  e^{- \beta_1 - \beta_3} + e^{- \beta_1- \beta_2} + e^{
- 2 (\beta_1 + \beta_2 + \beta_3) } -\notag
\\
&& e^{-2 \beta_1 - \beta_2 - \beta_3} - e^{- \beta_1 - 2 \beta_2 - \beta_3}
- e^{- \beta_1 - \beta_2 - 2\beta_3}
\cr (1-z_1) &=&
(1-e^{-\beta_1-
\beta_2})(1-e^{-\beta_2-\beta_3})(1-e^{-\beta_1-\beta_3})
\label{onemz} \eea We can now use the formula \be \label{qprod} -
\sum_{n=1}^\infty { \varphi(n) \over n} \log( 1 - q^n) = { q \over
(1-q) } \ee in order to  write \nref{singletrace} in terms of
\nref{onemz} to obtain \be \label{finresof} I_{s.t.~N_5=1} = { e^{
- \beta_2 - \beta_1 } \over 1-e^{-\beta_2 - \beta_1} } + { e^{ -
\beta_3 - \beta_1 } \over 1-e^{-\beta_3 - \beta_1} } + { e^{ -
\beta_3 - \beta_2 } \over 1-e^{-\beta_3 - \beta_2} } \ee In order
to read off which representations are contributing it is useful to
compute the index for the doubly atypical representations of the
form $(a_1,a_2,a_3|a_4|a_5) = (0,p,0|0|0)$. This notation refers
to the Dynkin labels, see figure \ref{young}(g) and \cite{markvr2}
for further details. We obtain \be \label{charprep}
I_{(0,p,0|0|0)} = e^{ -p ( \beta_1 + \beta_2) } { ( 1 - e^{-
\beta_2 - \beta_3} ) ( 1 - e^{- \beta_1 - \beta_3} ) \over ( 1 -
e^{ \beta_1 - \beta_3} )( 1 - e^{ \beta_2 - \beta_3} ) } + {\rm
cyclic } \ee We can see that if we sum this over $p$ we obtain \be
\label{sumtrunc} I_{s.t. ~N_5=1} = \sum_{p=1}^\infty  I_p \ee This
discussion implies that all the BPS representations that
contribute to the index for the $N_5=1$ case are the ones we
expect from the doubly atypical representations to which the
string ground state $tr[Z^J]$ belongs to. Of course, in order to
show that these representations are protected we do not need any
of this technology, since doubly atyical representations cannot be
removed \cite{markvr2}. All we are showing here is that we find no
evidence of further BPS representations for the $N_5=1$ vacuum.
 This result is not totally trivial since we can certainly
construct individual single trace states in other atypical
representations. These are singly atypical representations. But we
find that they always come in pairs that could combine into long
representations. Of course, the explicit analysis we described in
section \ref{furthersec} shows that they all do combine. Before we
leave this simple case, let us understand how we connect
 these results to the spectrum of the string theory in lightcone gauge.
It is convenient to focus on the $\su (2|2)$ subgroup in $\su
(2|4)$, the energy $\hat E $ in $\su (2|2)$ is the same as $\hat E
= E-J_3$ in $SU(2|4)$. So we are interested in taking a limit
 where $\beta_3$ is large and $\beta_1 +  \beta_2$ is small. We place no
 constraint on $\beta_1 - \beta_2$.
Actually, to be more precise,
 note that the choice of generator $J_3$, or field $Z=Y^5+iY^6$
leaves a subgroup $\su (2|2) \times SU(2)_G \subset \su (2|4)$
unbroken. The chemical potential $\beta_1 - \beta_2$ couples to
the generator in the global $SU(2)_G$ which is not part of the
$\su (2|2)$ supergroup. Our goal is to relate \nref{finresof}  to
an index we can compute on the string worldsheet of the form \be
\label{sutwoindap} {\cal I}(\gamma , \tilde \gamma) = Tr\left[ (-1)^F 2
S_3 e^{ - \hat \mu (\hat E - S_3 - \tilde S_3) } e^{ -\gamma \hat
E } e^{-\tilde \gamma J_3^G} \right] \ee which is the same as
\nref{sutwoind} except that we have added a chemical potential for
the generator $J_3^G$ in $SU(2)_G$. It is clear that we should
identify $\gamma = \beta_3$ and $\tilde \gamma = \beta_1 -
\beta_2$. Let us state the final result and then we will justify
it. We have \be \label{relatind} \lim_{\beta_1+\beta_2 \to 0}
\left[ I_{s.t.}(\beta_i) -{ q \over (1-q) } \right] = - {\cal
I}(\gamma = \beta_3 , \tilde \gamma = \beta_1 -\beta_2) \ee Let us
explain how we obtained this. The states giving rise to string
worldsheets in the plane wave limit have
  very large values of
$E$. So we need to isolate from \nref{finresof} the contribution
from states with large values of $E$. The first idea is to isolate
from \nref{finresof} terms with large powers of $q=e^{-\beta_1 -
\beta_2}$ but low powers of $e^{-\beta_3}$. The only such states
are the ones in the first term in \nref{finresof}. Unfortunately,
such states have no $\beta_3$ dependence at all and correspond to
the ground states. This is related to the fact that
\nref{sutwoindap} would vanish if we had not inserted $J_3$. In
\nref{finresof} the absence of high powers of $q$ in the $\beta_3$
dependent terms is due to the fact that each representation with
large $p$ contributes with a factor of $(1-q)$ to terms with
finite powers of $e^{-\beta_3}$. This factor arises as follows.
Among the supercharges with ${\cal U}=0$ we have one which has
zero $\hat E$. It has quantum numbers $Q^\dagger_{+,--+}$. This
supercharge does not annihilate $\beta_3$ dependent terms and
gives rise to the $(1-q)$ factor. This can be seen more explicitly
by rewriting the first term in \nref{charprep} as \bea
 && q^{-p} \left\{  1 +
(1-q) { q^{-1/2}e^{-\beta_3}  (e^{\tilde \gamma/2} + e^{-\tilde
\gamma/2 } ) - e^{-2 \beta_3}(1 + 1/q) \over ( 1 - q^{- 1/2}
e^{\tilde \gamma/2  - \beta_3} ) ( 1 - q^{-1/2} e^{ - \tilde
\gamma/2 - \beta_3} ) } \right\}
\label{othfm} \\
&& \tilde \gamma = \beta_1 - \beta_2 ~,~~~~~~~q = e^{-\beta_1 -
\beta_2} \eea The term independent of $\beta_3$ is the
contribution to the ground state of the string and is explicitly
subtracted in \nref{relatind}. The other terms in \nref{othfm} as
well as the second and third terms in \nref{charprep} contain a
factor of $(1-q)$. So these contributions would vanish if we took
the $q\to 1$ limit. In order to avoid this problem we introduce a
factor $w^{2S} $ when we compute the contribution of each BPS
representation. We then take a derivative with respect to $w$, set
$w=1$ and take $q\to 1$. This gives us a finite answer for each
$p$ in \nref{charprep}. In fact we get to strip off a factor of
$(1-wq)$ and replace it by $(-1)$, since  we are interested in
taking the large $p$ limit. We can equivalently obtain this limit
by simply starting from the full expression and taking the limit
in \nref{relatind} since terms which involve finite powers of $q$
will cancel out due to the $(1-q)$ factor, while the sum over many
terms involving a power of $q^p$ will give a $1/(1-q)$ cancelling
the explicit factor of $(1-q)$. In other words, if we were to
truncate the sum \nref{sumtrunc} to a finite number of terms and
then take $q\to 1$ then we would get zero for all $\beta_3$
dependent terms.

The limit \nref{relatind} gives us \be {\cal I} = - { e^{ -
\beta_3 -\tilde \gamma/2} \over 1 - e^{ - \beta_3 -\tilde
\gamma/2} } -{ e^{ - \beta_3 +\tilde \gamma/2} \over 1 - e^{ -
\beta_3 +\tilde \gamma/2} } \ee This is indeed the result we get
for ${\cal I}$ if we compute the contribution in the string theory
side for a $(4,4)_m$ worldsheet theory with fields in the
fundamental representation of $\su (2|2)$ and the fundamental of
$SU(2)_G$. These are precisely the fields coming from the first
four directions of the string worldsheet.

Now let us now consider vacua with $N_5>1$. In order to compute
the index in these cases it is useful to write a general formula
for the index for an arbitrary $SU(2|4)$ representation. If the
representation is typical then the index vanishes. We can
understand this as follows. On a typical $SU(2|4)$ representation
we find that, for the purposes of counting the states, the
supercharges act like fermionic creation and annihilation
operators. In fact, when we look at the expression for the
characters in \cite{kaccharac} we find that that there is a factor
of the form \be \label{fullfact} \prod_j ( 1 - e^{-\theta_i H^j_i}
) \ee where $j$ runs over half of the supercharges and $H^j_i$ are
the Cartan charges of this supercharge. The index we have defined
is simply a character evaluated for special values of $\theta_i$
which are such that a particular supercharge, $Q^\dagger_{-,+++}$,
gives a contribution of the form $(1-1) =0$. This ensures that the
index vanishes for long (or typical) representations. In atypical
representations one finds that the character does not contain the
full factor \nref{fullfact}. In fact, the index will receive
contributions only from states with ${\cal U}=0$ (see
\nref{anticommf}). So we can truncate the $SU(2|4)$ superalgebra
to the elements that have ${\cal U}=0$. This gives a $\su (1|3)$
superalgebra. So the states contributing to the index form
 $\su (1|3)$ representations. The index is the same as the
character of the $\su (1|3)$ representation. It turns out that if
we consider an atypical representation of $\su (2|4)$ of the form
$(a_1,a_2a_3|a_5+1|a_5)$
 then states with ${\cal U}=0$ form a typical representation of $\su (1|3)$.
Doubly atypical representations of $\su (2|4)$ give rise to
atypical representations of $\su (1|3)$. Let us be more explicit.
Let us start with the atypical $\su (2|4)$ representation $r$
labeled by $(a_1,a_2,a_3|a_5+1|a_5)$. The index evaluated on this
representation gives us \be \label{indgrep}
I_{(a_1,a_2,a_3|a_5+1|a_5)} = - (-1)^{a_5} e^{ - Q(\sum_i \beta_i
) } (1 - e^{-\beta_1 - \beta_2})(1- e^{-\beta_1 - \beta_3})(1 -
e^{-\beta_2 - \beta_3}) \chi_{(a_1,a_2)}( {\bf g } ) \ee where \be
Q = 2 + a_5 + a_3 + { 2 \over 3 } a_2  + {1 \over 3 } a_1 \ee and
$\chi_{(a_1,a_2)}$ is a character of an $SU(3)$ representation
with $(a_1,a_2)$ Dynkin labels and evaluated on an $SL(3)$ matrix
of the form $ {\bf g} = {\rm diag} \,  e^{{1 \over 3} \sum_i
\beta_i} ( e^{-\beta_1} , e^{-\beta_2}, e^{-\beta_3} )$. When we
derived \nref{indgrep} we used the fact that we obtain a typical
representation of $\su (1|3)$ and  we used the typical character
formulas in \cite{kaccharac} to write the character in terms of
$SU(3)$ representations. Notice that in \nref{indgrep} we see the
factor of the form \nref{fullfact}
 which comes from the supercharges in $\su (1|3)$ \cite{kaccharac}\footnote{
The formulas in \cite{kaccharac} say that the character of typical
representation of the $SU(n|m)$ supergroup are given by the
product of the characters of the $U(1)\times SU(n) \times SU(m)$
representations associated to
 the highest weight state  times a
factor of the form \nref{fullfact} which arises from half of the
supercharges (the supercharges are split into raising and lowering
and we get \nref{fullfact} from the lowering ones).}.

Returning to our problem, we  want to evaluate the single particle
contribution to the index from the additional Kaluza Klein
multiplets that we have for a fuzzy sphere. These multiplets
transform in the representations $(0,0,0|2(l-2) + 1| 2(l-2) )$,
$l\geq 2$, where $l=2$ correspond to the multiplet with a Young
supertableau with four vertical boxes as in figure \ref{young}(d).
The vacuum associated to $N_5$ fivebranes has multiplets with
$l=2,\cdots, N_5$, in addition to the multiplet present for the
trivial vacuum which is given by figure \ref{young}(a). Using
\nref{indgrep} we can evaluate the single particle contribution
from each of these multiplets as \be z_l = - e^{ - 2(l-1)(\sum_i
\beta_i ) } (1 - e^{-\beta_1 - \beta_2})(1- e^{-\beta_1 -
\beta_3})(1 - e^{-\beta_2 - \beta_3}) \ee where $l>1$. Then we see
that we can represent the full single particle contribution as \be
\label{singlpartr}
 1 - z_1 - \sum_{l=2}^{N_5} z_l =
 (1 - e^{-\beta_1 - \beta_2})(1- e^{-\beta_1 - \beta_3})(1 - e^{-\beta_2 - \beta_3})
 {(1-e^{- 2 N_5 (\beta_1 + \beta_2 + \beta_3)}) \over
 1-e^{- 2 (\beta_1 + \beta_2 + \beta_3)} }
 \ee
Inserting \nref{singlpartr} into \nref{singletrace} and using
\nref{qprod} we obtain \nref{bignf}. The final result \nref{bignf}
contains all the information about possible surviving BPS
representations for the single string case which can be obtained
by group theory alone. Notice that this was obtained purely from
representation theory and no assumptions were made on the
dynamics, other than the planar approximation, which implies that
single trace states do not mix with multiple trace states. It
could well be possible that by using more detailed properties of
the dynamics one might be able to obtain more detailed information
about BPS states. In other words, the index gives us a lower bound
on the number of BPS states, but the actual number could be
bigger.

We see that the structure of the index is such that we get one
contribution that is common to all vacua, which is what we had for
the trivial vacuum at
 $N_5=1$, plus some extra terms that arise only for $N_5>1$, which are
  written in \nref{bignf}.
 It is clear that these extra terms are the ones that contain the information
 about the extra four dimensions of the string. Focusing on these terms
 and taking the limit \nref{relatind} we obtain the value of the worldsheet index
 over the last four coordinates \nref{indestr}. As expected, we do not get any
 terms involving $\tilde \gamma$ since we expect that the $SU(2)_G$ symmetry only
 acts on the first four dimensions.

Let us be more explicit about the atypical representations that
contribute to the index. We find \be \label{lrep} I_{s.t. ~N_5} -
I_{s.t. ~N_5=1} = \sum_{n=1,~n\not = 0~mod(N_5)}^\infty  e^{ - 2
n( \beta_1 + \beta_2 + \beta_3) } \ee And each term can be written
as \be \label{consu} - e^{ - 2 n( \beta_1 + \beta_2 + \beta_3) } =
 \sum_{p=0}^\infty  I_{(0,p,0|2(n-1) +1| 2(n-1))}
\ee where \bea I_{(0,p,0|2(n-1) + 1|2(n-1))} &=& - e^{ - 2 n (\sum
\beta_i) } (1 - e^{-\beta_1 - \beta_2})(1- e^{-\beta_1 -
\beta_3})(1 - e^{-\beta_2 - \beta_3}) \times \notag
\\
&& \left[ { e^{ - p(\beta_1 +\beta_2) } \over(1- e^{\beta_1 -
\beta_3})(1-e^{\beta_2-\beta_3}) } + {\rm cyclic} \right] \eea is
the contribution of an atypical representation with the Young
supertableau in figure \ref{young}(f). We interpret the sum over
$p$ as indicating that we can add any number of $Z$s. So the
resulting BPS states could be matched by thinking that we have the
usual BPS states in the first four dimensions, the same we had for
$N_5=1$ plus
 BPS states along the other four dimensions with $\hat E = 2 n$.
In conclusion, we interpret each term of the form \nref{consu} as
giving rise to a BPS state in the second four dimensions with
$\hat E= 2n$ and $S_3 = \tilde S_3 = n $. This is the contribution
we would get from an $\su (2|2)$ supermultiplet with a single
column of $ 2n$ boxes.

Finally, let us explain why \nref{sutwoind} is an index that
counts BPS states for the theory on the string. We start by
defining \be \label{sutwoindt} \chi  = Tr[ (-1)^F w^{2 S_3} e^{-
\hat \mu (\hat E - S_3 - {\tilde S}_3)} e^{-\gamma \hat E} ] \ee
which is a character of $\su (2|2)$. Let us denote by ${\cal V}
\equiv  \hat E - S_3 - {\tilde S}_3$ the generator that is
conjugate to $\hat \mu$. Then we see that there are two
supercharges $Q^\dagger_{+,-}$, $Q^\dagger_{-,+}$ (and their
complex conjugates) which have ${\cal V}=0$ eigenvalue. In
addition, in $\su (2|2)$ all supercharges have $\hat E=0$
eigenvalues\footnote{In fact all generators have $\hat E =0$
eigenvalues, so one can truncate this algebra to $PSU(2|2)$. We
are not interested in doing this here.}. On a long representation
these two supercharges give rise to a factor of the form
$(1-w)(1-1/w)$. We see that for $w=1$ long representations do not
contribute. However, we also see that short representations do not
contribute either because they typically have one factor of
$\left. (1-w^{\pm 1 })\right|_{w=1}$. The solution to this problem
is to take the derivative of \nref{sutwoindt} with respect to $w$
and then set $w=1$. Then long representations will not contribute
but short representations will contribute. This proves that
\nref{sutwoind} is an index. Short representations of $SU(2|2)$
are, for example,
 those that have a single column or a single
row.

If we take the free theory that is associated to the second four
coordinates of the IIA pp wave we find that the single particle
excitations transform in a short representation of $\su (2|2)$
given by a single
column of two boxes (as in figure \ref{young}(a)).
This representation contains two BPS states,
with ${\cal V} =0$
 contributing to \nref{sutwoind}. These are a boson of spins $(S_3,\tilde S_3)=(1,0)$
 and a fermion with spins $(\half , \half)$. Only when these particles have zero momentum
 in the spatial dimension can they contribute to the index. We can
 thus evaluate the index in the
 Fock space by simply writing
 \be
 {\cal I}_{Fock} = \partial_w \left[ { 1- w e^{- 2 \gamma} \over 1 - w^2 e^{ -2 \gamma} }
 \right]_{w=1} = \sum_{n=1}^\infty e^{ - 2 n \gamma} = { e^{- 2 \gamma} \over
 (1- e^{- 2 \gamma}) }
 \ee
 where we used that only a single bosonic and fermionic oscillator contribute.
We see that this expression contains the contributions expected
from $\su (2|2)$ states with Young supertableaux with a column of
$2 n$ boxes. Such multiplets contain two BPS states with
$(S_3,\tilde S_3) = ( n,0) , ~(n-1/2,1/2) $ and energy $\hat E = 2
n$.

\section{ Formulas from \cite{llm} }
\renewcommand{\theequation}{H.\arabic{equation}}
\setcounter{equation}{0} \label{formulasllm}

For completeness we review the formulas for the general form of the solutions
in \cite{llm}.

The IIB ansatz is
\begin{eqnarray}
ds_{10}^{2} &=&-\frac{2y}{\sqrt{1-4z^{2}}}(dt+V)^{2}
+y\sqrt{\frac{1+2z}{1-2z}}d\Omega_{3}^{2}+y\sqrt{\frac{1-2z}{1+2z}}%
d\tilde{\Omega}_{3}^{2}+\frac{\sqrt{1-4z^{2}}}{2y}(dy^{2}+dx^{i}dx^{i})   \notag \\
F_{5} &=& -{1\over4}\left\{d[y^2 {1+2z \over 1-2z}(dt +V)]+y^{3}\ast
_{3}d({\frac{z+{\frac{1}{2}}}{y^{2}}}) \right\}\wedge d^3 \Omega \notag \\
&& -{1\over4}\left\{d[y^2 {1-2z \over 1+2z}(dt +V)]+y^{3}\ast
_{3}d({\frac{z-{\frac{1}{2}}}{y^{2}}})\right\}\wedge
d^{3}{\tilde{\Omega}}  \label{IIBans}
\end{eqnarray}%
where $dV =\frac{1}{y}\ast _{3}dz$,  $i=1,2$ and $*_3$ is the flat
space epsilon symbol in the three dimensions parametrized by
$y,x_1,x_2$. The
 function $z$  obeys the  equation
 \be
 \partial_i \partial_i z + y \partial_y ( { \partial_y z \over y} ) =0
\label{zequation} \ee

The M theory ansatz is
\begin{eqnarray}
ds_{11}^{2} &=&-{4e^{2\lambda }(1+y^{2}e^{-6\lambda })}(dt+V)^{2}+
4e^{2\lambda }d\Omega _{5}^{2}+y^{2}e^{-4\lambda
}d{\tilde{\Omega}}_{2}^{2}+ \frac{e^{-4\lambda
}}{1+y^{2}e^{-6\lambda }}(dy^{2}+e^{D}dx^{i}dx^{i})\notag \\
G_4 &=&\left\{ -4d[y^{3}e^{-6\lambda
}(dt+V)]+2\tilde{\ast}_{3}[y^{2}\partial
_{y}({\frac{1}{y}}\partial_{y}e^{D})dy+y\partial _{i}\partial
_{y}Ddx^{i}] \right\} \wedge d^{2}{\tilde{\Omega}}\label{Mansatz2}
\end{eqnarray}%
where $V_{i} ={\frac{1}{2}}\epsilon _{ij}\partial _{j}D$,
$e^{-6\lambda }= {\frac{\partial _{y}D}{y(1-y\partial _{y}D)}}$
and $\tilde{\ast}_{3}$ is the 3d flat space $\epsilon $ symbol.
The function $D$ obeys
\be
\partial_{i}\partial_{i} D + \partial_y^2 e^D =0
\ee


\begin{thebibliography}{99}



\bibitem{wittenindex}
  E.~Witten,
  Nucl.\ Phys.\ B {\bf 202}, 253 (1982).

\bibitem{bmn}
D.~Berenstein, J.~M.~Maldacena and H.~Nastase,
JHEP \textbf{0204}, 013 (2002), hep-th/0202021.

\bibitem{mssvr}
J.~Maldacena, M.~M.~Sheikh-Jabbari and M.~Van
Raamsdonk, 
JHEP \textbf{0301}, 038 (2003), hep-th/0211139.


\bibitem{markvr1}
K.~Dasgupta, M.~M.~Sheikh-Jabbari and M.~Van Raamsdonk,
  JHEP {\bf 0205}, 056 (2002)
  [arXiv:hep-th/0205185].


\bibitem{kimplefka}
  N.~Kim and J.~Plefka,
  Nucl.\ Phys.\ B {\bf 643}, 31 (2002)
  [arXiv:hep-th/0207034].

\bibitem{markvr2}
K.~Dasgupta, M.~M.~Sheikh-Jabbari and M.~Van Raamsdonk,
  JHEP {\bf 0209}, 021 (2002)
  [arXiv:hep-th/0207050].



\bibitem{kimpp}
  N.~Kim and J.~H.~Park,
  Phys.\ Rev.\ D {\bf 66}, 106007 (2002)
  [arXiv:hep-th/0207061].



\bibitem{plefka}
N.~Kim, T.~Klose and J.~Plefka,
Nucl.\ Phys.\ B \textbf{671}, 359 (2003) [arXiv:hep-th/0306054].
T.~Klose and J.~Plefka,
Nucl.\ Phys.\ B \textbf{679}, 127 (2004) [arXiv:hep-th/0310232].


\bibitem{fourloop}
T.~Fischbacher, T.~Klose and J.~Plefka,
JHEP \textbf{0502}, 039 (2005) [arXiv:hep-th/0412331].


\bibitem{llm}
  H.~Lin, O.~Lunin and J.~Maldacena,
  JHEP {\bf 0410}, 025 (2004)
  [arXiv:hep-th/0409174]. In appendix \ref{formulasllm} we reproduce
  the main formulas in \cite{llm}.


\bibitem{iiappwave1}
K.~Sugiyama and K.~Yoshida,
  Nucl.\ Phys.\ B {\bf 644}, 128 (2002)
  [arXiv:hep-th/0208029].

\bibitem{iiappwave2}
S.~Hyun and H.~Shin,
JHEP \textbf{0210}, 070 (2002) [arXiv:hep-th/0208074].
 S.~Hyun and H.~Shin,
  Nucl.\ Phys.\ B {\bf 654}, 114 (2003)
  [arXiv:hep-th/0210158].




\bibitem{Callan:1991at}
  C.~G.~Callan, J.~A.~Harvey and A.~Strominger,
  Nucl.\ Phys.\ B {\bf 359}, 611 (1991).
  C.~G.~Callan, J.~A.~Harvey and A.~Strominger,
  arXiv:hep-th/9112030.

\bibitem{Berenstein:2004kk}
  D.~Berenstein,
  JHEP {\bf 0407}, 018 (2004)
  [arXiv:hep-th/0403110].

\bibitem{Corley}
  S.~Corley, A.~Jevicki and S.~Ramgoolam,
  Adv.\ Theor.\ Math.\ Phys.\  {\bf 5}, 809 (2002)
  [arXiv:hep-th/0111222].


\bibitem{Haag:1974qh}
R.~Haag, J.~T.~Lopuszanski and M.~Sohnius,
Nucl.\ Phys.\ B \textbf{88}, 257 (1975). 


\bibitem{weinberg}
  S.~Weinberg,
  ``The quantum theory of fields.  Vol. 3: Supersymmetry.''


\bibitem{nahm}
  W.~Nahm,
  Nucl.\ Phys.\ B {\bf 135}, 149 (1978).


\bibitem{Blau:2001ne}
  M.~Blau, J.~Figueroa-O'Farrill, C.~Hull and G.~Papadopoulos,
  JHEP {\bf 0201}, 047 (2002)
  [arXiv:hep-th/0110242].




\bibitem{justin}
J.~Kinney, J.~Maldacena, S.~Minwalla and S. Raju, to appear.

\bibitem{minwalla}
  O.~Aharony, J.~Marsano, S.~Minwalla, K.~Papadodimas and M.~Van Raamsdonk,
  Adv.\ Theor.\ Math.\ Phys.\  {\bf 8}, 603 (2004)
  [arXiv:hep-th/0310285].



\bibitem{Frolov}
  S.~Frolov and A.~A.~Tseytlin,
  Nucl.\ Phys.\ B {\bf 668}, 77 (2003)
  [arXiv:hep-th/0304255].

\bibitem{MinahanVE}
  J.~A.~Minahan and K.~Zarembo,
  JHEP {\bf 0303}, 013 (2003)
  [arXiv:hep-th/0212208].


\bibitem{beisert}
N. Beisert, talk at Strings 2005,
http://www.fields.utoronto.ca/audio/05-06/strings/beisert.


\bibitem{gmmd}
 M.~R.~Douglas and G.~W.~Moore,
  arXiv:hep-th/9603167.

\bibitem{Horowitz:2001uh}
  G.~T.~Horowitz and T.~Jacobson,
  JHEP {\bf 0201}, 013 (2002)
  [arXiv:hep-th/0112131].



\bibitem{Boulatov:1991fp}
  D.~Boulatov and V.~Kazakov,
  Nucl.\ Phys.\ Proc.\ Suppl.\  {\bf 25A}, 38 (1992).

\bibitem{ward}
R.~S.~Ward,
Class.\ Quant.\ Grav.\ \textbf{7}, L95 (1990). 


\bibitem{null}
  D.~Bak, S.~Siwach and H.~U.~Yee,
  arXiv:hep-th/0504098.

\bibitem{Itzhaki:1998dd}
  N.~Itzhaki, J.~M.~Maldacena, J.~Sonnenschein and S.~Yankielowicz,
  Phys.\ Rev.\ D {\bf 58}, 046004 (1998)
  [arXiv:hep-th/9802042].


\bibitem{mm}
  J.~Maldacena and L.~Maoz,
  JHEP {\bf 0212}, 046 (2002)
  [arXiv:hep-th/0207284].


\bibitem{rt}
  J.~G.~Russo and A.~A.~Tseytlin,
  JHEP {\bf 0209}, 035 (2002)
  [arXiv:hep-th/0208114].

\bibitem{bs}
  I.~Bakas and J.~Sonnenschein,
  JHEP {\bf 0212}, 049 (2002)
  [arXiv:hep-th/0211257].

\bibitem{gmjmns}
  J.~M.~Maldacena, G.~W.~Moore and N.~Seiberg,
  JHEP {\bf 0110}, 005 (2001)
  [arXiv:hep-th/0108152].
J.~M.~Maldacena, G.~W.~Moore and N.~Seiberg,
  JHEP {\bf 0111}, 062 (2001)
  [arXiv:hep-th/0108100].

\bibitem{Polchinski}
  J.~Polchinski,
  ``String theory. Vol. 2: Superstring theory and beyond.''




\bibitem{lst1}
M.~Berkooz, M.~Rozali and N.~Seiberg,
  Phys.\ Lett.\ B {\bf 408}, 105 (1997)
  [arXiv:hep-th/9704089].
  N.~Seiberg,
  Phys.\ Lett.\ B {\bf 408}, 98 (1997)
  [arXiv:hep-th/9705221].


\bibitem{lst2}
  O.~Aharony, M.~Berkooz, D.~Kutasov and N.~Seiberg,
  JHEP {\bf 9810}, 004 (1998)
  [arXiv:hep-th/9808149].




\bibitem{Chong:2004ce}
  Z.~W.~Chong, H.~Lu and C.~N.~Pope,
  Phys.\ Lett.\ B {\bf 614}, 96 (2005)
  [arXiv:hep-th/0412221].





\bibitem{sz}
  A.~Santambrogio and D.~Zanon,
  Phys.\ Lett.\ B {\bf 545}, 425 (2002)
  [arXiv:hep-th/0206079].


\bibitem{Gross:2002su}
  D.~J.~Gross, A.~Mikhailov and R.~Roiban,
  Annals Phys.\  {\bf 301}, 31 (2002)
  [arXiv:hep-th/0205066].


\bibitem{3quaters}
  S.~S.~Gubser, I.~R.~Klebanov and A.~W.~Peet,
  Phys.\ Rev.\ D {\bf 54}, 3915 (1996)
  [arXiv:hep-th/9602135].

\bibitem{3loops}
  C.~G.~Callan, H.~K.~Lee, T.~McLoughlin, J.~H.~Schwarz, I.~Swanson and X.~Wu,
  Nucl.\ Phys.\ B {\bf 673}, 3 (2003)
  [arXiv:hep-th/0307032].
  D.~Serban and M.~Staudacher,
  JHEP {\bf 0406}, 001 (2004)
  [arXiv:hep-th/0401057].





\bibitem{zurab}
  M.~Bershadsky, Z.~Kakushadze and C.~Vafa,
  Nucl.\ Phys.\ B {\bf 523}, 59 (1998)
  [arXiv:hep-th/9803076].



\bibitem{Ooguri:1995wj}
  H.~Ooguri and C.~Vafa,
  Nucl.\ Phys.\ B {\bf 463}, 55 (1996)
  [arXiv:hep-th/9511164].





  \bibitem{S5orbifold}
  S.~Kachru and E.~Silverstein,
  Phys.\ Rev.\ Lett.\  {\bf 80}, 4855 (1998)
  [arXiv:hep-th/9802183].
   A.~E.~Lawrence, N.~Nekrasov and C.~Vafa,
  Nucl.\ Phys.\ B {\bf 533}, 199 (1998)
  [arXiv:hep-th/9803015].



\bibitem{Gregory:1997te}
  R.~Gregory, J.~A.~Harvey and G.~W.~Moore,
  Adv.\ Theor.\ Math.\ Phys.\  {\bf 1}, 283 (1997)
  [arXiv:hep-th/9708086].



\bibitem{pporbifold1}
  M.~Alishahiha and M.~M.~Sheikh-Jabbari,
  Phys.\ Lett.\ B {\bf 535}, 328 (2002)
  [arXiv:hep-th/0203018].
  N.~Kim, A.~Pankiewicz, S.~J.~Rey and S.~Theisen,
  Eur.\ Phys.\ J.\ C {\bf 25}, 327 (2002)
  [arXiv:hep-th/0203080].
  T.~Takayanagi and S.~Terashima,
  JHEP {\bf 0206}, 036 (2002)
  [arXiv:hep-th/0203093].
  E.~Floratos and A.~Kehagias,
  JHEP {\bf 0207}, 031 (2002)
  [arXiv:hep-th/0203134].

\bibitem{pporbifold2}
 M.~Alishahiha, M.~M.~Sheikh-Jabbari and R.~Tatar,
  JHEP {\bf 0301}, 028 (2003)
  [arXiv:hep-th/0211285].


\bibitem{Mukhi:2002ck}
  S.~Mukhi, M.~Rangamani and E.~P.~Verlinde,
  JHEP {\bf 0205}, 023 (2002)
  [arXiv:hep-th/0204147].



\bibitem{EH}
  R.~Clarkson and R.~B.~Mann,
  arXiv:hep-th/0508109; hep-th/0508200.

\bibitem{mannads}
  D.~Astefanesei, R.~B.~Mann and C.~Stelea,
  arXiv:hep-th/0508162.





\bibitem{aps}
  A.~Adams, J.~Polchinski and E.~Silverstein,
  JHEP {\bf 0110}, 029 (2001)
  [arXiv:hep-th/0108075].

\bibitem{eguchi}
  T.~Eguchi and A.~J.~Hanson,
  Phys.\ Lett.\ B {\bf 74}, 249 (1978).


\bibitem{david}
  D.~Berenstein,
  arXiv:hep-th/0507203.
 D.~Berenstein, D.~H.~Correa and S.~E.~Vazquez,
  arXiv:hep-th/0509015.




\bibitem{P-S}
J.~Polchinski and M.~J.~Strassler,
hep-th/0003136. 

\bibitem{Hai}
H.~Lin,
JHEP \textbf{0412}, 001 (2004) [arXiv:hep-th/0407250].


\bibitem{barsetal}
  A.~Baha Balantekin and I.~Bars,
  J.\ Math.\ Phys.\  {\bf 22}, 1149 (1981).
  A.~Baha Balantekin and I.~Bars,
  J.\ Math.\ Phys.\  {\bf 22}, 1810 (1981).
  A.~Baha Balantekin and I.~Bars,
  J.\ Math.\ Phys.\  {\bf 23}, 1239 (1982).
  I.~Bars, B.~Morel and H.~Ruegg,
  J.\ Math.\ Phys.\  {\bf 24}, 2253 (1983).

\bibitem{otherrep}
   V.~G.~Kac,
  Adv.\ Math.\  {\bf 26}, 8 (1977).
  V.~G.~Kac,
  Commun.\ Math.\ Phys.\  {\bf 53} (1977) 31.



\bibitem{kaccharac}
V.G. Kac, ``Representations of Classical Lie Superalgebras", in
{\it Differential Geometrical Methods in Mathematical Physics},
Edited by K. Bleuler, H.R. Petry and A. Reetz (Springer-Verlag,
Berlin, 1978).





\bibitem{fuzzysphere}
 B.~de Wit, J.~Hoppe and H.~Nicolai,
  Nucl.\ Phys.\ B {\bf 305}, 545 (1988).



\bibitem{beisertsusy}
  N.~Beisert,
  Nucl.\ Phys.\ B {\bf 659}, 79 (2003)
  [arXiv:hep-th/0211032].


\bibitem{beisertnfour}
  N.~Beisert,
  Nucl.\ Phys.\ B {\bf 676}, 3 (2004)
  [arXiv:hep-th/0307015].






\bibitem{freedmanag}
  L.~Alvarez-Gaume and D.~Z.~Freedman,
  Commun.\ Math.\ Phys.\  {\bf 91}, 87 (1983).

\bibitem{jourjine}
  A.~N.~Jourjine,
  Nucl.\ Phys.\ B {\bf 236}, 181 (1984).
  A.~N.~Jourjine,
  Annals Phys.\  {\bf 157}, 489 (1984).



\bibitem{Taylor}
W.~I.~Taylor,
Phys.\ Lett.\ B \textbf{394}, 283 (1997) [arXiv:hep-th/9611042].

\bibitem{BFSS}
 T.~Banks, W.~Fischler, S.~H.~Shenker and L.~Susskind,
  Phys.\ Rev.\ D {\bf 55}, 5112 (1997)
  [arXiv:hep-th/9610043].
 L.~Susskind,
  arXiv:hep-th/9704080.


\bibitem{banks}
T.~Banks and N.~Seiberg,
Nucl.\ Phys.\ B {\bf 497}, 41 (1997), hep-th/9702187.
S.~Sethi and L.~Susskind,
Phys.\ Lett.\ B {\bf 400}, 265 (1997), hep-th/9702101.

\bibitem{Sheikh-Jabbari:2004ik}
  M.~M.~Sheikh-Jabbari,
  JHEP {\bf 0409}, 017 (2004)
  [arXiv:hep-th/0406214].
  M.~M.~Sheikh-Jabbari and M.~Torabian,
  JHEP {\bf 0504}, 001 (2005)
  [arXiv:hep-th/0501001].



\bibitem{Pope:2003jp}
  C.~N.~Pope and N.~P.~Warner,
  JHEP {\bf 0404}, 011 (2004)
  [arXiv:hep-th/0304132].

\bibitem{Bena:2004jw}
  I.~Bena and N.~P.~Warner,
  JHEP {\bf 0412}, 021 (2004)
  [arXiv:hep-th/0406145].


\bibitem{ensemble}
  V.~Balasubramanian, J.~de Boer, V.~Jejjala and J.~Simon,
  arXiv:hep-th/0508023.
  V.~Balasubramanian, V.~Jejjala and J.~Simon,
  arXiv:hep-th/0505123.
P.~J.~Silva,
  arXiv:hep-th/0508081.
P.~G.~Shepard,
  arXiv:hep-th/0507260.
 P.~Horava and P.~G.~Shepard,
  JHEP {\bf 0502}, 063 (2005)
  [arXiv:hep-th/0502127].
  N.~V.~Suryanarayana,
  arXiv:hep-th/0411145.
M.~M.~Caldarelli, D.~Klemm and P.~J.~Silva,
  arXiv:hep-th/0411203.
  S.~S.~Gubser and J.~J.~Heckman,
  JHEP {\bf 0411}, 052 (2004)
  [arXiv:hep-th/0411001].
  A.~Buchel,
  arXiv:hep-th/0409271.
  G.~Milanesi and M.~O'Loughlin,
  arXiv:hep-th/0507056.
 M.~Alishahiha, H.~Ebrahim, B.~Safarzadeh and M.~M.~Sheikh-Jabbari,
  JHEP {\bf 0511}, 005 (2005)
  [arXiv:hep-th/0509160].




\bibitem{Khuri:1993ii}
  R.~R.~Khuri,
  Phys.\ Rev.\ D {\bf 48}, 2947 (1993)
  [arXiv:hep-th/9305143].



\bibitem{Gauntlett:1996pb}
  J.~P.~Gauntlett, D.~A.~Kastor and J.~H.~Traschen,
  Nucl.\ Phys.\ B {\bf 478}, 544 (1996)
  [arXiv:hep-th/9604179].

\bibitem{itzhakiseiberg}
  N.~Itzhaki, D.~Kutasov and N.~Seiberg,
  arXiv:hep-th/0508025.






\bibitem{levelrankdual}
  S.~G.~Naculich, H.~A.~Riggs and H.~J.~Schnitzer,
  Phys.\ Lett.\ B {\bf 246} (1990) 417.
  A.~Kuniba and T.~Nakanishi,
PRINT-90-0182 (KYUSHU)
{\it FGor Proc. of Int. Colloq. on Modern Quantum Field Theory,
Bombay, India, Jan 8-14,1990.}
 M.~Camperi, F.~Levstein and G.~Zemba,
  Phys.\ Lett.\ B {\bf 247}, 549 (1990).


\bibitem{liat}
  L.~Maoz and V.~S.~Rychkov,
  arXiv:hep-th/0508059.
  L.~Grant, L.~Maoz, J.~Marsano, K.~Papadodimas and V.~S.~Rychkov,
  arXiv:hep-th/0505079.





\bibitem{mandaletal}
G.~Mandal,
  JHEP {\bf 0508}, 052 (2005)
  [arXiv:hep-th/0502104].
  A.~Dhar,
  JHEP {\bf 0507}, 064 (2005)
  [arXiv:hep-th/0505084].
  A.~Dhar, G.~Mandal and N.~V.~Suryanarayana,
  arXiv:hep-th/0509164.
\bibitem{vazquez}
  H.~Lu, C.~N.~Pope and J.~F.~Vazquez-Poritz,
  Nucl.\ Phys.\ B {\bf 709}, 47 (2005)
  [arXiv:hep-th/0307001].
  D.~Astefanesei and G.~C.~Jones,
  JHEP {\bf 0506}, 037 (2005)
  [arXiv:hep-th/0502162].
  C.~h.~Ahn and J.~F.~Vazquez-Poritz,
  arXiv:hep-th/0508075.



\bibitem{nsnb}
  N.~Berkovits and N.~Seiberg,
  JHEP {\bf 0307}, 010 (2003)
  [arXiv:hep-th/0306226].

\bibitem{strass}
 E.~G.~Gimon, L.~A.~Pando Zayas, J.~Sonnenschein and M.~J.~Strassler,
  JHEP {\bf 0305}, 039 (2003)
  [arXiv:hep-th/0212061].


\bibitem{ggsh}
  G.~W.~Gibbons and S.~W.~Hawking,
  Phys.\ Lett.\ B {\bf 78} (1978) 430.

\bibitem{brfr}
  P.~Breitenlohner and D.~Z.~Freedman,
  Annals Phys.\  {\bf 144}, 249 (1982).



\bibitem{Duff:1999gh}
  M.~J.~Duff and J.~T.~Liu,
  Nucl.\ Phys.\ B {\bf 554}, 237 (1999)
  [arXiv:hep-th/9901149].


\bibitem{Cvetic:1999ne}
  M.~Cvetic and S.~S.~Gubser,
  JHEP {\bf 9904}, 024 (1999)
  [arXiv:hep-th/9902195].

\bibitem{Liu:1999ai}
  J.~T.~Liu and R.~Minasian,
  Phys.\ Lett.\ B {\bf 457}, 39 (1999)
  [arXiv:hep-th/9903269].


\bibitem{jeremy}
  S.~R.~Das, J.~Michelson and A.~D.~Shapere,
  Phys.\ Rev.\ D {\bf 70}, 026004 (2004)
  [arXiv:hep-th/0306270].



\bibitem{polyakov}
  A.~M.~Polyakov,
  Int.\ J.\ Mod.\ Phys.\ A {\bf 17S1}, 119 (2002)
  [arXiv:hep-th/0110196].












\end{thebibliography}
\end{document}